\documentclass{cernrep}
\usepackage{feynmp}
\usepackage{xspace}
\usepackage{a4wide}
\usepackage{booktabs}
\usepackage[small, nooneline, center]{subfig}
\usepackage{amsmath,amssymb,bm}
\usepackage{longtable}
\usepackage[colorlinks=true, pdftex, pdftitle={Quantum Chromodynamics for Collider Physics}, pdfauthor={Peter Zeiler Skands}, pdfsubject={Monte Carlo generators}, pdfkeywords={Monte Carlo, generators, QCD, Minimum Bias, HEP}]{hyperref}
\usepackage{graphicx}
\usepackage{color}
\usepackage{cite}
\definecolor{lightgray}{rgb}{0.85,0.85,0.87}
\definecolor{gray}{rgb}{0.5,0.5,0.5}
\definecolor{darkgray}{rgb}{0.36,0.36,0.36}
\def\lsim{\mathrel{\rlap{\lower4pt\hbox{\hskip1pt$\sim$}}
    \raise1pt\hbox{$<$}}}                
\def\gsim{\mathrel{\rlap{\lower4pt\hbox{\hskip1pt$\sim$}}
    \raise1pt\hbox{$>$}}}                
\renewcommand{\d}[1]{\ensuremath{\mrm{d}#1}\hspace*{0.2em} }
\newcommand{\drm}{\ensuremath{\mathrm{d}}}

\newcommand{\mrm}[1]{\ensuremath{\mathrm{#1}}}

\newcommand{\obs}{\ensuremath{\mathcal{O}}}

\newcommand{\pt}[1][]{\ensuremath{p_{\perp #1}}}
\newcommand{\pthat}[1][]{\ensuremath{\hat{p}_{\perp #1}}}
\newcommand{\ptmin}[1][]{\pt[\mrm{min}#1]\xspace}
\newcommand{\pdf}[2][]{\ensuremath{f_{#2}^{#1}}}

\newcommand{\pqcd}[2][]{\ensuremath{\sigma^{(#1)}_{#2}}}
\newcommand{\PS}[2][]{\ensuremath{\Phi_{#2}^{#1}}\xspace}
\newcommand{\dPS}[2][]{\ensuremath{\d{\PS[#1]{#2}}}\xspace}

\newcommand{\tsc}[1]{\textsc{#1}}
\newcommand{\ttt}[1]{\texttt{#1}}
\newcommand{\eqRef}[1]{equation~\eqref{#1}\xspace}

\newcommand{\eqsRef}[1]{equations~\eqref{#1}\xspace}

\newcommand{\secRef}[1]{section~\ref{#1}\xspace}
\newcommand{\SecRef}[1]{Section~\ref{#1}\xspace}
\newcommand{\secsRef}[1]{sections~\ref{#1}\xspace}

\newcommand{\TabRef}[1]{Table~\ref{#1}\xspace}
\newcommand{\figRef}[1]{figure~\ref{#1}\xspace}
\newcommand{\FigRef}[1]{Figure~\ref{#1}\xspace}

\newcommand{\figsRef}[1]{figures~\ref{#1}\xspace}
\newcommand{\FigsRef}[1]{Figures~\ref{#1}\xspace}

\newcommand{\Fw}{\tsc{Mc@nlo}\xspace}
\newcommand{\Hw}{\tsc{Herwig}\xspace}

\newcommand{\Py}{\tsc{Pythia}\xspace}
\newcommand{\Sh}{\tsc{Sherpa}\xspace}
\newcommand{\Vc}{\tsc{Vincia}\xspace}
\newcommand{\Pw}{\tsc{Powheg}\xspace}
\newlength{\tabcolsepsave}
\newenvironment{loopsnlegs}[1][t]{
\setlength{\tabcolsepsave}{\tabcolsep}
\setlength{\tabcolsep}{0pt}
\begin{tabular}{cc}\parbox[c]{1.1em}{\rotatebox{90}{\small $\ell$ (loops)}}&%
\begin{tabular}[#1]}{
\end{tabular}\\[-1mm]
 & \small $k$ (legs)
\end{tabular}%
\setlength{\tabcolsep}{\tabcolsepsave}
}
\newcommand{\cbox}[2]{%
\begin{minipage}[c]{1.4cm}%
\center%
{%
\parbox[c]{1.4cm}{\includegraphics*[width=1.4cm]{#1.pdf}}}%
\end{minipage}%
\hspace*{-1.4cm}%
\begin{minipage}[c]{1.4cm}
\center
#2%
\end{minipage}}
\newcommand{\cyanbox}[1]{\cbox{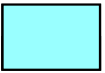}{#1}}
\newcommand{\eggbox}[1]{\cbox{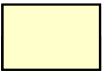}{#1}}
\newcommand{\gbox}[1]{\cbox{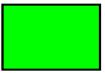}{#1}}
\newcommand{\wbox}[1]{\cbox{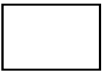}{#1}}
\newcommand{\gwbox}[1]{\cbox{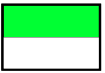}{#1}}
\newcommand{\gybox}[1]{\cbox{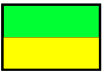}{#1}}
\newcommand{\gywbox}[1]{\cbox{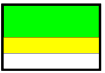}{#1}}
\newcommand{\ybox}[1]{\cbox{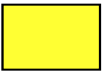}{#1}}
\newcommand{\ywbox}[1]{\cbox{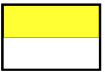}{#1}}
\newcommand{\rybox}[1]{\cbox{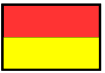}{#1}}
\newcommand{\wybox}[1]{\cbox{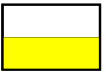}{#1}}
\newcommand{\wywbox}[1]{\cbox{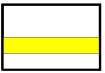}{#1}}
\newcommand{\wwybox}[1]{\cbox{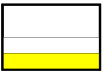}{#1}}
\begin{document}
\title{QCD for Collider Physics}
\author{P.Z. Skands\thanks
                 {peter.skands@cern.ch}}
\institute{Theoretical Physics, CERN, CH-1211 Geneva 23, Switzerland}
\maketitle
\begin{abstract}
These lectures are directed at a level suitable for 
graduate students in experimental and theoretical High Energy
Physics. They are intended to give an introduction to 
the theory and phenomenology of quantum  chromodynamics (QCD) as it
is used in collider physics applications. The aim is to bring the
reader to a level where informed decisions can be made 
concerning different approaches and their uncertainties. 
The material is divided into four main
areas:  1)~fundamentals, 2)~perturbative QCD, 
3)~soft QCD,  and 4)~Monte Carlo
event generators. \\
\end{abstract}
\tableofcontents
\clearpage%

\section{Introduction}

When
probed at very short wavelengths, QCD is essentially a theory of free `partons'
--- quarks and gluons --- which only scatter off one
another through relatively small quantum corrections, that can be
systematically calculated. 
At longer wavelengths, of order the size of the proton $\sim
1\mathrm{fm} = 10^{-15}\mathrm{m}$,  however, 
we see strongly bound towers of hadron resonances emerge, with string-like
potentials building up if we try to separate their partonic
constituents. Due to our
inability to solve strongly coupled field theories, QCD is therefore 
still only partially solved. Nonetheless,  all its features, across all
distance scales, are believed to be encoded in a single one-line
formula of alluring simplicity; the
Lagrangian of QCD.

The consequence for collider physics is that some parts of QCD can be
calculated in terms of the fundamental parameters of the Lagrangian,
whereas others must be expressed through models or functions whose effective 
parameters are not a priori calculable but which can be constrained
by fits to data. However, even in the absence of a
perturbative expansion, there are still several strong theorems which
hold, and which can be used to give relations between seemingly
different processes. (This is, e.g., the reason it makes sense to 
constrain parton distribution functions in $ep$ collisions and
then re-use the same ones for $pp$ collisions.) Thus, in the chapters
dealing with phenomenological models we shall emphasize that the loss
of a factorized perturbative expansion is not equivalent to a total
loss of predictivity.   

An alternative approach would be to give up on calculating QCD altogether 
and use leptons instead. Formally, this amounts to summing inclusively over
strong-interaction phenomena, when such are present. While such a
strategy might succeed in replacing what we do know about QCD by
``unity'', however, even the most adamant ``chromophobe'' must acknowledge
a few basic facts of collider physics for the next decade(s): 
1) At the Tevatron and LHC, the initial states \emph{are} 
unavoidably hadrons, and hence, at
the very least, well-understood and precise parton distribution
functions (PDFs) will be required; 2) high precision will mandate
 calculations to higher orders in perturbation theory, 
which in turn will involve more QCD; 3) the requirement of lepton
\emph{isolation} makes the very definition of a lepton
 depend implicitly on QCD, and 4) 
 the rate of jets that are misreconstructed as leptons in
 the experiment depends explicitly on it. Finally, 5) though many
 new-physics signals \emph{do} give observable signals in the lepton
 sector, this is far from guaranteed. 
 It would therefore be 
 unwise not to attempt to solve QCD to the best of our ability, the
 better to prepare ourselves for both the largest possible discovery
 reach and the highest attainable subsequent precision.  

In the following, we shall focus squarely on QCD for mainstream 
collider physics. This 
includes factorization, hard processes, infrared safety, parton showers and
matching, event generators, hadronization, and the so-called underlying event. 
While not covering everything, hopefully these topics can also serve
at least as stepping stones to more specialized
issues that have been left out, such as heavy flavours or forward
physics, or to topics more tangential to other fields, such as lattice
QCD or heavy-ion physics.  

\subsection{A First Hint of Colour}
\begin{figure}[t]
\begin{center}
\begin{tabular}{c}
\colorbox{gray}{\includegraphics*[scale=0.75]{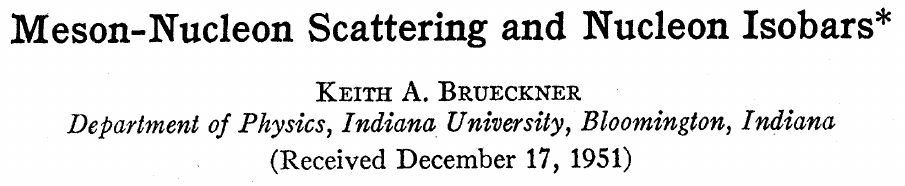}}\\[5mm]
\hspace*{2mm}\begin{minipage}{0.88\textwidth}
\small\sl  ``[...] It is concluded that the apparently anomalous features of the
scattering can be interpreted to be an indication of a resonant
meson-nucleon interaction corresponding to a nucleon isobar with spin
$\frac32$, isotopic spin $\frac32$, and with an excitation energy of
$277\,$MeV.''\\[1mm]
\end{minipage}
\end{tabular}
\caption{The title and part of the abstract of the 1951 paper
  \cite{Brueckner:1952zz} (published in 1952) in which the discovery
  of the $\Delta^{++}$ baryon was announced.\label{fig:Delta}}  
\end{center}
\end{figure}

Looking for new physics, as we do now at the LHC, it is instructive to 
consider the story of the discovery of colour. The first hint was
arguably the $\Delta^{++}$ baryon, found in 1951 
\cite{Brueckner:1952zz}. The title and part of the abstract from this
historical paper are reproduced in \figRef{fig:Delta}.
In the context of the quark model --- which first
had to be developed, successively joining together the notions of 
spin, isospin, strangeness, and 
the eightfold way --- the flavour and spin content of the $\Delta^{++}$
baryon is: 
\begin{equation}
\left\vert \Delta^{++} \right> = \left\vert
\,u_\uparrow\ u_\uparrow\ u_\uparrow \right>~,
\end{equation} 
clearly a highly symmetric configuration. However, since 
the $\Delta^{++}$ is a fermion, it must have an overall
antisymmetric wave function. In 1965, fourteen years after its
discovery, this was finally understood by the introduction of colour
as a new quantum number associated with the group SU(3)
\cite{Greenberg:1964pe,Han:1965pf}. The $\Delta^{++}$ wave function can now be made
antisymmetric by arranging its three quarks antisymmetrically 
in this new degree of freedom, 
\begin{equation}
\left\vert \Delta^{++} \right> = \epsilon^{ijk} \left\vert
\,u_{i\uparrow}\ u_{j\uparrow}\ u_{k\uparrow}\right>~,
\end{equation} 
hence solving the mystery.

More direct experimental tests of the number of colours were provided first by
measurements of the decay width of $\pi^0\to \gamma\gamma$ decays, which 
is proportional to $N_C^2$, and later by the famous ``R'' ratio in
$e^+e^-$ collisions. Below, in \SecRef{sec:L} we shall see how to
calculate such colour factors. 

\subsection{The Lagrangian of QCD \label{sec:L}}
Quantum Chromodynamics is based on the gauge group $\mrm{SU(3)}$, the
Special Unitary group in 3 (complex) dimensions. 
In the context of QCD, we represent this group as a set
of unitary $3\times 3$ matrices with determinant one. This is called 
the \emph{adjoint} representation and can be used to represent gluons in
colour space. Since there are 9 linearly independent unitary complex
matrices, one of which has determinant $-1$, there are a total of 8
independent directions in the adjoint colour space, i.e., the gluons
are \emph{octets}. In QCD,  
these matrices can operate both on each other (gluon
self-interactions) and on a set of complex $3$-vectors (the  \emph{fundamental}
representation), the latter of which represent quarks in colour
space. The fundamental representation has one linearly
independent basis vector per degree of $\mrm{SU(3)}$, and hence 
the quarks are \emph{triplets}. 

The Lagrangian of QCD is 
\begin{equation}
{\cal L} = \bar{\psi}_q^i(i\gamma^\mu)(D_\mu)_{ij}\psi_q^j - m_q
\bar{\psi}_q^i\psi_{qi} - \frac14 F^a_{\mu\nu}F^{a\mu\nu}~,\label{eq:L}
\end{equation}
where $\psi_q^i$ denotes a quark field with
colour index $i$, 
$\psi_q = ({\textcolor{red}{\psi_{qR}}},{\color{green}\psi_{qG}}, 
{\color{blue}\psi_{qB}})^T$, 
$\gamma^\mu$ is a Dirac matrix that expresses the
vector nature of the strong interaction, with $\mu$ being a Lorentz
vector index, $m_q$ allows for the
possibility of non-zero quark masses (induced by the standard Higgs
mechanism or similar), $F^a_{\mu\nu}$ is the gluon field strength
tensor for a gluon with
colour index $a$ (in the adjoint representation, i.e.,
$a\in[1,\ldots,8]$), 
and $D_\mu$ is the covariant derivative in QCD,
\begin{equation}
(D_{\mu})_{ij} = \delta_{ij}\partial_\mu - i g_s t_{ij}^a A_\mu^a~,\label{eq:D}
\end{equation}
with $g_s$ the strong coupling (related to $\alpha_s$ by $g_s^2 = 4\pi
\alpha_s$; we return to the strong coupling in more detail below), 
$A^a_\mu$  the gluon field with (adjoint-representation)
colour index $a$, and $t_{ij}^a$ proportional to the hermitean and
traceless Gell-Mann matrices of $\mrm{SU(3)}$, 
\begin{equation}
\mbox{\includegraphics*[scale=1.0]{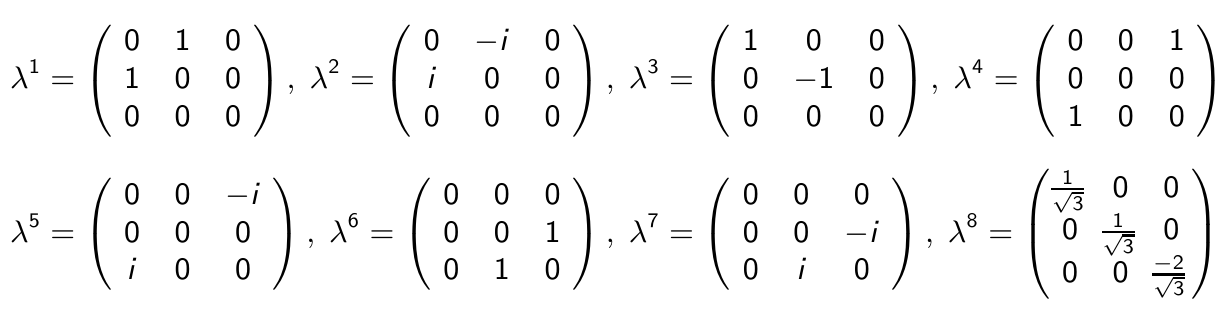}}~.
\end{equation}
These generators are just the $\mrm{SU(3)}$ analogs of the
Pauli matrices in 
$\mrm{SU(2)}$. By convention, the constant of proportionality is normally
taken to 
be\footnote{Another choice that is occasionally (though rarely) seen in the
literature is $t = \lambda / \sqrt{2}$. This gives a more intuitive colour
counting, but since it also implies a different normalization for the
coupling and since most text material uses the convention 
defined by \eqRef{eq:t}, we shall stick to that choice for the
remainder of these lectures.} 
\begin{equation}
t^a_{ij} = \frac12 \lambda^a_{ij}~. \label{eq:t}
\end{equation}
This choice in turn determines the normalization of the coupling
$g_s$, via \eqRef{eq:D}, and
fixes the values of the $\mrm{SU(3)}$ Casimirs and structure constants, to
which we return below. 

\begin{figure}[t]
\begin{center}
\begin{minipage}[h]{4.6cm}
\begin{center}
$A^1_\mu$\\
\includegraphics*[scale=0.75]{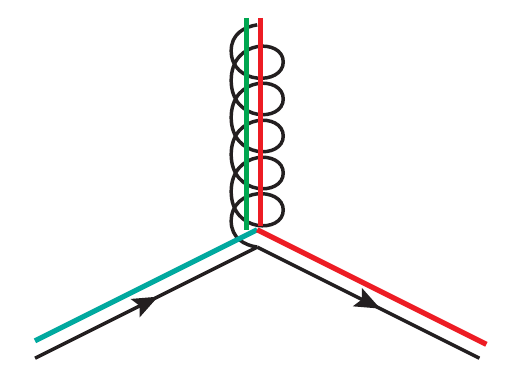}\\[-3mm]
$\psi_{q\textcolor{red}{R}}$\hfill$\psi_{q\textcolor{green}{G}}$
\end{center}
\end{minipage}~~~
\parbox{0.4\textwidth}{
$
\begin{array}{ccccc}
\propto & - \frac{i}{2} g_s & \bar{\psi}_{q\color{red}R}  & \lambda^{1} & \psi_{q\color{green}G} 
\\[2mm]
= & -\frac{i}{2}g_s & \left(\begin{array}{ccc} \textcolor{red}{1} & \color{green} 0 &
  \color{blue} 0 
\end{array}\right) & 
\left(\begin{array}{ccc}
0 & 1 & 0  \\
1 & 0 & 0 \\
0 & 0 & 0
\end{array}\right) & 
 \left(\begin{array}{c}
\textcolor{red}{0} \\
\color{green}1 \\
\color{blue}0
\end{array}\right) \end{array}
$}
\caption{Illustration of a $qqg$ vertex in QCD, before
  summing/averaging over colours: a gluon in a state represented by $\lambda^1$
  interacts with quarks in the states $\psi_{qR}$ and
  $\psi_{qG}$. \label{fig:qg}}
\end{center}
\end{figure}

An example of the colour flow for a quark-gluon interaction in colour
space is given in \figRef{fig:qg}.
Typically, however, we do not measure colour in the final state ---
instead we average over all possible incoming colours and sum over all
possible outgoing ones, wherefore QCD scattering amplitudes (squared) in
practice always contain sums over quark fields contracted with
Gell-Mann matrices. These contractions in turn produce traces 
which yield the \emph{colour factors} that are associated to each QCD
process, and which basically count the number of ``paths through
colour space'' that the process at hand can take, modulo that the
convention choice represented by \eqRef{eq:t} introduces a ``spurious''
factor of 2 for each power of the coupling $\alpha_s$, as we shall see\footnote{Again,
although one could in principle absorb that factor into a redefinition
of the coupling, effectively redefining the normalization of ``unit
colour charge'', the standard definition of $\alpha_s$ is now so
entrenched that alternative choices would be counter-productive, at
least in the context of a supposedly pedagogical review.}.

A very simple example of a colour factor is given by the decay process $Z\to
q\bar{q}$. This vertex contains a simple $\delta_{ij}$ in colour
space; the outgoing quark and antiquark must have identical 
(anti-)col\-ours. Squaring the corresponding matrix element and summing over
final-state colours yields a colour factor of
\begin{equation}
Z \to q\bar{q}~~~:~~~\sum_{\mrm{colours}}|M|^2 \propto
\delta_{ij}\delta^*_{ji} = \mrm{Tr}\{\delta\} = N_C = 3~,
\end{equation}
since $i$ and $j$ are quark (i.e., 3-dimensional
fundamental-representation) indices. 

A next-to-simplest example is given by the Drell-Yan process, $q\bar{q}\to
\gamma^*/Z$, i.e., just a crossing of the previous one. By crossing
symmetry, the squared matrix element, including the colour factor, is
exactly the same as before, but since the quarks are here incoming, we
must \emph{average} rather than sum over their colours, leading to
\begin{equation}
q\bar{q}\to Z~~~:~~~\frac{1}{9}\sum_{\mrm{colours}}|M|^2 \propto \frac19\delta_{ij}\delta^*_{ji} = \frac19 \mrm{Tr}\{\delta\} = \frac13~,
\end{equation}
where the colour factor now expresses a \emph{suppression} which can
be interpreted as due to the fact that only quarks of matching colours
are able to collide and produce a $Z$ boson, effectively reducing 
the incoming quark-antiquark flux by a factor 1/$N_C$.

To illustrate what happens when we insert (and sum over) quark-gluon
vertices, such as the one depicted in \figRef{fig:qg}, we take
the process $Z\to3\,$jets. The colour factor for this process can be
computed as follows, with the accompanying illustration showing a
corresponding diagram (squared) with explicit colour-space indices on
each vertex:\\
\begin{equation}
\mbox{
\begin{tabular}{cc}
\parbox{5.2cm}{
$Z \to qg\bar{q}$~~~:~~~\\
\[
\begin{array}{rcl}
\displaystyle\sum_{\mrm{colours}}|M|^2 & \propto & \displaystyle
\delta_{ij}t_{jk}^a\left(t_{\ell
    k}^a\delta^*_{i\ell}\right)^* \\
& = & \displaystyle
\mrm{Tr}\{t^at^a\}\\[4mm] & = & \displaystyle
  \frac12\mrm{Tr}\{\delta\} = 4~,
\end{array}
\]}
&
\parbox{8.5cm}{\includegraphics*[scale=0.6]{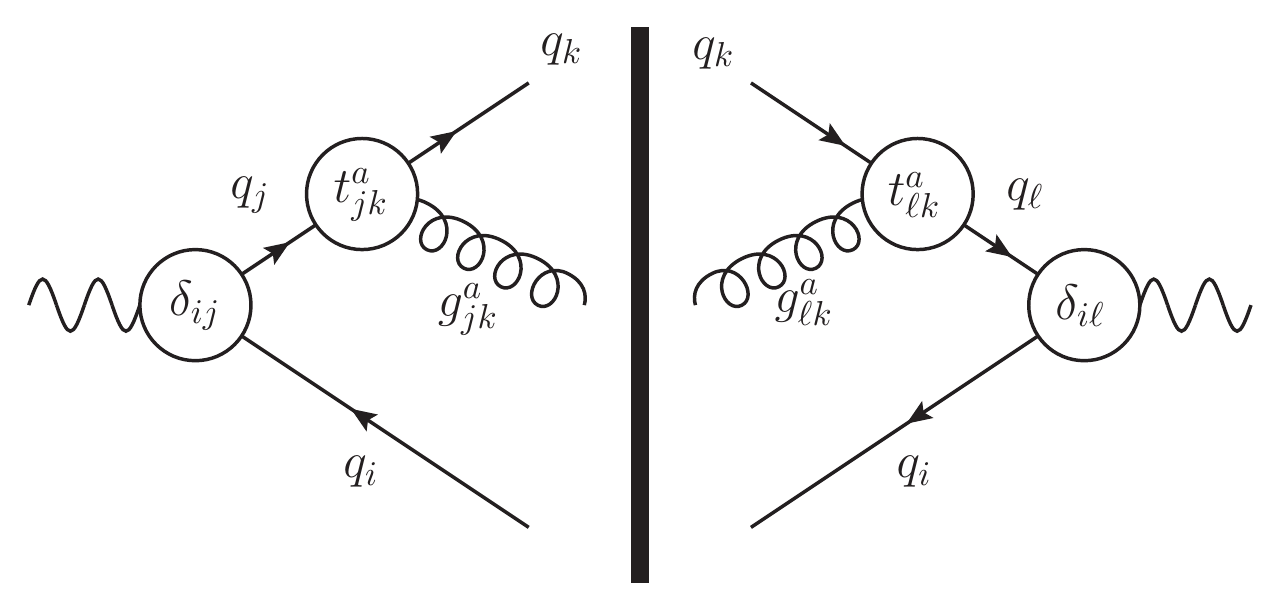}
}
\end{tabular}}
\end{equation}
where the last $\mrm{Tr}\{\delta\} = 8$, since the trace runs over
indices in the 8-dimensional adjoint representation.

The tedious task of taking traces over $\mrm{SU(3)}$ matrices can be greatly
alleviated by use of the relations given in \TabRef{tab:lambda}. 
\begin{table}
\begin{center}
\scalebox{1.04}{\begin{tabular}{ccc}
\toprule
Trace Relation & Indices & Occurs in Diagram Squared
\\
\midrule
$\mrm{Tr}\{t^at^b\} = T_R\, \delta^{ab}$ & $a,b\in[1,\ldots,8]$
& \parbox[c]{4cm}{\includegraphics*[scale=0.5]{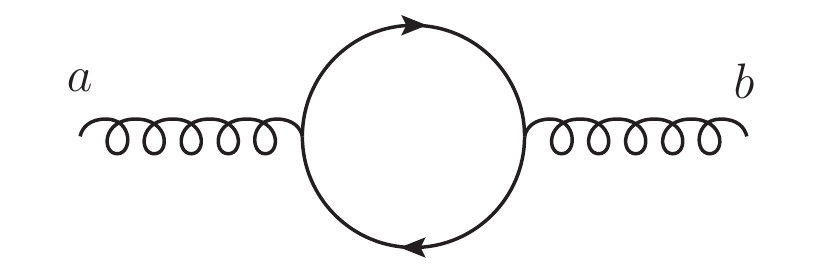}}\\
$\sum_a t^a_{ij}t^a_{jk} = C_F\, \delta_{ik}$ &%
\parbox[c]{3cm}{\begin{center}
$a\in[1,\ldots,8]$\\
$i,j,k\in[1,\ldots,3]$\end{center}}
& \parbox[c]{4cm}{\includegraphics*[scale=0.5]{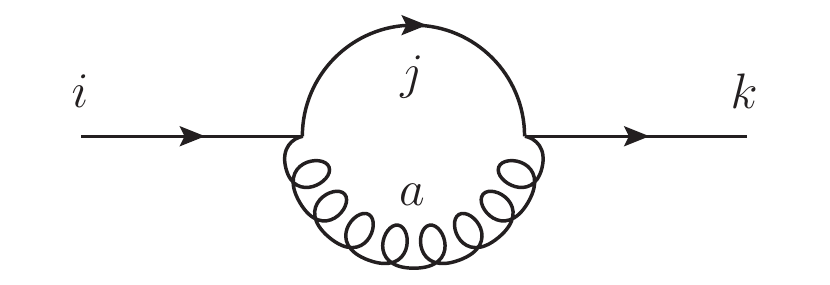}}\\
$\sum_{c,d} f^{acd} f^{bcd} = C_A\, \delta^{ab}$ & $a,b,c,d\in[1,\ldots,8]$
& \parbox[c]{4cm}{\includegraphics*[scale=0.5]{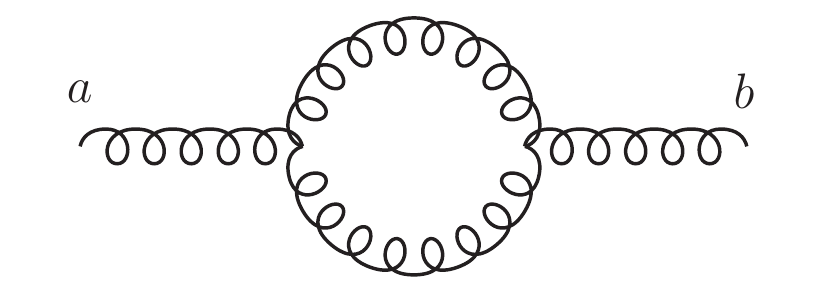}}\\
$ t^a_{ij}t^a_{k\ell} = T_R \left(\delta_{jk}\delta_{i\ell}
- \frac{1}{N_C}\delta_{ij}\delta_{k\ell}\right)$ & $i,j,k,\ell\in[1,\ldots,3]$
& \parbox[c]{4cm}{\includegraphics*[scale=0.5]{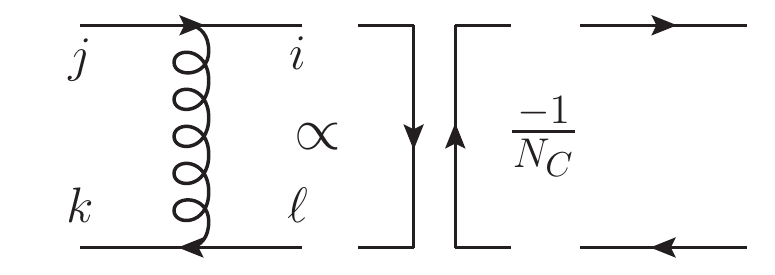}}\hspace*{-0.2cm}(Fierz)\\
\bottomrule
\end{tabular}}
\caption{Trace relations for $t$ matrices. These relations are
  convention-independent as they stand. Relations for a specific
  normalization convention for the $t$ matrices are obtained by
  inserting the specific values of $T_R$, $C_F$, and $C_A$ pertaining
  to that convention choice, as discussed in the text. More relations
  can be found in \cite[Section 1.2]{Ellis:1991qj} and in 
  \cite[Appendix A.3]{Peskin:1995ev}.
\label{tab:lambda}}
\end{center}
\end{table}
In the standard normalization convention for the $\mrm{SU(3)}$ generators,
\eqRef{eq:t}, the Casimirs of $\mrm{SU(3)}$ appearing in
\TabRef{tab:lambda} are\footnote{See, e.g., \cite[Appendix
    A.3]{Peskin:1995ev} for how to obtain the Casimirs in other
  normalization conventions.} 
\begin{equation}
T_R = \frac12 \hspace*{2cm} C_F = \frac43 \hspace*{2cm} C_A = N_C = 3~.
\end{equation}
In addition, the gluon self-coupling on the third line in
\TabRef{tab:lambda} involves factors of $f^{abc}$. These
are called the \emph{structure constants} of QCD and they enter due to
the non-Abelian term in the gluon field strength tensor appearing in
\eqRef{eq:L}, 
\begin{equation}
F^a_{\mu\nu} = \partial_\mu A_\nu^a - \partial_\nu A^a_\mu + g_s
f^{abc} A_\mu^b A_\nu^c~. \label{eq:F}
\end{equation}

\noindent\begin{minipage}[t]{0.46\textwidth}
The structure constants of $\mrm{SU(3)}$ are listed in the table to the
right. Expanding the $F_{\mu\nu}F^{\mu\nu}$ term of the
Lagrangian using \eqRef{eq:F}, we see that there is a 3-gluon and a
4-gluon vertex that involve $f^{abc}$, the latter of which has two
powers of $f$ and two powers of the coupling. 

Finally, the last line of \TabRef{tab:lambda} is not really a trace
relation but instead a useful so-called Fierz transformation. 
It is often used, for instance, in shower Monte Carlo
applications, to assist in mapping between colour flows in $N_C = 3$,
in which cross sections and splitting probabilities are calculated, 
and those in $N_C\to\infty$, used to represent colour flow in
the MC ``event record''.
\end{minipage}%
\hfill%
\colorbox{darkgray}{%
\colorbox{lightgray}{%
\begin{minipage}[t]{0.46\textwidth}
\begin{center}
Structure Constants of $\mrm{SU(3)}$
\begin{equation}
f_{123} = 1
\end{equation}
\begin{equation}
f_{147} = f_{246} = f_{257} = f_{345} = \frac12
\end{equation}
\begin{equation}
f_{156} = f_{367} = -\frac12
\end{equation}
\begin{equation}
f_{458} = f_{678} = \frac{\sqrt{3}}{2}
\end{equation}
Antisymmetric in all indices\\
All other $f_{ijk}=0$\\
\end{center}
\begin{center}
\tiny (valid for the convention $t=\frac{\lambda}{2}$)\\
\tiny (for the alternative convention $t=\frac{\lambda}{\sqrt{2}}$, multiply
all $f_{ijk}$ by $\sqrt{2}$)
\end{center}
\end{minipage}%
}}\vskip1mm

\begin{figure}[t]
\begin{center}
\begin{minipage}[h]{4.6cm}
\begin{center}
$A_\nu^4(k_2)$\\
\includegraphics*[scale=0.75]{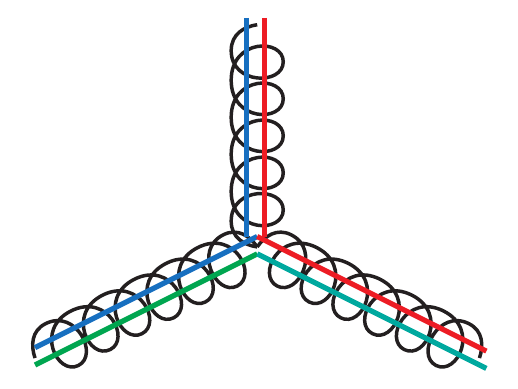}\\[-3mm]
$A^6_\rho(k_1)$\hfill$A_\mu^2(k_3)$
\end{center}
\end{minipage}~~~
\parbox{0.35\textwidth}{
$
\begin{array}{cccc}
\propto & - g_s \ f^{246} \!\! & \!\! [ (k_3 - k_2)^\rho g^{\mu\nu}  \\ 
& & +(k_2 - k_1)^\mu g^{\nu\rho} \\ 
& &+(k_1 - k_3)^\nu g^{\rho\mu}]
\end{array}
$}\vspace*{1mm}
\caption{Illustration of a $ggg$ vertex in QCD, before
  summing/averaging over colours: interaction between gluons in the 
  states $\lambda^2$, $\lambda^4$, and $\lambda^6$ is represented by
  the structure constant $f^{246}$. 
\label{fig:gg}}
\end{center}
\end{figure}
A gluon self-interaction vertex is
illustrated in \figRef{fig:gg}, to be compared with the quark-gluon
one in \figRef{fig:qg}. We remind the reader that gauge boson
self-interactions are a hallmark of non-Abelian theories and that their
presence leads to some of the main differences between QED and
QCD. One should also keep in mind 
that the colour factor for the vertex in \figRef{fig:gg}, $C_A$, 
is roughly twice as large as that for a quark, $C_F$.

\subsection{The Strong Coupling \label{sec:coupling}}
To first approximation, QCD is \emph{scale invariant}. That is, if one
``zooms in'' on a QCD jet, one will find a repeated self-similar 
pattern of jets within jets within jets, reminiscent of
fractals such as the famous Mandelbrot set in mathematics, or the
formation of frost crystals in physics. 
In the context of QCD, this property was originally 
called light-cone scaling, or 
Bjorken scaling after the famous physicist James D.\ Bjorken. 
It has since been rebranded by a new
generation as \emph{conformal invariance}, a mathematical property of several
QCD-``like'' theories which are now being studied. It is also closely
related to the physics of so-called ``unparticles'', though that is a
relation that goes beyond the scope of these lectures.

Regardless of the labeling, 
if the strong coupling did not run (we shall return to the running
of the coupling below), Bjorken scaling would be absolutely true. QCD
would be a theory with a fixed coupling, the same at all scales. 
This simplified picture already captures some of the most important
properties of QCD, as we shall discuss presently.  

In the limit of exact Bjorken scaling --- QCD at fixed coupling
--- properties of high-energy interactions are determined 
only by \emph{dimensionless} kinematic quantities, such as scattering
angles (pseudorapidities) and ratios of energy
scales\footnote{Originally, the observed approximate agreement with
this was used as a powerful argument
for pointlike substructure in hadrons; since measurements at different
energies are sensitive to different resolution scales, independence of the absolute
energy scale is indicative of the absence of other fundamental
scales in the problem and hence of pointlike constituents.}.
For applications of QCD to high-energy collider physics, an important
consequence of Bjorken scaling is thus that the rate of bremsstrahlung
jets with a given transverse momentum scales in direct proportion to the hardness
of the fundamental partonic scattering process they are produced in
association with. For instance, in the limit of exact scaling, a
measurement of the rate of 5-GeV jets produced in association with an ordinary $Z$
boson could be used as a direct prediction of the rate of 50-GeV jets that would be
produced in association with a 900-GeV $Z'$ boson, and so 
forth. Our intuition about how many bremsstrahlung jets a given type of
process is likely to have should therefore be governed first and
foremost by the \emph{ratios} of scales that appear in that particular
process, as has been  highlighted in a number of studies focusing on
the mass and $p_\perp$ scales appearing, e.g., in Beyond-the-Standard-Model (BSM)
physics processes
\cite{Plehn:2005cq,Skands:2005bj,Alwall:2008qv,Papaefstathiou:2009hp,Krohn:2011zp}. 
Bjorken scaling 
is also fundamental to the understanding of jet substructure in QCD, see, e.g.,
\cite{Vermilion:2011nm}.  

In real QCD, the coupling runs logarithmically with the energy,
\begin{equation}
Q^2 \frac{\partial \alpha_s}{\partial Q^2} = \frac{\partial
  \alpha_s}{\partial \ln Q^2} =
\beta(\alpha_s)~, \label{eq:running}
\end{equation}
where the function driving the energy dependence, the \emph{beta
  function}, is defined as
\begin{equation}
\beta(\alpha_s) = -\alpha_s^2(b_0 +
b_1\alpha_s + b_2\alpha_s^2 + \ldots)~,\label{eq:beta}
\end{equation}
with LO (1-loop) and NLO (2-loop) coefficients
\begin{eqnarray}
b_0 & = & \frac{11C_A - 4 T_R n_f}{12\pi}~,\\[3mm]
b_1 & = & \frac{17C_A^2 - 10 T_R C_A n_f - 6 T_R C_F n_f}{24\pi^2} ~=~
\frac{153-19 n_f}{24\pi^2}~.\label{eq:b}
\end{eqnarray}
Numerically, the value of the strong coupling is usually specified by
giving its value at the specific 
reference scale $Q^2=M^2_Z$, from which we can obtain its
value at any other scale by solving \eqRef{eq:running}, 
\begin{equation}
\alpha_s(Q^2) = \alpha_s(M_Z^2) \frac{1}{1+b_0
  \alpha_s(M_Z^2)\ln\frac{Q^2}{M_Z^2} + {\cal O}(\alpha_s^2)}~,
\label{eq:alphaq2}
\end{equation}
with relations including the ${\cal O}(\alpha_s^2)$ terms 
available, e.g., in \cite{Ellis:1991qj}. 
Relations between scales 
not involving $M_Z^2$ can obviously be obtained by just replacing $M_Z^2$
by some other scale $Q'^2$ everywhere in \eqRef{eq:alphaq2}. 
As an application, let us prove that the 
logarithmic running of the coupling implies that an intrinsically 
multi-scale problem can be converted to a single-scale one, up to
corrections suppressed by two powers of $\alpha_s$, 
by taking the geometric mean of the scales involved. This follows from
expanding an arbitrary product of individual $\alpha_s$ factors around an
arbitrary scale $\mu$, using \eqRef{eq:alphaq2}, 
\begin{eqnarray}
\alpha_s(\mu_1)\alpha_s(\mu_2)\cdots\alpha_s(\mu_n) & = &
\prod_{i=1}^{n} \alpha_s(\mu) \left(1 +
b_0\,\alpha_s\ln\left(\frac{\mu^2}{\mu_i^2}\right) + {\cal O}(\alpha_s^2)\right)
\nonumber\\[2mm]
& = & \alpha_s^n(\mu) \left(1 + b_0\, \alpha_s \ln \left(
 \frac{\mu^{2n}}{\mu_1^2\mu_2^2\cdots\mu_n^2}\right) +  {\cal
   O}(\alpha_s^2) \right)~,
\end{eqnarray}
whereby the specific single-scale choice $\mu^n =
\mu_1\mu_2\cdots\mu_n$ (the geometric mean) can
be seen to push the difference between the two sides of the equation one order higher
than would be the case for any other combination of scales\footnote{In
  a fixed-order calculation, the individual scales $\mu_i$,
would correspond, e.g., to the $n$ hardest scales appearing in an infrared
safe sequential clustering algorithm applied to the given momentum
configuration.}. 

The appearance of the number of flavours, $n_f$, in $b_0$ implies that the
slope of the running depends on the number of contributing
flavours. Since full QCD is best approximated by $n_f=3$ below the charm
threshold, by $n_f=4$ from there to the $b$ threshold, and by $n_f=5$
above that, it is therefore important to be aware that 
the running changes slope across quark flavour
thresholds. Likewise, it would change across the threshold for top or
for any coloured
new-physics particles that might exist, with a magnitude depending on
the particles' colour and spin quantum numbers.

The negative overall sign of \eqRef{eq:beta}, combined with the fact
that $b_0 > 0$, leads to the famous
 result\footnote{
Perhaps the highest pinnacle of fame for \eqRef{eq:beta} was reached
when the sign of it ``starred'' in an episode of the TV series ``Big Bang
Theory''.} 
that the QCD coupling effectively \emph{decreases} with
 energy, called asymptotic 
freedom, for the discovery of which the Nobel prize in physics was
awarded to D.~Gross, H.~Politzer, and F.~Wilczek in 2004. An extract
of the prize announcement runs as follows:
\begin{center}
\begin{minipage}{0.84\textwidth}
\sl  What this year's Laureates discovered was something that, at
first sight, seemed completely contradictory. The interpretation of
their mathematical result was that the closer the quarks are to each
other, the \emph{weaker} is the ``colour charge''. When the quarks are
really close to each other, the force is so weak that they behave
almost as free particles\footnote{More correctly, it is the coupling
  rather than the  
  force which becomes weak as the distance decreases. 
  The $1/r^2$ Coulomb singularity of the force is only dampened, not removed, 
  by the diminishing coupling.}. 
This phenomenon is called ``asymptotic
freedom''. The converse is true when the quarks move apart: the force
becomes stronger when the distance increases\footnote{More correctly,
 it is the potential which grows, linearly, while the force becomes
 constant.}. 
\end{minipage}
\end{center}

\begin{figure}[t]
\begin{center}
\parbox[c]{2.5cm}{\includegraphics*[scale=0.65]{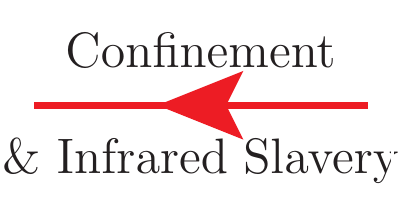}}
\parbox[c]{8cm}{\includegraphics*[scale=0.5]{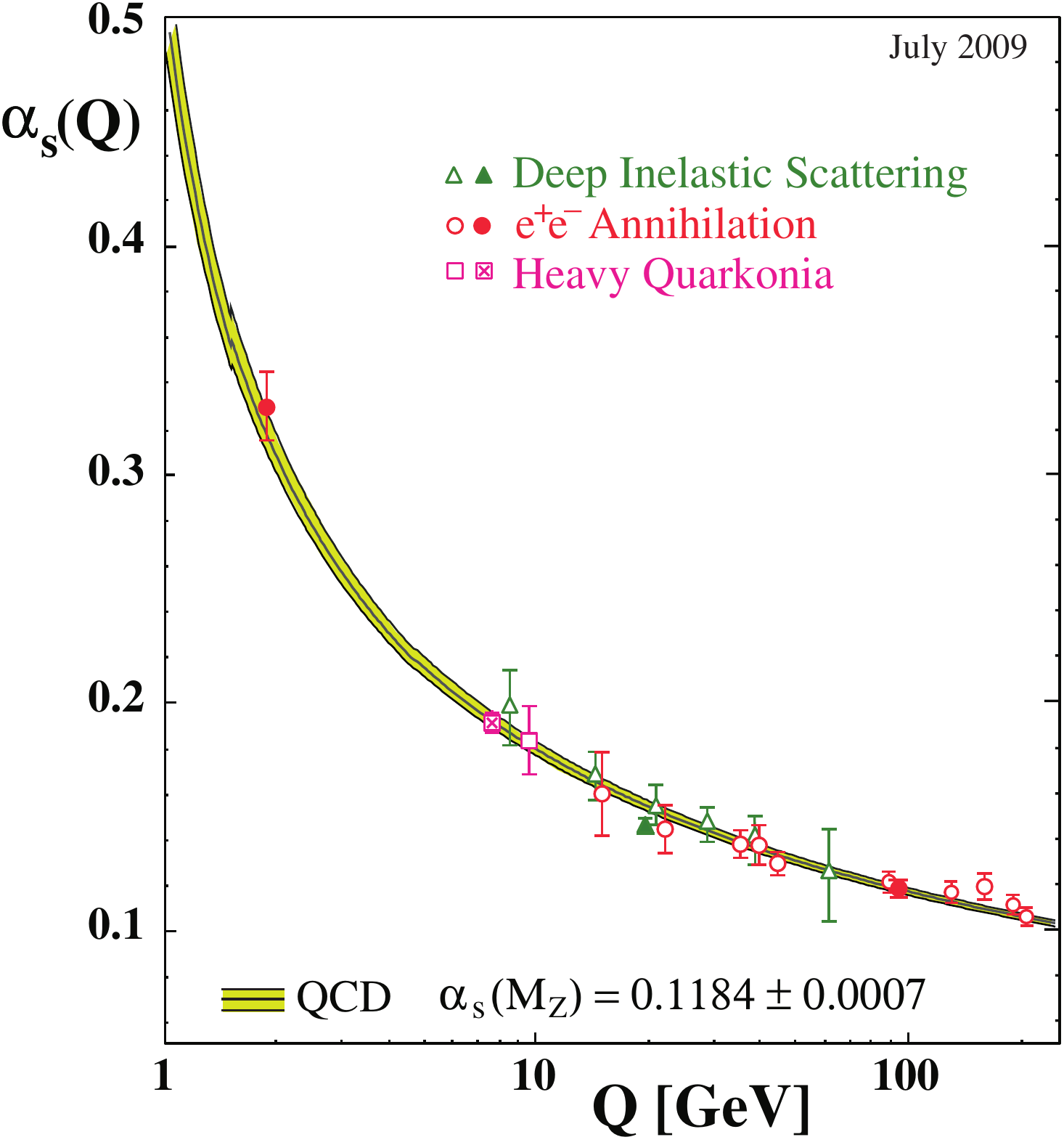}}
\parbox[c]{3.cm}{\includegraphics*[scale=0.65]{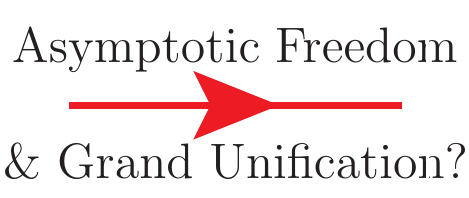}}
\caption{Illustration of the running of $\alpha_s$ in a theoretical
  calculation (yellow shaded band) and in physical processes at
  different characteristic scales, from \cite{Bethke:2009jm}. \label{fig:alphas}}
\end{center}           
\end{figure}

Among the consequences of asymptotic freedom is that perturbation
theory becomes better behaved at higher absolute energies, due to the
effectively decreasing coupling. Perturbative calculations for our
900-GeV $Z'$ boson from before should therefore be slightly faster
converging than equivalent calculations for the 90-GeV one. 
Furthermore, since the running of $\alpha_s$ explicitly
breaks Bjorken scaling, we also expect to see small changes in jet
shapes and in jet production ratios as we vary the energy. For
instance, since high-$p_\perp$ jets
start out with a smaller effective coupling, their intrinsic shape
(irrespective of boost effects) is
somewhat narrower than for low-$p_\perp$ jets, an issue which can be
important for jet calibration. Our current understanding of the
running of the QCD coupling is summarized by the plot in
\figRef{fig:alphas}, taken from a recent comprehensive review by S.\ Bethke
\cite{Bethke:2009jm}. 

As a final remark on asymptotic freedom, note that the decreasing
value of the strong coupling with energy must eventually cause it to
become comparable to the electromagnetic and weak ones, at some energy
scale. Beyond that point, which may lie at energies of order
$10^{15}-10^{17}\,$GeV (though it may be lower if as yet undiscovered
particles generate large corrections to the running), 
we do not know  what the further evolution of the combined theory will 
actually look like, or whether it will continue to exhibit asymptotic
freedom. 

Now consider what happens when we run the coupling in the other
direction, towards smaller energies. 
Taken at face value, the numerical value of the coupling diverges
rapidly at scales below 1 GeV, as illustrated by the curves
disappearing off the left-hand edge of the plot in
\figRef{fig:alphas}. To make this divergence
explicit, one can rewrite
\eqRef{eq:alphaq2} in the following form, 
\begin{equation}
\alpha_s(Q^2) = \frac{1}{b_0 \ln \frac{Q^2}{\Lambda^2}}~,\label{eq:alphasLam}
\end{equation}
where 
\begin{equation}
\Lambda \sim 200\, \mbox{GeV}
\end{equation}
specifies the energy scale at which the perturbative coupling would nominally become
infinite, called the Landau pole. (Note, however, that this only
parametrizes the purely \emph{perturbative} result, which is not
reliable at strong coupling, so \eqRef{eq:alphasLam} should 
not be taken to imply that the physical behaviour of full QCD should
exhibit a divergence for $Q\to \Lambda$.) 

Finally, one should be aware that there is a multitude of different
ways of defining both $\Lambda$ and $\alpha_s(M_Z)$. At the very
least, the numerical value one obtains depends both on the
renormalization scheme used (with the dimensional-regularization-based
``modified minimal subtraction'' scheme, $\overline{\mbox{MS}}$, being the
most common one) and on the perturbative order of the calculations 
used to extract them. As a rule of thumb, fits to experimental data typically yield 
smaller values for $\alpha_s(M_Z)$ the higher the order of the
calculation used to extract it (see, e.g.,
\cite{Bethke:2009jm,Dissertori:2009ik}), with  $
\alpha_s(M_Z)\vert_{\mrm{LO}} \gsim \alpha_s(M_Z)\vert_{\mrm{NLO}}
\gsim \alpha_s(M_Z)\vert_{\mrm{NNLO}}$. 
Further, since the number of flavours changes the slope
of the running, the location of the Landau pole for fixed
$\alpha_s(M_Z)$ depends explicitly on the number of flavours used in
the running. Thus each value of $n_f$ is associated with its own
value of $\Lambda$, with the following matching relations across
thresholds guaranteeing continuity of the coupling at one loop,
\begin{eqnarray}
n_f = 4 \leftrightarrow 5 ~~~:~~~~~~\Lambda_5 = \Lambda_4
  \left(\frac{\Lambda_4}{m_b}\right)^{\frac{2}{23}} & & 
\Lambda_4 = \Lambda_5
  \left(\frac{m_b}{\Lambda_5}\right)^{\frac{2}{25}} ~, \\[2mm]
n_f = 3 \leftrightarrow 4 ~~~:~~~~~~\Lambda_4 = \Lambda_3 
  \left(\frac{\Lambda_3}{m_c}\right)^{\frac{2}{25}} & &
\Lambda_3 = \Lambda_4 
  \left(\frac{m_c}{\Lambda_4}\right)^{\frac{2}{27}} ~.
\end{eqnarray}

It is sometimes stated that QCD only has a single free
parameter, the strong coupling. Appealing as this may be, it is a bit
of an overstatement. Even in the perturbative
region, the beta function depends explicitly on the number of
quark flavours, as we have seen, and thereby also on the quark
masses. Furthermore, in the non-perturbative region around or below
$\Lambda_{\mrm{QCD}}$, the value of the 
perturbative coupling, as obtained, e.g., from \eqRef{eq:alphasLam},
gives little or no insight into the behaviour of the full theory. 
Instead, universal functions (such as parton densities, form factors,
fragmentation functions, etc), effective theories (such as the
Operator Product Expansion, Chiral Perturbation Theory, or Heavy Quark
Effective Theory), or phenomenological models (such as Regge Theory or
the String and Cluster Hadronization Models) must be used, which in
turn depend on additional non-perturbative parameters whose relation to, e.g.,
$\alpha_s(M_Z)$, is not a priori known. For some of these questions,
such as hadron masses, lattice QCD can furnish important
additional insight, but for multi-scale and/or time-evolution
problems, the applicability of lattice methods is still severely
restricted. 

\section{Perturbative QCD \label{sec:pQCD}}
Our main tool for solving QCD for high-energy collider physics
is perturbative quantum field theory, the starting point for which is
Matrix Elements (MEs) which can be calculated systematically at fixed
orders in the strong coupling $\alpha_s$. 
At least at lowest order (LO), the procedure is
standard textbook material \cite{Peskin:1995ev} and it has also by now been
highly automated, by the advent of tools like \textsc{CalcHep} \cite{Pukhov:2004ca},
\textsc{CompHep} \cite{Boos:2004kh}, \textsc{MadGraph} \cite{Alwall:2007st}, 
and others
\cite{Kanaki:2000ey,Krauss:2001iv,Moretti:2001zz,Bahr:2008pv,Gleisberg:2008fv}. 
Here, we  require only that the reader has a
basic familiarity with the methods involved from graduate-level particle
physics courses based, e.g., on \cite{Peskin:1995ev,Dissertori:2003pj}. 
Our main concern are the uses to which these calculations 
are put, their limitations, and ways to improve on the results obtained
with them.

For illustration, take one of the
most commonly occurring processes in hadron collisions ---
Rutherford scattering of two quarks via a $t$-channel gluon exchange
--- which has the differential cross section
\begin{equation}
qq'\to qq' ~~~:~~~\frac{\d{\sigma}}{\d{\hat{t}}} =
\frac{\pi}{\hat{s}^2}\, \frac{4}{9}\,
\alpha_s^2\, \frac{\hat{s}^2 + \hat{u}^2}{\hat{t}^2}~,
\end{equation}
with the $2\to 2$ Mandelstam variables (``hatted'' to emphasize that
they refer to a partonic $2\to 2$ scattering rather than the full
$pp\to\mbox{jets}$
process)
\begin{eqnarray}
\hat{s} & = & (p_1+p_2)^2 ~,\\[1.5mm]
\hat{t} & = &(p_3-p_1)^2 = -\hat{s}\frac{(1-\cos\hat\theta)}{2} ~,\\
\hat{u} & = &(p_4-p_1)^2 = -\hat{s}\frac{(1+\cos\hat\theta)}{2}~.
\end{eqnarray}
\begin{figure}
\center
\begin{tabular}{ccc}
\begin{minipage}[c]{5.5cm}
$p_1$\hfill $p_3$\\[-1mm]
\includegraphics*[scale=0.5]{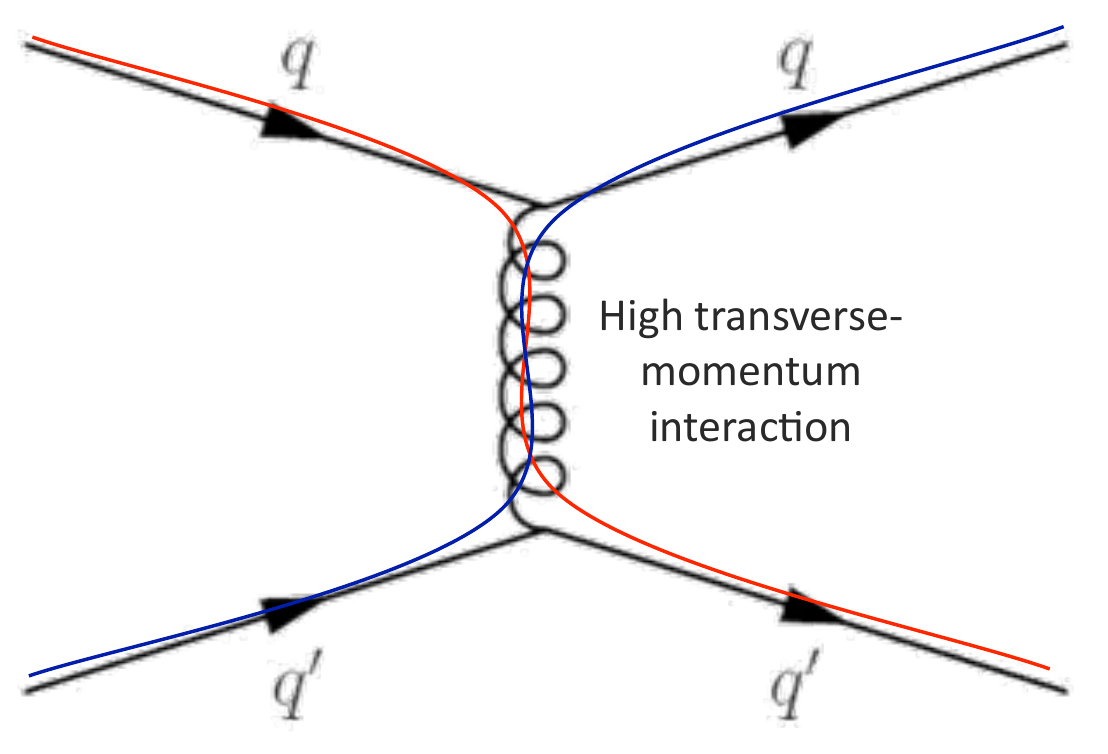}\\[-4.5mm]
$p_2$\hfill $p_4$
\end{minipage}
& ~~~&
\begin{minipage}[c]{7cm}
\includegraphics*[scale=0.3]{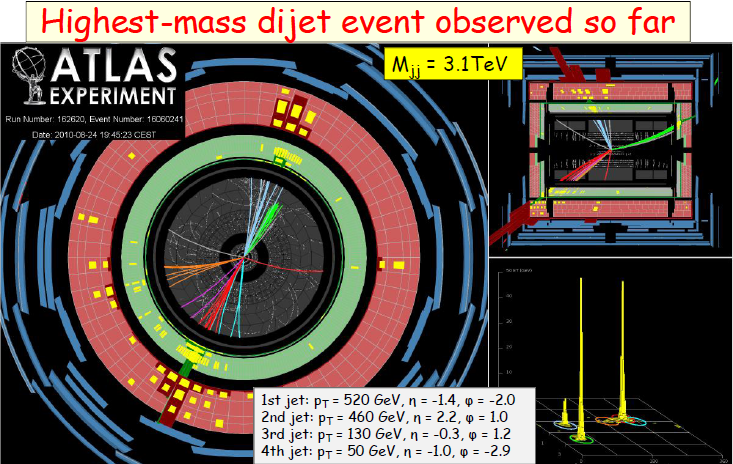}
\end{minipage}
\end{tabular}
\caption{{\sl Left:} Rutherford scattering of quarks in QCD, exemplifying the type
  of process that dominates the short-distance interaction cross section at
  hadron colliders. {\sl Right:} an example of what such a reaction
  may look like in a detector, in this case the ATLAS experiment.
\label{fig:rutherford} }
\end{figure}
This process is illustrated in the left-hand pane of 
\figRef{fig:rutherford}, including a
rough (formally leading-$N_C$) representation of the ``colour
transfer'' mediated by the gluon (as was discussed in \SecRef{sec:L}). 

Reality, however, is more complicated; the picture on the right-hand pane of
\figRef{fig:rutherford} shows a real dijet event, as recorded by the
ATLAS experiment. 
The complications to be addressed when going from left to right in
\figRef{fig:rutherford} are: 
firstly, additional jets, a.k.a.\ real-emission corrections, 
which significantly change the topology of the final state, potentially
shifting jets in or out of an experimentally defined acceptance
region. Secondly, loop factors, a.k.a.\ virtual corrections, change the number of
 available quantum paths through phase space, and hence modify 
the normalization of the cross section (total \emph{and}
differential). And finally, additional corrections to the simple
factorized perturbative picture are generated by components such as
hadronization and the underlying event. These corrections 
must be taken into account to complete our understanding of QCD and
connect the short-distance physics with macroscopic experiments.  
Apart from the perturbative expansion itself, the most powerful tool
we have to organize this vast calculation, is factorization. 

\subsection{Factorization \label{sec:factorization}}
When applicable, factorization allows us to
subdivide the calculation of an observable into a perturbatively calculable
short-distance part and an approximately universal long-distance part,
the latter of which may be modeled and constrained by fits to data. Factorization
can also be applied multiple times, to break up a complicated calculation
into simpler  pieces that can be treated as approximately independent, 
such as when dealing with successive emissions in a parton shower, or
when factoring off  decays of long-lived particles from a hard
production process. 

Using collinear factorization (see, e.g., \cite{Ellis:1991qj,Brock:1993sz}),
the differential cross section for an observable $\cal O$ 
in hadron-hadron collisions can be computed as:
\begin{equation}
\frac{\d{\sigma}}{\d{\obs}} = 
\sum_{a,b}\int_{0}^{1}\d{x_a}\d{x_b}\sum_F\int\!\dPS{F}\,
\pdf[h_1]{a}(x_a,\mu_F)\pdf[h_2]{b}(x_b,\mu_F)\,\frac{\d{\hat{\sigma}_{ab\to
      F}}}{\d{\hat\obs}}D_F(\hat{\obs}\to\obs,\mu_F)
\label{eq:factorization}
\end{equation}
where the outer sum runs over all partonic constituents, $a$ and $b$ 
of the colliding hadrons, $h_{1,2}$, respectively, and the inner sum runs
over all possible final states, $ab\to F$ (with the standard
final-state phase-space differential denoted $\dPS{F}$). 

Before we discuss the integrand --- composed of the 
factors $f_{a,b}$, $\d{\hat{\sigma}}$, and
$D_F$ --- let us first re-emphasize \emph{the} crucial
feature of \eqRef{eq:factorization}; it separates
the calculation of the cross section into two independent pieces, 
one of which is the perturbatively calculable short-distance cross
section, $\d{\hat\sigma}$, and the other of which is the product of parton
distribution functions (PDFs), $f_a f_b$, with a fragmentation
function (FF), $D_F$, with the latter components being 
universal functions\footnote{At least, they are universal
within the framework of collinear factorization. In full QCD, there
are several types of corrections, including also some perturbative ones, 
that go beyond this framework, such as small-$x$ effects and multiple
parton interactions, both of which mandate the introduction of objects
that go beyond the scope of collinear-factorized PDFs. In the case of
small-$x$ evolution, these more general objects are so-called
\emph{unintegrated} PDFs, which have an explicit dependence on the
parton transverse momentum in addition to the factorization scale, 
while multi-parton interactions require
explicit \emph{multi-parton} and/or \emph{generalized}
(impact-parameter-dependent) PDFs.}
whose forms are a priori unknown
but which can be constrained in one process and then reused in
another.
The dividing line between
the two is drawn at an arbitrary (``user-defined'') scale $\mu_F$,
called the \emph{factorization scale}. 

Returning now to the integrand, the parton density functions,
$\pdf[h_i]{j}(x_j,\mu_F)$,  
parametrize the distribution of partons of type $j$ carrying momentum
fraction $x_j$ inside a hadron of type $h_i$ when probing the latter
at the factorization scale $\mu_F$. (Note: issues specific to 
PDFs in the context of Monte Carlo event generators will be covered in 
\SecRef{sec:pdfs}.)
The partonic scattering cross section $\d{\hat{\sigma}_{ab\to F}}$ is 
calculable in fixed-order perturbation theory as
\begin{equation}
\d{\hat{\sigma}_{ab\to F}} = 
\frac{1}{2\hat s_{ab}}|{\cal M}_{ab\to F}|^2(\PS{F};\mu_F,\mu_R)\,,
~, \label{eq:dsigma}
\end{equation}
with $|{\cal M}|^2$ the matrix element
squared for the process $ab\to F$, appropriately summed and averaged
over helicities and/or colours, and evaluated at the factorization and
renormalization scales $\mu_F$ and $\mu_R$, respectively.
The fragmentation functions (FFs), $D_F(\hat{\obs}\to\obs,\mu_F)$
parametrize the transition from  
partonic final state to the hadronic observable (bremsstrahlung,
hadronization, jet definition, etc). 

There is some arbitrariness involved in this division of the
calculation into a short-distance and a long-distance part. Firstly,
one has to choose a value for the dividing scale,
$\mu_F$. Some heuristic arguments to guide in the choice of factorization
scale are the following. On the
long-distance side, the PDFs include a (re)summation of multiple
emissions (bremsstrahlung) all the way up to the scale $\mu_F$. 
It would therefore not make much sense to take $\mu_F$ 
significantly larger than the scales characterizing resolved particles
on the short-distance side of the calculation (i.e., the particles
appearing explicitly in $\PS{F}$); 
otherwise the PDFs would be including sums over radiations as hard as or
harder than those included explicitly in the matrix element which
would result in double-counting. On the other hand, it should not be
taken much lower than the scales appearing in the matrix element
either, since, as we shall see in subsequent chapters, fixed-order
matrix elements are at most able to include \emph{part} of such
multiple-bremsstrahlung emissions, and hence a low choice of
factorization scale would lead to problems with ``undercounting'' of
such corrections. 

For matrix elements characterized by a single well-defined scale,
such as the $Q^2$ scale in deeply inelastic scattering (DIS) or the
invariant-mass scale $\hat{s}$ in Drell-Yan production
($q\bar{q}\to Z/\gamma^*\to \ell^+\ell^-$), such arguments essentially
fix the preferred scale choice, which may then be varied by a factor
of 2 (or larger) around the nominal value in order to estimate
uncertainties. For multi-scale problems, however, such as $pp\to
Z/W+n\,$jets, there are several a priori equally good choices
available, from the lowest to the highest QCD scales that can be
constructed from the final-state momenta, usually with several
dissenting groups of theorists arguing over which particular choice is
best. Suggesting that one might simply \emph{measure} the scale
would not really be an improvement, as the factorization
scale is fundamentally unphysical and therefore
unobservable (similarly to gauge or convention choices). 
One plausible strategy is to
look at higher-order (NLO or NNLO) calculations, in which correction
terms appear that explicitly remove the over- or under-counting
introduced by the initial scale choice up to the given order, 
thus reducing the overall dependence on it and stabilizing the final
result. From such comparisons, a
``most stable'' initial 
scale choice can in principle be determined, which then furnishes a
reasonable starting point, but we  emphasize that the
question \emph{is} intrinsically ambiguous, and no ``golden recipe''
is likely to magically give all the right answers. The best we can do is
to vary the value of $\mu_F$ not only by an overall factor, but also by
exploring different possible forms for its 
functional dependence on the momenta appearing in
\PS{F}. In this way, one could hope to provide a more complete
uncertainty estimate for multi-scale problems. 

Secondly, and more technically, at NLO and beyond one also has to settle on a
\emph{factorization scheme} in which to do the calculations. 
For all practical
purposes, students focusing on LHC physics are only likely to
encounter one such scheme, the modified minimal subtraction
($\overline{\mrm{MS}}$) one already mentioned in the discussion of the
definition of the strong coupling in \SecRef{sec:coupling}. At the
level of these lectures, we shall therefore not elaborate further on this choice
here.  

\subsection{Infrared Safety}
The second perturbative tool, infrared safety, provides us 
with a special class of observables which have \emph{minimal sensitivity}
to long-distance physics, and which can be consistently computed in
perturbative QCD (pQCD). By ``infrared'', we here mean any  limit
that involves a low scale (i.e., any non-UV limit), 
without regard to  whether it is collinear or soft\footnote{This
  distinction will be discussed further in \SecRef{sec:parton-showers}.}.
An observable is infrared safe if:
\begin{enumerate}
\item {\sl(Safety against soft radiation):} Adding any number of infinitely soft particles should not change
  the value of the observable. 
\item {\sl(Safety against collinear radiation):} Splitting an existing particle up into two comoving particles, with
arbitrary fractions $z$ and $1-z$, respectively, of 
the original momentum, should not change the value of the
observable.
\end{enumerate}
If both of these conditions are satisfied, any long-distance non-perturbative
corrections will be suppressed by the ratio of the 
long-distance scale to the short-distance one to some
(observable-dependent) power, typically
\begin{equation}
\mbox{IR Safe Observables: IR corrections~~~$\propto$~~~} \frac{Q_{\mathrm{IR}}^2}{Q_{\mathrm{UV}}^2} 
\end{equation}
where $Q_\mathrm{UV}$ denotes
a generic hard scale in the problem, and $Q_\mrm{IR} \sim
\Lambda_\mrm{QCD} \sim \mathcal{O}(\mrm{1\ GeV})$. 

Due to this \emph{power suppression}, 
IR safe observables are not so sensitive to our lack
of ability to solve the strongly coupled IR physics, unless of course
we go to processes for which the relevant hard scale, $Q_{\mrm{UV}}$, is small
(such as minimum-bias, soft jets, or small-scale jet substructure). 
Even when a high scale is present, however, as in resonance decays, jet
fragmentation, or underlying-event-type studies, infrared safety only
guarantees us that infrared corrections are small, not that they are zero. 
Thus, ultimately, we run into 
a precision barrier even for IR safe observables, 
which only a reliable understanding of the
long-distance physics itself can address. 

To constrain models of long-distance physics, one needs 
infrared \emph{sensitive} observables\footnote{
Hence it is not always the case that infrared safe
observables are preferable --- the purpose decides the tool.}. 
Instead of the suppressed corrections above, the perturbative
prediction for such observables contains logarithms
\begin{equation}
\mbox{IR Sensitive Observables: IR Corrections~~~$\propto$~~~} 
\alpha_s^n\log^{m}\left(\frac{Q_{\mathrm{UV}}^2}{Q_{\mathrm{IR}}^2}\right)~~~,~~~m
\le 2n ~~~,
\end{equation}
 which grow
increasingly large as $Q_{\mathrm{IR}}/Q_{\mathrm{UV}}\to 0$.   
As an example, consider such a fundamental quantity as particle 
multiplicities; in the absence of nontrivial infrared
effects, the number of partons that would be
mapped to hadrons in a na\"ive local-parton-hadron-duality
\cite{Azimov:1984np} picture would tend logarithmically to infinity 
as the IR cutoff is lowered. Similarly, the distinction between
a charged and a neutral pion only occurs in the very last phase of
hadronization, and hence observables that only include charged tracks,
for instance, are always IR sensitive\footnote{This remains true in principle 
even if the tracks are clustered into jets, although the energy
clustered in this way does provide a lower bound on $Q_{\mrm{UV}}$ in
the given event, since ``charged + neutral $>$ charged-only''.}.

Two important categories of infrared safe observables that are widely used are
\emph{event shapes} and \emph{jet algorithms}. Jet algorithms are
perhaps nowhere as pedagogically described as in last year's ESHEP lectures by 
Salam \cite[Chapter 5]{Salam:2010zt}. Event shapes in the context of
hadron colliders have not yet been as widely explored, but the basic
phenomenology is introduced also by Salam and collaborators in
\cite{Banfi:2010xy}, with a first measurement reported by CMS 
\cite{Khachatryan:2011dx} and a proposal to use them also for the
characterization of minimum-bias events put forth in \cite{Wraight:2011ej}.

Let us here merely emphasize that the real reason to prefer infrared safe
jet algorithms over unsafe ones 
is not that they  necessarily give very different
or ``better'' answers in the experiment --- experiments are infrared safe by
definition, and the difference between infrared safe and unsafe
algorithms may not even be visible when running the
algorithm on experimental data --- but that it is only possible to
compute perturbative QCD predictions for the infrared safe ones. Any
measurement performed with an infrared unsafe algorithm can only be
compared to calculations that include a detailed hadronization model. This
both limits the number of calculations that can be compared to and
also adds an a priori unknown sensitivity to the details of the 
hadronization description, details which one would rather investigate and
constrain separately, in the framework of more dedicated fragmentation
studies. 

\subsection{Fixed-Order QCD: Matrix Elements \label{sec:fixed-order}}

Schematically, we express the all-orders
differential cross section for an observable \obs, in the
production of $F$ + anything ($\equiv$ \emph{inclusive} $F$ production, with $F$ an
arbitrary final state), in the following way:
\begin{equation}
\left.\frac{\d{\sigma_F}}{\d{\obs}}\right\vert_{\textcolor{blue}{\mrm{ME}}}
= \underbrace{\sum_{k=0}^\infty\int \dPS{F+k} }_{\Sigma~\mbox{legs}} 
 \Big\vert \underbrace{\sum_{\ell=0}^{\infty} {\cal
   M}_{F+k}^{(\ell)}}_{\Sigma~\mbox{loops}} \Big\vert^2 \,
\delta\left(\obs-\obs(\PS{F+k})\right)~,\label{eq:fixed-order}
\end{equation}
where, for compactness, we have suppressed all PDF and luminosity
normalization factors. The sum over $k$ represents a sum over
additional ``real-emission'' corrections, called legs, and the sum
over $\ell$ runs over additional virtual corrections, loops. Without the 
$\delta$ function, the formula would give the total integrated
cross section, instead of the cross section differentially in $\obs$. 
The purpose of the $\delta$ function is
thus to project out hypersurfaces of constant \obs\ 
in the full $\dPS{F+k}$ phase space, with $\obs(\PS{F+k})$ a function that
defines $\obs$ evaluated on each specific momentum configuration, 
$\PS{F+k}$. 

We recover the various fixed-order truncations of pQCD by limiting the
nested sums in \eqRef{eq:fixed-order} to include only specific values
of $k+\ell$. Thus, 
\begin{center}
\begin{tabular}{lcp{9.5cm}}
$k=0$, $\ell=0$ &$\implies$& Leading Order  (usually tree-level) 
for inclusive $F$ production\\
$k=n$, $\ell=0$ &$\implies$& Leading Order for $F+n\,$jets\\
$k+\ell\le n$,  &$\implies$& N$^n$LO for $F$ {\small (includes N$^{n-1}$LO for
  $F+1\,$jet, N$^{n-2}$LO for $F+2\,$jets, and so on up to LO for $F+n\,$jets)}~.\\
\end{tabular}
\end{center}
Already at this stage, before entering into the details of the
calculations, we can make several observations on how 
numerical values of cross sections and decay widths must be computed
in fixed-order perturbation theory.

Firstly, the dimensionality of the phase space to be integrated
increases by $d=3$ for each leg we add. In dimensions higher than 5,
the fastest converging numerical integration algorithm is Monte Carlo
integration  \cite{James:1980yn}, 
whose purely stochastic error $\propto {\cal O}(1/\sqrt{\cal N})$,
with $\cal N$ the number of generated points, is independent of
dimension, while all other algorithms scale 
with powers of the dimension. Therefore, virtually
all numerical cross section calculations are based on Monte Carlo
techniques in one form or another, the simplest being the 
\textsc{Rambo} algorithm \cite{Kleiss:1985gy} which can be expressed
in about half a page of code and generates a flat scan over $n$-body
phase space\footnote{Strictly speaking, \textsc{Rambo} is only truly
  uniform for massless particles. Its massive variant makes up for
  phase-space biases by returning weighted momentum configurations.}. 

Secondly, due to
the infrared singularities in perturbative QCD, the functions to be
integrated, $|{\cal{M}}|^2$, are highly non-uniform for large $k$, 
which implies that we will have to be clever in the way
we sample phase space if we want the integration to converge in any
reasonable amount of time --- simple algorithms like \textsc{Rambo} 
quickly become inefficient for $k$ greater than a few. 
To address this bottleneck, 
the simplest step up
from \textsc{Rambo} is to introduce generic (i.e., automated) 
importance-sampling methods, such as offered by the  
\textsc{Vegas} algorithm \cite{Lepage:1977sw,Lepage:1980dq}. This is still the
dominant basic technique, although most modern codes do employ several 
additional refinements, such as several different copies of \textsc{Vegas}
running in parallel (multi-channel integration), to further optimize the sampling. 
Alternatively, a few algorithms incorporate the singularity structure of QCD
explicitly in their phase-space sampling, either by directly generating momenta
distributed according to the leading-order QCD singularities, in a
sort of ``QCD-preweighted'' analog of \textsc{Rambo}, called
\textsc{Sarge} \cite{Draggiotis:2000gm}, or by using all-orders
Markovian parton showers to generate them (\textsc{Vincia}
\cite{Giele:2011cb,LopezVillarejo:2011ap}).  

Thirdly, for $k\ge1, \ell=0$, we are really not considering inclusive
$F$ production anymore; instead, we are considering the LO 
contribution to the process $F+k\,$jets. However, if we simply 
integrate over all momenta, as implied by the integration over \dPS{F+k}
in \eqRef{eq:fixed-order}, we would be including configurations in
which one or more of the $k$ partons become collinear or soft, leading to
singularities in the integration region. At the LO level, 
this problem can only be mitigated 
by restricting the integration region to only include
``hard, well-separated'' momenta. As discussed above, due to the 
approximate Bjorken scaling of QCD, it would be meaningless to express
this requirement in dimensionful terms, as an absolute scale. 
Instead, it is the \emph{ratios} of scales present in any given
process that determine whether such enhancements are present or
absent: a 50-GeV jet
would be considered hard and well-separated if produced in association with
an ordinary $Z$ boson, while it would be considered soft if produced
in association with a 900-GeV $Z'$ boson
\cite{Plehn:2005cq,Skands:2005bj,Alwall:2008qv}. Thus, for example, 
it would be a complete
disaster to use the same dimensionful phase-space
cuts for $Z'+$jets as one uses for $Z+$jets (unless of course the $Z'$ happens
to have a mass scale very close to the $Z$ one). A good rule of thumb
is that if $\sigma_{k+1} \approx \sigma_{k}$ (at whatever order you
are calculating), then you are integrating
over a region in which the perturbative series is no longer
converging, or is converging too slowly for a fixed-order truncation
of it to be reliable. For fixed-order perturbation theory to be
applicable, you must have $\sigma_{k+1} \ll \sigma_{k}$. In the
discussion of parton showers and resummations in
\SecRef{sec:parton-showers}, we shall see how the region of
applicability of perturbation theory can be extended. 

And finally, the virtual amplitudes, for $\ell \ge 1$, are divergent
for any point in phase space. However, as encapsulated by the famous KLN theorem
\cite{Kinoshita:1962ur,Lee:1964is}, unitarity (which essentially
expresses probability conservation) puts a powerful
constraint on the IR divergences\footnote{The loop integrals also
  exhibit UV divergences, but these are dealt with by
  renormalization.}, 
forcing them to cancel exactly 
against those coming from the unresolved emissions 
that we had to cut out above, order by order, 
making the complete answer for fixed
$k+\ell = n$ finite. 
Nonetheless, since this cancellation happens
between contributions that formally live in different phase spaces, 
a main aspect of loop-level higher-order calculations is how to
arrange for this cancellation in practice, either analytically or 
numerically, with many different methods currently on the market.

\begin{figure}[t]\vspace*{-0.15cm}
\begin{center}
\scalebox{0.90}{
\begin{tabular}{l}
\large\bf F @ LO\\[2mm]
\begin{loopsnlegs}[c]{p{0.25cm}|ccccc}
 \small 2&~\wbox{\pqcd[2]{0}} & \wbox{\pqcd[2]{1}} & \ldots \\[2mm]
 \small 1&~\wbox{\pqcd[1]{0}} & \wbox{\pqcd[1]{1}}  
   & \wbox{\pqcd[1]{2}} & \ldots \\[2mm]
 \small 0&~\gbox{\pqcd[0]{0}} & \wbox{\pqcd[0]{1}} 
   & \wbox{\pqcd[0]{2}} 
   & \wbox{\pqcd[0]{3}} & \ldots \\
\hline
& \small 0 & \small 1 & \small 2 & \small 3 & \ldots
 \end{loopsnlegs}\end{tabular}\hspace*{-1.0cm}
\raisebox{0.9cm}{\begin{minipage}{1.9cm}\center\includegraphics*[scale=0.23]{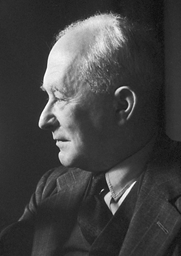}
\tiny Max Born, 1882-1970\\
Nobel 1954\end{minipage}}\hspace*{-0.15cm}
\begin{tabular}{l}
\large\bf F + 2 @ LO\\[2mm]
\begin{loopsnlegs}[c]{p{0.25cm}|ccccc}
 \small 2&~\wbox{\pqcd[2]{0}} & \wbox{\pqcd[2]{1}} & \ldots
&  \multicolumn{2}{l}{\tiny 
\colorbox{red}{\parbox[b]{1.75cm}{\raggedright\textcolor{white}{LO for $F+2$\\
      $\to \infty$ for $F+1$\\
      $\to \infty$ for $F+0$\\
}}
}} \\[2mm]
 \small 1&~\wbox{\pqcd[1]{0}} & \wbox{\pqcd[1]{1}}  
   & \wbox{\pqcd[1]{2}} & \ldots \\[2mm]
 \small 0&~\wbox{\pqcd[0]{0}} & \wbox{\pqcd[0]{1}} 
   & \gwbox{\pqcd[0]{2}} 
   & \wbox{\pqcd[0]{3}} & \ldots \\
\hline
& \small 0 & \small 1 & \small 2 & \small 3 & \ldots
 \end{loopsnlegs}
\end{tabular}}
\caption{Coefficients of the perturbative series covered by LO calculations. 
{\sl Left:} $F$ production at lowest order. {\sl Right:} $F+2\,$jets at LO, with the
  half-shaded box illustrating the restriction to the region of phase
  space with exactly 2 resolved jets.
  The total power of $\alpha_s$ for each coefficient is $n = k+\ell$.
\label{fig:loopsnlegs}}
\end{center}
\end{figure}

A convenient way of illustrating the terms of the perturbative series that
a given matrix-element-based calculation includes is given in
\figRef{fig:loopsnlegs}. 
In the left-hand pane, the shaded box corresponds to 
the lowest-order ``Born-level''\footnote{Photo from
    \ttt{nobelprize.org}} matrix element squared. This coefficient 
is non-singular and hence can be integrated over all of phase space,
which we illustrate by letting the shaded area fill all of the
relevant box. 
A different kind of leading-order calculation is illustrated in
the right-hand pane of \figRef{fig:loopsnlegs}, where the shaded box
corresponds to the lowest-order matrix element squared for
$F+2\,$jets. This coefficient diverges in the part of phase space
where one or both of the jets are unresolved (i.e., soft or collinear), 
  and hence integrations can only cover the hard part of
  phase space, which we reflect by only shading the upper half of
  the relevant box. 

\begin{figure}[t]
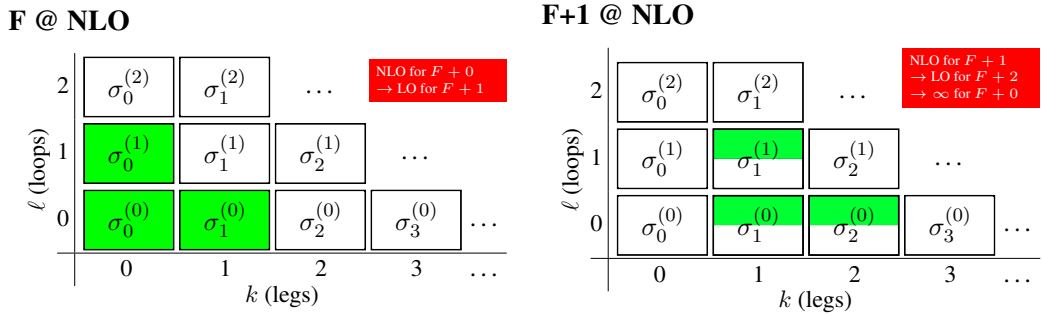

\begin{center}%
\scalebox{0.90}{
\begin{tabular}{l}
\large\bf F @ NLO\\[2mm]
\begin{loopsnlegs}[c]{p{0.25cm}|ccccc}
 \small 2&~\wbox{\pqcd[2]{0}} & \wbox{\pqcd[2]{1}} & \ldots &
 \multicolumn{2}{l}{\tiny 
\colorbox{red}{\parbox[b]{1.75cm}{\raggedright\textcolor{white}{NLO for
      $F+0$\\
$\to$ LO
      for $F+1$}}
}}
\\[2mm]
 \small 1&~\gbox{\pqcd[1]{0}} & \wbox{\pqcd[1]{1}}  
   & \wbox{\pqcd[1]{2}} & \ldots \\[2mm]
 \small 0&~\gbox{\pqcd[0]{0}} & \gbox{\pqcd[0]{1}} 
   & \wbox{\pqcd[0]{2}} &\wbox{\pqcd[0]{3}} & \ldots \\
\hline
& \small 0 & \small 1 & \small 2 & \small 3 & \ldots
 \end{loopsnlegs}
\end{tabular}\hspace*{-1mm}
\begin{tabular}{l}
\large\bf F+1 @ NLO\\[2mm]
\begin{loopsnlegs}[c]{p{0.25cm}|ccccc}
 \small 2&~\wbox{\pqcd[2]{0}} & \wbox{\pqcd[2]{1}} & \ldots &
 \multicolumn{2}{l}{\tiny 
\colorbox{red}{\parbox[b]{1.75cm}{\raggedright\textcolor{white}{NLO for
      $F+1$\\
     $\to$ LO
      for $F+2$ \\$\to \infty$ for $F + 0$}}
}}
\\[2mm]
 \small 1&~\wbox{\pqcd[1]{0}} & \gwbox{\pqcd[1]{1}}  
   & \wbox{\pqcd[1]{2}} & \ldots \\[2mm]
 \small 0&~\wbox{\pqcd[0]{0}} & \gwbox{\pqcd[0]{1}} 
   & \gwbox{\pqcd[0]{2}} 
   & \wbox{\pqcd[0]{3}} & \ldots \\
\hline
& \small 0 & \small 1 & \small 2 & \small 3 & \ldots
 \end{loopsnlegs}
\end{tabular}}
\caption{Coefficients of the perturbative series covered by NLO calculations. 
{\sl Left:} 
   $F$ production at NLO. {\sl Right:} $F+1\,$jet at NLO, with
  half-shaded boxes illustrating the restriction to the region of phase
  space with exactly 1 resolved jet.
  The total power of $\alpha_s$ for each coefficient is $n = k+\ell$.
\label{fig:loopsnlegs2}}
\end{center}
\end{figure}

\FigRef{fig:loopsnlegs2} illustrates the inclusion of NLO
virtual corrections. 
To prevent confusion, first a point on notation: by 
$\sigma_0^{(1)}$, we intend
\begin{equation}
\sigma_0^{(1)} = \int \dPS{0} \, 2\mrm{Re}[{\cal M}_{0}^{(1)} {\cal M}_0^{(0)*}]~,
\end{equation}
which is of order $\alpha_s$ relative to the Born
level. Compare, e.g., with the expansion of
\eqRef{eq:fixed-order} to order $k+\ell = 1$. In particular, $\sigma_0^{(1)}$
should \emph{not} be confused with the integral over the 1-loop matrix
element squared (which would be of relative order $\alpha_s^2$ and
hence forms part of the NNLO coefficient $\sigma_0^{(2)}$). Returning
to \figRef{fig:loopsnlegs2}, the unitary cancellations between real
and virtual singularities imply that we can now extend the
integration of the real correction in the left-hand pane over all of
phase space, while retaining a finite total cross section,
\begin{equation}
\begin{array}{rclll}
\displaystyle\sigma^\mrm{NLO}_0 & = &\displaystyle\int \dPS{0} |{\cal M}^{(0)}_0|^2 &\displaystyle + \int \dPS{0} \,
2\mrm{Re}[{\cal M}_{0}^{(1)} {\cal M}_0^{(0)*}] &\displaystyle + \int \dPS{1}
 |{\cal M}^{(0)}_1|^2\\[6mm]
 & = &\displaystyle \sigma^{(0)}_0
 & \displaystyle+\ \sigma^{(1)}_{0}&\displaystyle +\ \sigma^{(0)}_{1}
~,\end{array}
\end{equation}
where the divergence
caused by integrating the third term over all of phase space 
is canceled by that coming from the integration over loop momenta in
the second term. 
However,  if our starting point for the NLO
calculation is a process which already has a non-zero number of hard jets, we must
continue to impose that at least that number of jets must still be resolved in the
final-state integrations,
\begin{equation}
\begin{array}{rclll}
\displaystyle\sigma^\mrm{NLO}_1(\ptmin) & = 
 &\displaystyle
  \int_{\pt>\ptmin}\hspace*{-11mm} \dPS{1}\, |{\cal M}^{(0)}_1|^2
 &\displaystyle 
   + \int_{\pt>\ptmin}\hspace*{-11mm} \dPS{1} \, 
   2\mrm{Re}[{\cal M}_{1}^{(1)} {\cal M}_1^{(0)*}] 
 &\displaystyle 
   + \int_{\pt[1]>\ptmin}\hspace*{-11mm} \dPS{2}\,
   |{\cal M}^{(0)}_2|^2\\[6mm]
 & = &\displaystyle \sigma^{(0)}_1(\pt>\ptmin)
 & \displaystyle+\ \sigma^{(1)}_{1}(\pt>\ptmin)&\displaystyle 
   +\ \sigma^{(0)}_{2}(\pt[1]>\ptmin)
~,\end{array}\label{eq:sigmanlo1}
\end{equation}
where the restriction to at least one jet having 
$\pt>\ptmin$ has been illustrated in the right-hand 
pane of \figRef{fig:loopsnlegs2} by shading only the upper part of the
relevant boxes. In the last term in \eqRef{eq:sigmanlo1}, 
the notation $\pt[1]$ is used to
denote that the integral runs over the phase space in which at least
one ``jet'' (which may consist of one or two partons) must be resolved
with respect to $\ptmin$. Here, therefore, an explicit dependence on
the algorithm used to define ``a jet'' enters for the first time. This
is discussed in more details in the ESHEP lectures by Salam
\cite{Salam:2010zt}. 

\begin{figure}[t]
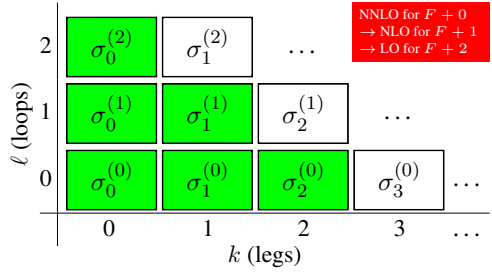

\scalebox{0.90}{\large\bf F @ NNLO}
\begin{center}
\scalebox{0.90}{
\begin{loopsnlegs}[c]{p{0.25cm}|ccccc}
 \small 2&~\gbox{\pqcd[2]{0}} & \wbox{\pqcd[2]{1}} & \ldots &
 \multicolumn{2}{l}{\tiny 
\colorbox{red}{\parbox[b]{1.75cm}{\raggedright\textcolor{white}{NNLO for
      $F+0$\\
     $\to$ NLO
      for $F+1$ \\$\to$ LO for $F + 2$}}
}}
\\[2mm]
 \small 1&~\gbox{\pqcd[1]{0}} & \gbox{\pqcd[1]{1}}  
   & \wbox{\pqcd[1]{2}} & \ldots \\[2mm]
 \small 0&~\gbox{\pqcd[0]{0}} & \gbox{\pqcd[0]{1}} 
   & \gbox{\pqcd[0]{2}} &\wbox{\pqcd[0]{3}} & \ldots \\
\hline
& \small 0 & \small 1 & \small 2 & \small 3 & \ldots
 \end{loopsnlegs}}
\caption{Coefficients of the perturbative series covered by an NNLO
  calculation. 
  The total power of $\alpha_s$ for each coefficient is $n =
  k+\ell$. Green  shading represents the full perturbative 
  coefficient at the respective $k$ and $\ell$.
\label{fig:loopsnlegs3}}
\end{center}
\end{figure}

To extend the integration to cover also the case of 2 unresolved jets,
we must combine the left- and right-hand parts of
\figRef{fig:loopsnlegs2} and add the new coefficient
\begin{equation}
\sigma_0^{(2)} = |{\cal M}_0^{(1)}|^2 + 2\mrm{Re}[ {\cal
    M}_0^{(2)}{\cal M}_0^{(0)*} ]~,
\end{equation}
 as illustrated by the diagram in \figRef{fig:loopsnlegs3}. 

\subsection{Infinite-Order QCD: Parton Showers\label{sec:parton-showers}}

In the preceding section, we noted two
conditions that had to be valid for fixed-order truncations of the
perturbative series to be valid: firstly, the strong coupling
$\alpha_s$ must be small for perturbation theory to be valid at
all. This restricts us to the region in which all scales $Q_i\gg
\Lambda_\mrm{QCD}$. We shall maintain this restriction in this
section, i.e., we are still considering \emph{perturbative QCD}. 
Secondly, however, in order to be allowed to \emph{truncate} the
perturbative 
series, we had to require  $\sigma_{k+1} \ll \sigma_k$, i.e., the
corrections at successive orders must become successively smaller, which --- due
to the enhancements from soft/collinear singular (conformal) dynamics
--- effectively restricted us to consider only the phase-space 
region in which all jets
are ``hard and well-separated'', equivalent to requiring all $Q_i/Q_j \approx
1$. In this section, we shall see how to lift this restriction,
extending the applicability of perturbation theory into regions that include scale
hierarchies, $Q_i \gg Q_j \gg \Lambda_\mrm{QCD}$, such as occur for
soft jets, jet substructure, etc.

In fact, the simultaneous restriction to all resolved scales being
larger than $\Lambda_{\mrm{QCD}}$ \emph{and} no large hierarchies is
extremely severe, if taken at face value. 
Since we collide and observe  hadrons ($\to$ low
scales) while
simultaneously wishing to study short-distance physics processes ($\to$ high
scales), it would appear trivial to conclude that fixed-order pQCD is not 
applicable to collider physics at all. So why do we still use it?

The answer lies in the fact that we actually never truly perform a fixed-order
calculation in QCD. Let us repeat the factorized formula for the cross
section,  \eqRef{eq:factorization}, 
\begin{equation}
\frac{\d{\sigma}}{\d{\obs}} = 
\sum_{a,b}\int_{0}^{1}\d{x_a}\d{x_b}\sum_F\int\!\dPS{F}\,
\pdf[h_1]{a}(x_a,\mu_F)\pdf[h_2]{b}(x_b,\mu_F)\,\frac{\d{\hat{\sigma}_{ab\to
      F}}}{\d{\hat\obs}}D_F(\hat{\obs}\to\obs,\mu_F)~.
\label{eq:factorizationrepeated}
\end{equation}
Although $\d{\sigma_{ab\to F}}$ does represent a fixed-order
calculation, the parton densities, $f_a^{h_1}$ and
$f_b^{h_2}$,  
include so-called resummations of perturbative corrections \emph{to
  all orders} from the initial scale of
order the mass of the proton, up to the factorization scale, $\mu_F$. 
Note that the oft-stated mantra that the PDFs are purely 
non-perturbative functions is therefore misleading.
 True, they are defined as essentially non-perturbative functions at some very low
scale, but, if $\mu_F$ is taken large, they necessarily incorporate a
significant amount of perturbative physics as well.
On the ``fixed-order side'', all we have left to ensure in 
$\d{\sigma_{ab\to F}}$ is then
that there are no large hierarchies remaining between $\mu_F$ and the
QCD scales appearing in \PS{F}. 
Likewise, in the final
state, the fragmentation functions, $D_F$, include infinite-order
resummations of perturbative corrections all the way \emph{from}
$\mu_F$ down to some low scale, with
similar caveats concerning mantras about their non-perturbative 
nature as for the PDFs. 

\subsubsection{Step One: Infinite Legs}
\begin{figure}[t]
\begin{center}
\small
\begin{tabular}{ccc}
\hspace*{0.5cm}\includegraphics*[scale=0.55]{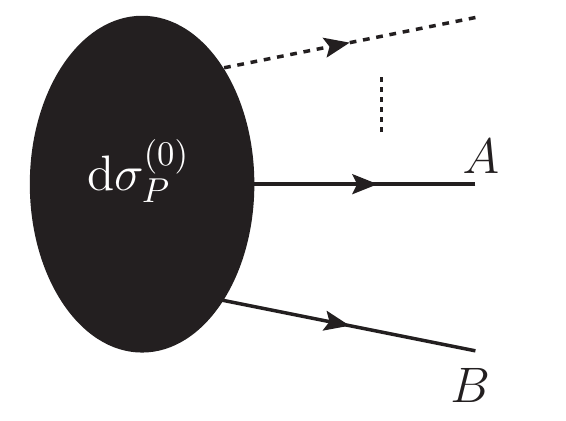}\hspace*{0.5cm}&
\hspace*{0.5cm}\includegraphics*[scale=0.55]{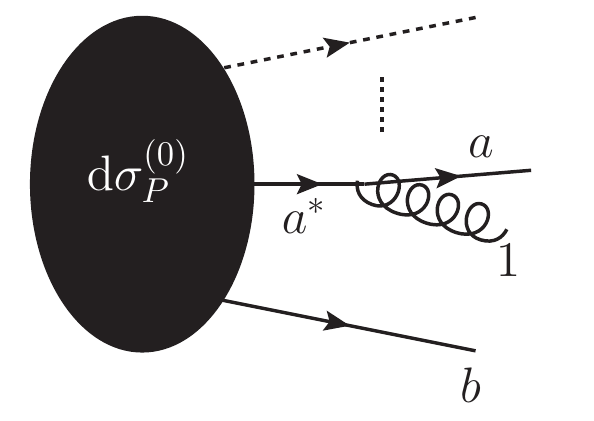}\hspace*{0.5cm}&
\hspace*{0.5cm}\raisebox{3mm}{\includegraphics*[scale=0.55]{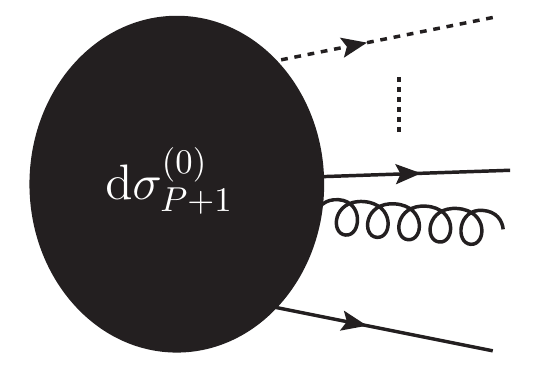}}\hspace*{0.5cm}\\[1mm]
\sl a) \mbox{Original Configuration: $\d{\sigma_{F}^{(0)}}$} & 
\sl b) \mbox{A contribution to $\d{\sigma_{F+1}^{(0)}}$} &
\sl c) \mbox{Recursion $\to$ $\d{\sigma^{(0)}_{F+n}}$}
\end{tabular}
\caption{{\sl a)} and {\sl b)} Illustration of the QCD
  singularities induced by on-shell 
  propagators. {\sl c)} The approximation obtained in
  the first step can be iterated to add additional legs. \label{fig:eikonal}}
\end{center}
\end{figure}

The infinite-order resummations that are 
included in objects such as the PDFs and FFs in
\eqRef{eq:factorizationrepeated} (and in their parton-shower equivalents)
rely on some very simple and powerful
properties of gauge field theories. One way to arrive at them is the
following; assume we have 
computed the Born-level cross section for some process,
$F$, and that this process contains some number of coloured
partons\footnote{Assume further that octet colour charges (gluons) may
  be represented as the sum of a colour triplet and an antitriplet charge --- 
compare, e.g., with the illustrations of
gluon colour flow, \FigsRef{fig:qg} and \ref{fig:gg}. This picture
of octets is correct up to corrections of order $1/N_C^2$, which will
be good enough for our purposes here.}. 
For each pair of (massless) colour-anticolour charges $A$ and $B$ in $F$, 
it is then a universal property of
QCD that the cross sections for $F+1\,$partons, 
$\d{\sigma^{(0)}_{F+1}}$  will include a factor
\begin{equation}
\d{\sigma^{(0)}_{F+1}} = g_s^2\left({\cal N}_{AB\to a1b} \,
\frac{\d{s_{a1}}}{s_{a1}}\frac{\d{s_{1b}}}{s_{1b}}~+~\mbox{less singular
  terms}\right)\times{\d{\sigma^{(0)}_F}},\label{eq:eikonal}
\end{equation}
where, for compactness, we have lumped some uninteresting normalization
factors\footnote{
I.e., ${\cal N}_{AB\to a1b}$ contains colour and phase-space 
normalization factors. Up to mildly non-universal corrections of order
$1/N_C^2$ (which  
depend on whether the emitting particles are quarks or gluons), it is ${\cal
  N}_{AB\to a1b} = 2C_A/(16\pi^2)$ } 
into ${\cal N}_{AB\to a1b}$,  $g_s^2 = 4\pi\alpha_s$ is the strong coupling,
$a$ and $b$ represent partons $A$ and $B$ after the branching (i.e.,
they include possible recoil effects) and $s_{i1}$ is
the invariant between parton $i$ and the emitted ``+1''
parton. Intuitively, this structure follows from the simple
observations illustrated by the left and middle panes of 
\figRef{fig:eikonal}; 
the Feynman diagram in which parton ``1'' is emitted
from the ``$a$'' (or ``$b$'') leg has a pole for
$s_{a1}\to 0$ ($s_{1b}\to 0$), corresponding to the 
intermediate propagator ``$a^*$'' (``$b^*$'') going on shell (middle
pane). Summing
the two and squaring them, i.e., including their mutual interference,
one obtains the structure in \eqRef{eq:eikonal}, which is called the
\emph{Eikonal} factor. 

The leading part of the singularity structure to which we have already
referred many times is clearly visible here: if we integrate over the
entire phase space including the region $s_{a1}\to 0$, $s_{1b}\to 0$,
we end up with a double pole. If we instead regulate the divergence by 
cutting off the integration at some minimal \emph{perturbative cutoff
  scale} $\mu_\mrm{IR}^2$, we end up with a logarithm squared of that
scale\footnote{The precise definition of $\mu_\mrm{IR}^2$ is not
  unique. Any scale choice that properly isolates the singularities
  from the rest of phase space will do, with some typical choices being, for
  example, invariant-mass and/or transverse-momentum scales.}. This is
a typical example of ``large logs'' being generated by the presence of scale
hierarchies.

Before we continue, it is worth noting that  \eqRef{eq:eikonal} is often
rewritten in other forms to emphasize specific aspects of it. 
One such rewriting is thus to reformulate the invariants
$s_{i1}$ appearing in \eqRef{eq:eikonal} 
in terms of energies and angles, 
\begin{equation}
s_{ij} = 2 E_iE_j\left(1-\cos\theta_{ij}\right)~. 
\end{equation}
Rewritten in this way, the differentials in \eqRef{eq:eikonal} become
\begin{equation}
\frac{\d{s_{a1}}}{s_{a1}}\frac{\d{s_{1b}}}{s_{1b}} \propto
\frac{\d{E_1}}{E_1}\frac{\d{\theta_{a1}}}{\theta_{a1}} + 
\frac{\d{E_1}}{E_1}\frac{\d{\theta_{1b}}}{\theta_{1b}} ~. 
\end{equation}
This kind of rewriting 
enables an intuitively appealing categorization of the singularities 
as related to vanishing energies and angles, called \emph{soft}
and \emph{collinear} limits, respectively. Although such formulations
have undeniably been helpful in obtaining many important results in
QCD, one should still keep in mind 
that Lorentz non-invariant formulations come with similar caveats and
warnings as do  gauge non-invariant formulations of quantum field
theory: while they can be practical to work with at 
intermediate stages of a calculation, one should be careful with
any physical conclusions that rely explicitly on them.
We shall therefore here restrict ourselves to a 
Lorentz invariant formalism based directly on
\eqRef{eq:eikonal}. The collinear limit is then replaced by a more
general \emph{single-pole} limit in which a single 
parton-parton invariant vanishes (as, \emph{for instance}, when 
a pair of partons become collinear),
while the soft limit is replaced by one in which two (or more) 
invariants involving the
same parton vanish simultaneously (as, for instance by that parton
becoming soft in a frame defined by two or more hard partons). This
avoids frame-dependent ambiguities from entering into the language,
at the price of a slight reinterpretation of what
is meant by collinear and soft.

Independently of rewritings and philosophy, 
the real power of \eqRef{eq:eikonal} lies in the fact
that it is \emph{universal}. Thus, for
\emph{any} process $F$, we can apply \eqRef{eq:eikonal} in order to
get an approximation for $\d{\sigma_{F+1}}$. We may then, for instance,  
take our newly obtained expression for $F+1$ as our arbitrary process
and crank \eqRef{eq:eikonal} again, to obtain an approximation for
$\d{\sigma_{F+2}}$, and so forth. What we have here is therefore a
very simple recursion relation that can be used to generate approximations to
leading-order cross sections with arbitrary numbers of additional legs. The quality
of this approximation is governed by how many terms besides the leading
one shown in \eqRef{eq:eikonal} are included in the game. Including
all possible terms, the most general form for the cross section at 
$F+n\,$ jets, restricted to the phase-space region above some 
infrared cutoff scale $\mu_{\mrm{IR}}$, has the following algebraic structure,
\begin{equation}
\sigma_{F+n}^{(0)} = \alpha_s^n \left( 
   \ln^{2n} 
 + \ln^{2n-1} 
 + \ln^{2n-2} 
 + \ldots 
 + \ln{}  
 + {\cal R} \right)  \label{eq:transcend}
\end{equation}
where we use the notation $\ln^\lambda$ without an argument  to denote 
generic functions of \emph{transcendentality} $\lambda$ (the logarithmic function
to the power $\lambda$ being a ``typical'' example of a function 
with transcendentality $\lambda$ appearing in cross section expressions, but
also dilogarithms and higher logarithmic functions\footnote{Note: 
due to the theorems 
that allow us, for instance, to rewrite dilogarithms in different
ways with logarithmic and lower ``spillover'' terms, the coefficients at each 
$\lambda$ are only well-defined up to reparametrization ambiguities
involving the  terms with transcendentality greater than $\lambda$.} of
transcendentality $>1$ should be implicitly understood to belong to
our notation $\ln^\lambda$). The last term, $\cal R$, represents a rational
function of transcendentality 0. We shall also use the nomenclature 
\emph{singular} and \emph{finite} for  the $\ln^\lambda$ and 
${\cal R}$ terms, respectively,  a terminology which reflects their respective
behaviour in the limit $\mu_{\mrm{IR}}\to 0$. 

The simplest approximation one can build on \eqRef{eq:transcend}, dropping
all but the leading $\ln^{2n}$ term in the parenthesis, 
is thus the \emph{leading-transcendentality} approximation. This
approximation is better known as 
the DLA (double logarithmic approximation), since it generates the
correct coefficient for terms which have two powers of logarithms for
each power of $\alpha_s$, while terms of lower transcendentalities are not
guaranteed to have the correct coefficients. In so-called LL
(leading-logarithmic) parton shower algorithms, one generally expects
to reproduce the correct coefficients for the $\ln^{2n}$ and
$\ln^{2n-1}$ terms. In addition, several formally subleading
improvements are normally also introduced in such algorithms 
(such as explicit momentum
conservation, gluon polarization and other 
spin-correlation effects
\cite{Richardson:2001df},
higher-order coherence effects, renormalization scale choices
\cite{Catani:1990rr}, finite-width effects \cite{Gigg:2008yc}, etc), 
as a means to improve the agreement
with some of the more subleading coefficients as well, if not in every
phase-space point then at least on average. 
Though LL showers do not magically acquire NLL
(next-to-leading-log) precision from such procedures, one therefore
still expects a significantly better average performance from them 
than from corresponding ``strict'' LL analytical resummations. A side
effect of this is that it is often possible to ``tune'' shower algorithms to give
better-than-nominal agreement with experimental distributions, by
adjusting the parameters controlling the treatment of subleading
effects. One should remember, however, that there is a limit to how
much can be accomplished in this way --- at some point, agreement with
one process will only come at the price of disagreement with another,
and at this point further tuning would be meaningless.  

\begin{figure}
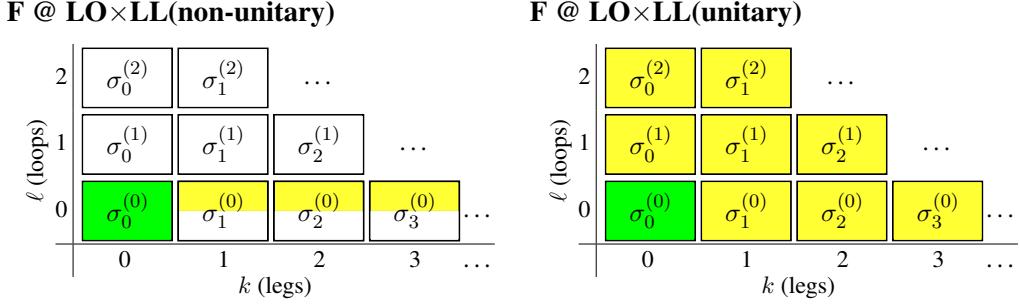

\begin{center}%
\scalebox{0.90}{
\begin{tabular}{l}
\large\bf F @ LO$\times$LL(non-unitary) \\[2mm]
\begin{loopsnlegs}[c]{p{0.25cm}|ccccc}
 \small 2&~\wbox{\pqcd[2]{0}} & \wbox{\pqcd[2]{1}} & \ldots &
\\[2mm]
 \small 1&~\wbox{\pqcd[1]{0}} & \wbox{\pqcd[1]{1}}  
   & \wbox{\pqcd[1]{2}} & \ldots \\[2mm]
 \small 0&~\gbox{\pqcd[0]{0}} & \ywbox{\pqcd[0]{1}} 
   & \ywbox{\pqcd[0]{2}} &\ywbox{\pqcd[0]{3}} & \ldots \\
\hline
& \small 0 & \small 1 & \small 2 & \small 3 & \ldots
 \end{loopsnlegs}
\end{tabular}\hspace*{-1mm}
\begin{tabular}{l}
\large\bf F @ LO$\times$LL(unitary)\\[2mm]
\begin{loopsnlegs}[c]{p{0.25cm}|ccccc}
 \small 2&~\ybox{\pqcd[2]{0}} & \ybox{\pqcd[2]{1}} & \ldots & 
\\[2mm]
 \small 1&~\ybox{\pqcd[1]{0}} & \ybox{\pqcd[1]{1}}  
   & \ybox{\pqcd[1]{2}} & \ldots \\[2mm]
 \small 0&~\gbox{\pqcd[0]{0}} & \ybox{\pqcd[0]{1}} 
   & \ybox{\pqcd[0]{2}} &\ybox{\pqcd[0]{3}} & \ldots \\
\hline
& \small 0 & \small 1 & \small 2 & \small 3 & \ldots
 \end{loopsnlegs}
\end{tabular}}
\caption{Coefficients of the perturbative series covered by LO + LL 
  calculations, {\sl Left:} without imposing unitarity, and {\sl
    Right:} imposing unitarity order by order for each $n =
  k+\ell$. Green (darker) shading represents the full perturbative
  coefficient at the respective $k$ and $\ell$. Yellow (lighter)
  shading represents an LL approximation to it. Half-shaded boxes
  indicate phase spaces in which we are prohibited from integrating
  over the IR singular region, as discussed in
  \secsRef{sec:fixed-order} and \ref{sec:matching}.
\label{fig:LL}}
\end{center}
\end{figure}
Applying such an iterative process on a Born-level cross section, one
obtains the description of the full perturbative series illustrated in
the left-hand pane of \figRef{fig:LL}. The yellow (lighter) shades
used here for $k\ge 
1$ indicate that the coefficient obtained is not the exact one, but rather an 
approximation to it that only gets its leading singularities
right. However, since this is still only an approximation to
infinite-order \emph{tree-level} cross sections 
(we have not yet included any virtual corrections), 
we cannot yet integrate this approximation over all of
phase space, as illustrated by the yellow boxes being only
half filled on the left-hand side of \figRef{fig:LL}; the summed total cross section
would still be infinite. This particular approximation would therefore
still appear to be  
very useless indeed --- on one hand, 
it is only guaranteed to get the singular terms right, but on the
other, it does not actually allow us to  integrate over the singular
region. In order to obtain a truly \emph{all-orders} calculation, the
constraint of unitarity must also be explicitly imposed, which
furnishes an approximation to all-orders loop corrections as well.
Let us therefore emphasize that \figRef{fig:LL} is included for
pedagogical purposes only; all resummation calculations, whether
analytical or parton-shower based, include virtual corrections as well
and consequently yield finite total cross sections, as will now be described.

\subsubsection{Step Two: Infinite Loops}

Order-by-order unitarity, such as used in the KLN theorem, implies
that the singularities caused by integration over unresolved radiation
in the tree-level matrix elements must be canceled, order by order, by
equal but opposite-sign singularities in the virtual corrections at
the same order. That is, from \eqRef{eq:eikonal}, we immediately 
know that the 1-loop correction to $\d{\sigma_F}$ \emph{must} contain a
term, 
\begin{equation}
\d{\sigma^{(1)}_{F}} = - g_s^2 \, {\cal N}_{AB\to a1b} \,
\d{\sigma^{(0)}_F}\int
\frac{\d{s_{a1}}}{s_{a1}}\frac{\d{s_{1b}}}{s_{1b}}~+~\mbox{less singular
  terms},\label{eq:eikonalv}
\end{equation}
that cancels the divergence coming from \eqRef{eq:eikonal}
itself. Further, since this is universally true, we may apply
\eqRef{eq:eikonalv} again to get an approximation to the 
corrections generated by \eqRef{eq:eikonal} at the next order and so
on. By adding such terms explicitly, order by order, we may now
bootstrap our way around the entire perturbative series, using
\eqRef{eq:eikonal} to move horizontally and \eqRef{eq:eikonalv} to
move along diagonals of constant $n=k+\ell$. 
Since real-virtual cancellations are now explicitly restored, we may
finally extend the integrations over all of phase space, resulting in
the picture shown on the right-hand pane of \figRef{fig:LL}. 

The right-hand pane, not the left-hand one, 
corresponds to what is actually done in
\emph{resummation} calculations, both of the analytic and
parton-shower types\footnote{In the way these calculations are 
formulated in practice, they
in fact rely on one additional property, called exponentiation, that allows us
to move along straight vertical lines in the loops-and-legs diagrams. However, 
since the two different directions furnished by \eqsRef{eq:eikonal} and
\eqref{eq:eikonalv} are already sufficient to move freely in the
full 2D coefficient space, we shall use exponentiation
without extensively justifying it here.}.
Physically, there is a significant and intuitive meaning to the imposition of
unitarity, as follows. 

Take a jet algorithm, with some measure of jet resolution, $Q$, and
apply it to an arbitrary sample of events, say dijets. At a very crude
resolution scale, corresponding to a high value for $Q$, 
you find that everything is clustered back to a dijet configuration,
and the 2-jet cross section is equal to the total inclusive cross
section,
\begin{equation}
\sigma_\mrm{tot} = \sigma_{F;\mrm{incl}} ~.
\end{equation}
At finer resolutions, decreasing $Q$, you see that 
some events that were previously classified as 2-jet events contain
additional, lower-scale jets, that you can now resolve, and hence
those events now migrate to the 3-jet bin, while the total inclusive
cross section of course remains unchanged,
\begin{equation}
\sigma_\mrm{tot} = \sigma_{F;\mrm{incl}} = \sigma_{F;\mrm{excl}}(Q)
+ \sigma_{F+1;\mrm{incl}}(Q)~,
\end{equation}
where ``incl'' and ``excl'' stands for inclusive and exclusive cross
sections\footnote{$F$ \emph{inclusive} $=$ $F$ plus
  anything. $F$ \emph{exclusive} $=$ $F$ and only
  $F$. Thus, $\sigma_{F;\mathrm{incl}}=\sum_{\mrm{k}=
    0}^{\infty}\sigma_{F+k;\mrm{excl}}$},  
respectively, 
and the $Q$-dependence in the two terms on the 
right-hand side must cancel so that the total inclusive cross
section is independent of $Q$. Later, some 3-jet events now migrate 
further, to 4 and higher jets, 
while still more 2-jet events migrate \emph{into} the 3-jet
bin, etc. For arbitrary $n$ and $Q$, we have
\begin{equation}
\sigma_{F+n;\mrm{incl}}(Q) = \sigma_{F;\mrm{incl}} - \sum_{m=0}^{n-1} \sigma_{F+m;\mrm{excl}}(Q)~.
\end{equation}
This equation expresses the trivial fact that the cross section for
$n$ or more jets can be computed as the total inclusive cross section for $F$
minus a sum over the cross sections for $F$ + exactly $m$ jets including
all $m<n$. On the theoretical side, it is these negative terms which must
be included in the calculation, for each order $n=k+\ell$, to restore
unitarity. Physically, they 
express that, at a given scale $Q$, a given event will be classified
as having \emph{either} 0, 1, 2, or whatever jets. Or, equivalently,
for each event we gain in the 3-jet bin as $Q$ is lowered, we must loose one
event in the 2-jet one; the negative contribution to the 2-jet bin
is exactly minus the integral of the positive contribution to the
3-jet one, and so on. We may perceive of this \emph{detailed balance}
 as an \emph{evolution} of the event structure with $Q$, for each 
event, which is effectively what is done in parton-shower
algorithms, to which we shall return in \SecRef{sec:Markov}.

\section{Soft QCD}
In a complete high-energy collision, many different physics
(sub-)processes contribute to the total observed activity. 
We here give a very brief overview of the main aspects of soft
QCD that are relevant for hadron-hadron collisions, such as parton
distribution functions, minimum-bias and soft-inclusive physics, and
the so-called ``underlying event''. This will be kept at a
strictly pedestrian level and is largely based on the review in
\cite{Buckley:2011ms}. 
A discussion of the \emph{modeling} of
these components, as well as a discussion of the process of hadronization, 
is deferred to the relevant parts of \SecRef{sec:MC} on Monte Carlo
event generators.  

\subsection{Parton Densities \label{sec:pdfs}}
\begin{figure}[t]
\begin{center}
\includegraphics*[scale=0.5]{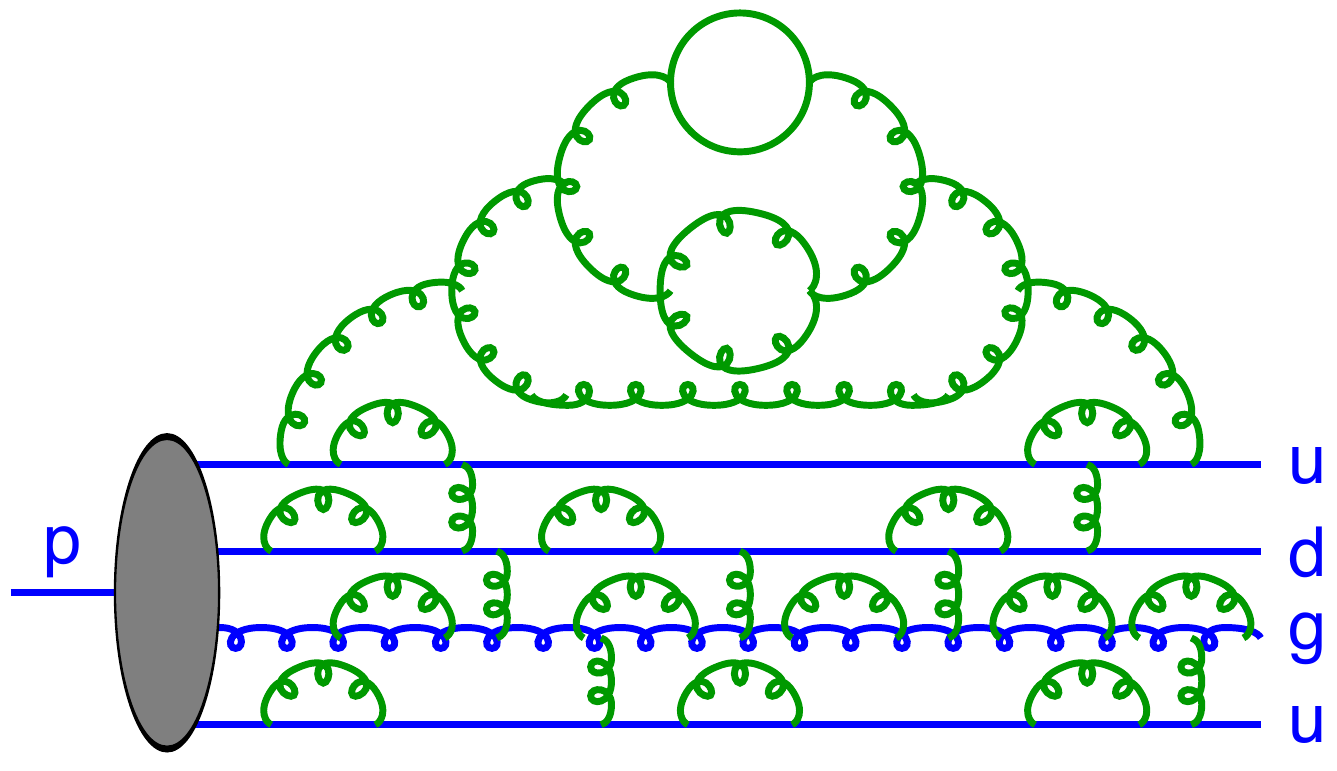}
\caption{Illustration (from \cite{Sjostrand:2006su}) 
  of partonic fluctuations inside a proton beam. \label{fig:pdfs}}
\end{center}
\end{figure}

Physically, parton densities express the fact that hadrons are
composite, with a time-dependent structure, illustrated in
\figRef{fig:pdfs}. 
More formally, they are defined
by the factorization theorem discussed in \SecRef{sec:factorization}. 
Occasionally, 
the words \emph{structure functions} and \emph{parton densities} are used
interchangeably. However, there is a very important distinction
between the two, which we find often in (quantum) physics: one is a
physical observable, the other is a ``fundamental'' quantity extracted
from it. 

Structure functions, such as $F_2$, are completely unambiguous 
physical observables, which can be measured,  
for instance, in DIS processes. (For a definition, see, e.g.,
\cite{Nakamura:2010zzi}.)
From these, and other observables, a set of more fundamental and
theoretically useful objects, parton density functions (PDFs), can be
extracted, but there is a price; since the parton densities are not, themselves,
physically observable,  they can only be defined within a
specific \emph{factorization scheme}, order by order in perturbation
theory. The only exception is at leading order, at which they have a very
simple physical interpretation, as the probability of finding a quark
of a given flavor and carrying a given momentum fraction, $x$, inside a
hadron of a given type, probed at a specific scale, $Q^2$. They are
then related to the structure function $F_2$ by their charge-weighted
momentum sum, 
\begin{equation}
\underbrace{F_2(x,Q^2)}_{\mbox{\small Physical Observable}} = \sum_i
e_i^2 x \hspace*{-5mm} \underbrace{f_i(x,Q^2)}_{\mbox{\small Extracted Quantity}}~,
\end{equation}
where $f_i$ denotes the parton density for a parton of flavor/type
$i$. When going to higher orders, we tend to keep the simple intuitive
picture from leading order in mind, but one should be aware that the
fundamental relationship is now more complicated, and that the parton
densities no longer have a clear probabilistic interpretation.

The reader should also be aware that there is currently a significant
amount of debate concerning many aspects of PDF definitions and usage:
\begin{itemize}
\item The ``initial condition'' for the PDFs, i.e., their shape in $x$
  at some low value of $Q_F^2$, and other constraints imposed on their 
  evolution, such as positivity, flavour symmetries, treatment of
  mass effects, and extrapolation beyond the fit region. 
  Each PDF group has its own particular ideology when it comes to
  these issues, and while the differences caused by these choices in
  well-constrained regions may appear small, the 
  user should be warned that large differences can occur when
  extrapolating, e.g., to small $x$, or for observables that are
  particularly sensitive, e.g., to flavour symmetries, etc.
\item Using PDFs extracted using higher-order matrix elements in
  lower-order calculations, as, e.g., when using NLO PDFs as input to
  an LO calculation. In principle, the higher-order PDFs are better
  constrained and the difference between, e.g., an NLO and an LO set
  should formally be beyond LO precision, so that one might be tempted
  to simply use the highest-order available PDFs for any calculation. 
  However, as described in 
  \secRef{sec:parton-showers}, it is often possible to partly
  absorb higher-order terms into lower-order  coefficients. In the
  context of PDFs, the fit parameters of lower-order PDFs will 
  effectively attempt to ``compensate'' for missing higher-order
  contributions in the matrix 
  elements. To the extent those higher-order contributions are
  \emph{universal}, this is both desirable and
  self-consistent. However, this will only give an improvement when used
  with matrix elements \emph{at the same order} as those used to
  extract the PDFs. It is therefore quite possible that NLO PDFs used
  in conjunction with LO matrix elements give a \emph{worse} agreement
  with data than LO PDFs do.
\item PDF uncertainties. Uncertainty estimates for 
  PDF determinations is a highly delicate procedure,
  owing in part to the diversity of the data sets that enter into the
  fitting procedures (especially since some data sets appear to have
  ``tensions'', i.e., mutual incompatibilities, between them),
  but also the differences in philosophy mentioned above (e.g., on 
  parametrizations and evolution constraints) can cause
  apparent incompatibilities between different sets which are hard to
  give precise uncertainty estimates for. Currently,
  a consensus on meaningful uncertainty estimates is slowly building,
  though future years are likely to see continued active discussions on how
  best to address this topic. 
\item How to use PDFs in conjunction with parton-shower Monte Carlo
  codes. The initial-state showers in a Monte Carlo model are
  essentially supposed to mimic the evolution in the PDFs, and vice
  versa. However, since PDF fits are not done with MC codes, but
  instead use analytical resummation models that are not identical to
  their MC counterparts, the PDF fits are essentially ``tuned'' to a
  slightly different resummation than that incorporated in a given MC
  model. Since both types of calculations are supposed to be accurate
  at least to LL, any difference between them should in principle be
  subleading. In practice, not much is known about the size and impact
  of this ambiguity, so we mention it mostly to make sure the reader
  is aware that it exists. Known differences include: 
  the size of phase space (purely collinear massless PDF evolution
  vs.\ the finite-transverse-momentum massive  MC phase space),  
  the treatment of momentum conservation and recoil effects,
  additional higher-order effects explicitly or implicitly included in
  the MC evolution, choice of renormalization scheme
  and scale, and, for those MC algorithms that do not rely on
  collinear (DGLAP, see \cite{Dissertori:2003pj}) 
  splitting kernels (e.g., the various kinds of dipole
  evolution algorithms, see \cite{Bern:2008ef}), differences in the
  effective factorization 
  scheme. 
\end{itemize}

\subsection{Elastic and Inelastic Components of $\sigma_{\mrm{tot}}$ \label{sec:total}} 
Elastic scattering 
consists of  all reactions of the type 
\begin{equation}
A(p_A)B(p_B)\to A(p_A')B(p_B')~,
\end{equation}
where $A$ and $B$ are particles
carrying momenta $p_A$ and $p_B$, respectively. Specifically, 
the only exchanged quantity is momentum; all quantum numbers and
masses remain unaltered, and no new particles are produced. 
Inelastic scattering covers everything else, i.e., 
\begin{equation}
 AB\to X \ne AB~,
\end{equation} 
where $X\ne AB$ signifies that one 
or more quantum numbers are changed, and/or more particles are
produced. The total hadron-hadron cross section can thus be written as
a sum of these two physically distinguishable components, 
\begin{equation}
\sigma_{\mathrm{tot}}(s) = 
\sigma_{\mathrm{el}}(s) +
\sigma_{\mathrm{inel}}(s)~, 
\end{equation}
where $s=(p_A+p_B)^2$ is the beam-beam centre-of-mass energy squared. 

If $A$ and/or $B$ are not elementary, the inelastic final states may be
further divided into ``diffractive'' and ``non-diffractive''
topologies. This is a qualitative classification, usually based on
whether the final state looks like
the decay of an excitation of the beam particles 
(diffractive\footnote{An example of a process that would be labeled as
  diffractive would be if one the protons is excited to a
$\Delta^+$ which then decays back to $p^++\pi^0$, without anything else
  happening in the event.  In
  general, a whole tower of possible diffractive excitations are
  available, which in the continuum limit can be described by a mass
  spectrum falling roughly as $\d{M^2}/M^2$.}), 
or not (non-diffractive), or upon the presence of a
large rapidity gap somewhere in the final state which would separate
such excitations. 

Given that an event has been labeled as diffractive, either within the
context of a theoretical model, or by a final-state observable, 
we may distinguish
between three different classes of diffractive topologies, which it is
possible to distinguish between physically, at least in principle. 
In double-diffractive (DD) events, both of the beam particles are
diffractively excited and hence none of them survive the collision
intact. In single-diffractive (SD) events, only one of the beam
particles gets excited and the other survives intact. The last 
diffractive topology is  central diffraction (CD),  in which 
both of the beam particles survive intact, leaving an excited system
in the central region between them. (This latter topology includes
``central exclusive production'' where a single particle is produced
in the central region.)  
That is,  
\begin{equation}
\sigma_{\mathrm{inel}}(s) = 
\sigma_{\mathrm{SD} }(s)
+
\sigma_{\mathrm{DD}}(s) +
\sigma_{\mathrm{CD}}(s) + 
\sigma_{\mathrm{ND}}(s) ~, \label{eq:diff}
\end{equation}
where ``ND'' (non-diffractive, here understood not to include elastic
scattering) contains no gaps in the event
consistent with the chosen definition of diffraction. Further, 
each of the diffractively excited systems in the events labeled SD,
DD, and CD, respectively, may in principle consist of
several subsystems with gaps between them. Eq.~(\ref{eq:diff}) may 
thus be defined to be exact, within a specific definition of
diffraction, even in the presence of multi-gap events. 
Note, however, that different
theoretical models almost always use different (model-dependent) definitions of
diffraction, and therefore the individual components in one model are
in general not directly comparable to those of another. It is
therefore important that data be presented at the level of physical
observables if unambiguous conclusions are to be drawn from them.

\subsection{Minimum-bias and soft inclusive physics}
The term ``minimum-bias'' (MB) is an experimental term, used to define a
certain class of events that are selected with the minimum
possible trigger bias, to ensure they are as inclusive as
possible\footnote{A typical min-bias trigger would thus be the requirement
  of at least one measured particle in a given rapidity region, so
  that  all events which produce at least one observable particle would be
  included, which must, indeed, be considered the minimal possible bias. 
  In principle, \emph{everything} is a subset of
minimum-bias, including both hard and soft processes. 
However, compared to the total minimum-bias cross section, 
the fraction that is made up of hard processes is only a very small
tail. Since only a tiny fraction of the total
minimum-bias rate can normally be stored, the minimum-bias sample
would give quite poor statistics if used for hard physics studies. Instead,
separate dedicated 
hard-process triggers are typically included in addition to the
minimum-bias one, in order to ensure maximal statistics also for hard
physics processes.}. 
In theoretical contexts, 
the term ``minimum-bias'' is often used with a slightly different
meaning; to denote specific (classes of) inclusive soft-QCD
subprocesses in a given model. 
Since these two usages are not exactly identical, in these lectures 
we have chosen to reserve the term ``minimum bias''  to pertain strictly to
definitions of experimental measurements, and instead use  
the term ``soft inclusive'' physics as a generic descriptor for the
class of processes which generally dominate the various experimental
``minimum-bias'' measurements in theoretical models. This parallels
the terminology used in the review \cite{Buckley:2011ms}, from which
most of the discussion here has been adapted. 
See \eqRef{eq:diff} above for a compact overview of the types of
physical processes that contribute to minimum-bias data samples.
For a more detailed description of Monte Carlo models of this physics,
in particular ones based on Multiple Parton Interactions (MPI), see
\SecRef{sec:mc-ue}.  

\subsection{Underlying event and jet pedestals} 
In events containing a hard parton-parton interaction,
the underlying event (UE) can be roughly conceived of as
the \emph{difference} between QCD with and without including the
remnants of the original beam hadrons.  Without such ``beam
remnants'', only the hard interaction itself, and its  
associated parton showers and hadronization, would contribute to the
observed particle production. In reality, after the partons that
participate in the hard interaction have been taken out, 
the remnants still contain whatever is left of the incoming beam hadrons,
including also a partonic substructure, which leads to the possibility
of  ``multiple parton interactions''
(MPI), as  will be discussed  in \secRef{sec:mc-ue}.  
Due to the simple fact that the remnants are not empty, an
``underlying event'' will always be there --- but how much additional energy
does it deposit in a given measurement region? A quantifation of this
can be obtained, for instance, by comparing measurements of the UE to the
average activity in minimum-bias events at the same
$\sqrt{s}$. Interestingly, it turns out that the underlying event is 
much more active, with larger fluctuations, than the average MB event.  
This is called the jet pedestal effect (hard jets sit on top of a
higher-than-average ``pedestal'' of underlying activity), and is
interpreted as follows. When two hadrons collide at non-zero impact
parameter, high-$p_\perp$ interactions can only take place  
inside the overlapping region. 
Imposing a hard trigger therefore statistically
biases the event sample toward more central collisions, which will also
have more underlying activity.  See \SecRef{sec:mc-ue} for a more
detailed description of Monte Carlo models of this physics, based on
MPI. 

\section{Monte Carlo Event Generators \label{sec:MC}}

In this section, we discuss the physics of Monte Carlo generators and
their mathematical foundations, at an introductory level. We shall attempt
to convey the main ideas as clearly as possible without burying them
in an avalanche of technical details. References to
more detailed discussions are included where applicable.
We assume the reader is already
familiar with the contents of the preceding sections of this report,
in particular \secRef{sec:fixed-order} on matrix
elements and  \secRef{sec:parton-showers} on parton showers. 
Several of the discussions rely on material from the recent more comprehensive 
review in \cite{Buckley:2011ms}, which also contains brief descriptions of the
physics implementations of each of the main general-purpose event generators on the
market, together with a guide on how to use (and not use) generators
in various connections, and a collection of comparisons to important experimental
distributions. We highly recommend readers to obtain a copy of that
review, as it is the most comprehensive and up-to-date review of event
generators currently available. Another useful and pedagogical review
on event generators is contained in the 2006 ESHEP lectures
by Sj\"ostrand \cite{Sjostrand:2006su}, with a more recent update in
\cite{Sjostrand:2009ad}. 

\subsection{Perturbation Theory with Markov Chains \label{sec:Markov}}
Consider again the Born-level cross section for an arbitrary hard process,
$F$, differentially in an arbitrary infrared-safe observable $\cal O$,
as obtained from \eqRef{eq:fixed-order}:
\begin{equation}
\left.   \frac{\d{\sigma^{(0)}_F}}{\d{\cal O}}\right\vert_{\mbox{\textcolor{black}{Born}}}
 = \int \dPS{F} \ |{\cal M}_F^{(0)}|^2 \ \delta({\cal
   O}-{\cal O}(\PS{F}))~,
\label{eq:starting}
\end{equation}
where the integration runs over the full final-state on-shell phase space of
$F$ (this expression and those below would also apply to hadron collisions
were we to include integrations over the parton distribution functions
in the initial state), and the $\delta$ function projects out a
1-dimensional slice defined by $\cal O$ evaluated on the 
set of final-state momenta which we denote $\PS{F}$.
 
To make the connection to parton showers, 
we insert an operator, ${\cal S}$, that acts on the Born-level
final state \emph{before} the observable is evaluated, i.e., 
\begin{equation}
\left.   \frac{\d{\sigma_F}}{\d{\cal
    O}}\right\vert_{\mbox{\textcolor{black}{${\cal S}$}}}
 = \int \d{\Phi_F} \ |{\cal M}_F^{(0)}|^2 \ {\cal S}(\PS{F},{\cal O})~.
\end{equation}
Formally, this operator --- the evolution operator --- will be
responsible for generating all (real and virtual) 
higher-order corrections to the Born-level expression.
The measurement $\delta$ function appearing explicitly in
\eqRef{eq:starting} is now implicit in ${\cal S}$.

Algorithmically, parton showers cast $\cal S$ as an iterative Markov
(i.e., history-independent) chain, 
with an evolution parameter, $Q_E$, that formally 
represents the factorization scale of the event, below which all
structure is summed over inclusively. Depending on the particular
choice of shower algorithm, $Q_E$ may be defined as a parton
virtuality (virtuality-order showers), as a transverse-momentum scale
($p_\perp$-ordered showers), or as a combination of
energies times angles (angular ordering). Regardless of the specific
form of $Q_E$, 
the evolution parameter will go towards zero as the Markov chain
develops, and the event structure
will become more and more exclusively resolved. A transition from a
perturbative evolution to a non-perturbative one can also be
inserted, when the evolution reaches an appropriate scale, typically 
around $1$~GeV. This scale thus 
represents the lowest perturbative scale that can appear in the
calculations, with all perturbative corrections below it
summed over inclusively.

Working out the precise form that $\cal S$ must have in order to give the
correct expansions discussed in \secRef{sec:parton-showers} takes a
bit of algebra, and is beyond the scope we aim to cover in these
lectures. Heuristically, the procedure is as follows. 
We noted that the singularity structure of QCD is universal
and that at least its first few terms are known to us. We also saw
that we could iterate 
that singularity structure, using universality and unitarity, 
thereby bootstrapping our way around the entire perturbative
series. This was illustrated by the right-hand pane of 
\figRef{fig:LL} in \secRef{sec:parton-showers}. 

Skipping intermediate steps, the form of the all-orders pure-shower
Markov chain, for the evolution of an event between two scales $Q_{E1}
> Q_{E2}$, is,  
\begin{equation}\begin{array}{rcl}
\displaystyle \hspace*{-1mm} {\cal S}(\PS{F},Q_{E1},Q_{E2},\obs) \hspace*{-1mm}
&  \hspace*{-1mm}= \hspace*{-1mm} &\displaystyle \hspace*{-1mm}
  \underbrace{
\Delta(\PS{F},Q_{E1},Q_{E2})\ 
\delta\left(\obs-\obs(\PS{F})\right) 
}_{\mbox{$F+0$ exclusive above $Q_{E2}$}}\\[9mm]
& & \hspace*{-1.2cm}+ \displaystyle 
\underbrace{
\sum_r \int_{Q_{E2}}^{Q_{E1}}
\frac{\dPS[r]{F+1}}{\dPS{F}} 
\ S_r(\PS{F+1}) \ \Delta(\PS{F},Q_{E1},Q_{F+1})
 \ {\cal S}(\PS{F+1},Q_{F+1},Q_{E2},\obs)}_{\mbox{$F+1$ inclusive above $Q_{E2}$}}
~,\label{eq:markov} \hspace*{-1mm}
\end{array}
\end{equation}
with the so-called \emph{Sudakov factor},
\begin{equation} 
\Delta(\PS{F},Q_{E1},Q_{E2}) = \exp\left[-\sum_r\int_{Q_{E2}}^{Q_{E1}} 
\frac{\dPS[r]{F+1}}{\dPS{F}} S_r(\PS{F+1}) \right]~, \label{eq:sudakov}
\end{equation}
defining the probability that there is \emph{no evolution} (i.e., no
emissions) between the scales $Q_{E1}$ and $Q_{E2}$, according to the
\emph{radiation functions} $S_r$ to which we shall return below. 
The term on the first line of
\eqRef{eq:markov} thus represents all events that \emph{did not}
evolve as the resolution scale was lowered from $Q_{E1}$ to $Q_{E2}$,
while the second line contains a sum and phase-space integral over
those events that \emph{did} evolve --- including the insertion of ${\cal
  S}(\PS{F+1})$ representing the possible further evolution of the
event and completing the iterative definition of the Markov chain. 

The factor $\dPS[r]{F+1}/\dPS{F}$
defines the chosen phase space factorization. Our favourite is the
so-called dipole-antenna factorization, whose principal virtue is that
it is the simplest Lorentz invariant 
factorization which is simultaneously exact over
all of phase space while only involving on-shell momenta. For
completeness, its form is
\begin{equation}  
\frac{\dPS[r]{F+1}}{\dPS{F}} = \frac{\dPS[r]{3}}{\dPS{2}} =
\d{s_{a1}} \d{s_{1b}} \frac{\d{\phi}}{2\pi} \frac{1}{16\pi^2
  s_r}~, \label{eq:phasespace} 
\end{equation}
which involves just one colour-anticolour pair for each $r$, with
invariant mass squared $s_r = (p_a+p_1+p_b)^2$. 
Other choices, such as purely collinear ones (only exact in the
collinear limit \emph{or} involving explicitly off-shell momenta), 
more global ones involving all partons in the event (more complicated,
in our opinion), or less global ones with a single parton playing the
dominant role as emitter, are also possible, again depending 
on the specific algorithm considered.

The radiation functions $S_r$ obviously play a crucial role in these
equations, driving the emission probabilities. For example, if 
$S_r \to 0$, then  $\Delta \to \exp(0) = 1$ and all events stay in the
top line of \eqRef{eq:markov}. Thus, in regions of phase  
space where $S_r$ is small, there is little or no evolution. 
Conversely, for $S_r\to \infty$, we have $\Delta\to 0$, implying that \emph{all}
events evolve. 
One possible choice for the radiation functions $S_r$ was
implicit in \eqRef{eq:eikonal}, in which we took them to include only the leading
(double) singularities, with $r$ representing colour-anticolour
pairs. In general, the shower
may exponentiate the entire set of universal singular terms, or only
a subset of them (for example, the terms leading in the number of colours
$N_C$), which is why we here let the explicit form of $S_r$ be
unspecified. Suffice it to say that 
in traditional parton showers, $S_r$ would simply be the DGLAP
splitting kernels (see, e.g., \cite{Dissertori:2003pj}), 
while they would be so-called dipole or antenna radiation
functions in the various dipole-based approaches to QCD (see, e.g.,
\cite{Gustafson:1987rq,Catani:1996vz,GehrmannDeRidder:2005cm,Schumann:2007mg,Giele:2011cb,LopezVillarejo:2011ap}).   

The procedure for how to technically ``construct'' a shower algorithm
of this kind, using random numbers to generate scales distributed according to
\eqRef{eq:markov}, is described more fully in \cite{Giele:2011cb},
using a notation that closely parallels the one used here. The
procedure is also described at a more technical level in the review
\cite{Buckley:2011ms}, though using a slightly different
notation. Finally, a pedagogical introduction to Monte Carlo
methods in general can be found in \cite{James:1980yn}. 

\subsection{Matching \label{sec:matching}}
The essential problem that leads to matrix-element/parton-shower
matching can be illustrated in a very simple way.
\begin{figure}
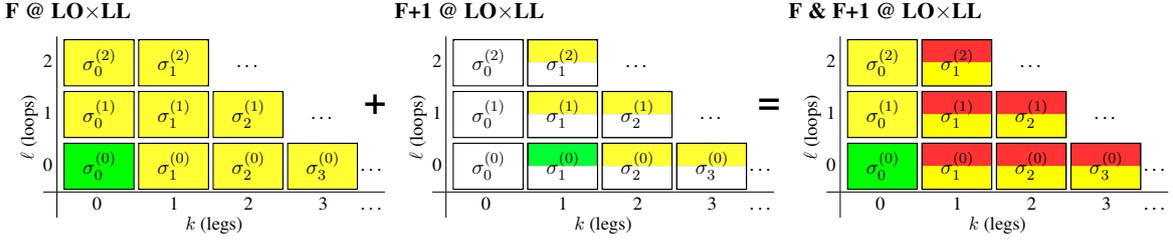

\begin{center}%
\scalebox{0.70}{\begin{tabular}{l}
\large\bf F @ LO$\times$LL \\[2mm]
\begin{loopsnlegs}[c]{p{0.25cm}|ccccc}
 \small 2&~\ybox{\pqcd[2]{0}} & \ybox{\pqcd[2]{1}} & \ldots &
\\[2mm]
 \small 1&~\ybox{\pqcd[1]{0}} & \ybox{\pqcd[1]{1}}  
   & \ybox{\pqcd[1]{2}} & \ldots \\[2mm]
 \small 0&~\gbox{\pqcd[0]{0}} & \ybox{\pqcd[0]{1}} 
   & \ybox{\pqcd[0]{2}} &\ybox{\pqcd[0]{3}} & \ldots \\
\hline
& \small 0 & \small 1 & \small 2 & \small 3 & \ldots
 \end{loopsnlegs}
\end{tabular}\hspace*{-7mm}\raisebox{0.2cm}{\huge\bf +}\hspace*{-1.5mm}
\begin{tabular}{l}
\large\bf F+1 @ LO$\times$LL\\[2mm]
\begin{loopsnlegs}[c]{p{0.25cm}|ccccc}
 \small 2&~\wbox{\pqcd[2]{0}} & \ywbox{\pqcd[2]{1}} & \ldots & 
\\[2mm]
 \small 1&~\wbox{\pqcd[1]{0}} & \ywbox{\pqcd[1]{1}}  
   & \ywbox{\pqcd[1]{2}} & \ldots \\[2mm]
 \small 0&~\wbox{\pqcd[0]{0}} & \gwbox{\pqcd[0]{1}} 
   & \ywbox{\pqcd[0]{2}} &\ywbox{\pqcd[0]{3}} & \ldots \\
\hline
& \small 0 & \small 1 & \small 2 & \small 3 & \ldots
 \end{loopsnlegs}
\end{tabular}
\hspace*{-7mm}\raisebox{0.2cm}{\huge\bf =}\hspace*{-1.5mm}
\begin{tabular}{l}
\large\bf F \& F+1 @ LO$\times$LL\\[2mm]
\begin{loopsnlegs}[c]{p{0.25cm}|ccccc}
 \small 2&~\ybox{\pqcd[2]{0}} & \rybox{\pqcd[2]{1}} & \ldots & 
\\[2mm]
 \small 1&~\ybox{\pqcd[1]{0}} & \rybox{\pqcd[1]{1}}  
   & \rybox{\pqcd[1]{2}} & \ldots \\[2mm]
 \small 0&~\gbox{\pqcd[0]{0}} & \rybox{\pqcd[0]{1}} 
   & \rybox{\pqcd[0]{2}} &\rybox{\pqcd[0]{3}} & \ldots \\
\hline
& \small 0 & \small 1 & \small 2 & \small 3 & \ldots
 \end{loopsnlegs}
\end{tabular}}
\caption{Illustration of the double-counting problem caused by naively
  adding cross sections involving matrix elements with different
  numbers of legs.\label{fig:doublecounting}}
\end{center}
\end{figure}
 Assume we have computed the LO cross section for some process, $F$,
 and that we have added an LL shower to it, as in the left-hand pane
 of \figRef{fig:doublecounting}. We know that this only gives us
 an LL description of $F+1$. We now wish to improve this from LL to LO  
 by adding the actual LO matrix element for $F+1$. Since we also want to
 be able to hadronize these events, etc, we again add an LL shower off
 them. However, since the matrix element for $F+1$ is divergent, we must
 restrict it to cover only the phase-space region with at least one
 hard resolved jet, illustrated by the half-shaded boxes in the middle
 pane of \figRef{fig:doublecounting}. Adding these two samples,
 however, we end up counting the LL terms of the inclusive cross
 section for $F+1$ twice, since we are now getting them once from the shower
 off $F$ and once from the matrix element for $F+1$, illustrated by
 the dark shaded (red) areas of the right-hand pane of
 \figRef{fig:doublecounting}. This \emph{double-counting} problem
 would grow worse if we attempted to add more matrix elements, with
 more legs. The cause is very simple. Each such calculation
 corresponds to an \emph{inclusive} cross section, and hence naive
 addition would give
\begin{equation}
\sigma_{\mrm{tot}} =\sigma_{0;\mathrm{incl}} +
  \sigma_{1;\mathrm{incl}} =  \sigma_{0;\mathrm{excl}} +  2\,\sigma_{1;\mathrm{incl}}~.
\end{equation}
 Instead, we must \emph{match} the coefficients calculated
 by the two parts of the full calculation --- showers and matrix
 elements --- more systematically, for each order in perturbation
 theory, so that the nesting of inclusive and exclusive cross sections
 is respected without overcounting.

\begin{figure}
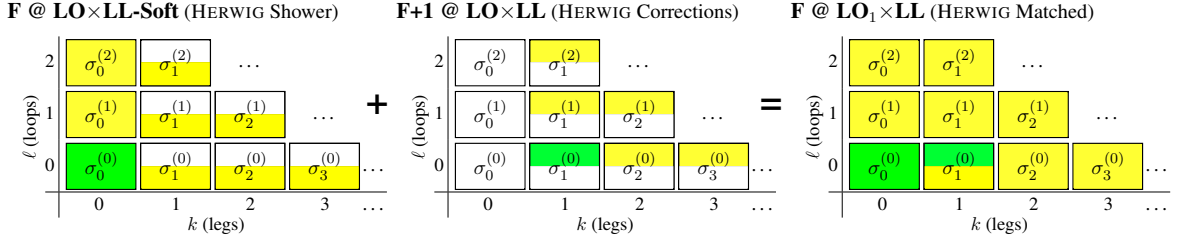

\begin{center}%
\scalebox{0.70}{\begin{tabular}{l}
{\large\bf F @ LO$\times$LL-Soft} (\Hw\ Shower)\\[2mm]
\begin{loopsnlegs}[c]{p{0.25cm}|ccccc}
 \small 2&~\ybox{\pqcd[2]{0}} & \wybox{\pqcd[2]{1}} & \ldots &
\\[2mm]
 \small 1&~\ybox{\pqcd[1]{0}} & \wybox{\pqcd[1]{1}}  
   & \wybox{\pqcd[1]{2}} & \ldots \\[2mm]
 \small 0&~\gbox{\pqcd[0]{0}} & \wybox{\pqcd[0]{1}} 
   & \wybox{\pqcd[0]{2}} &\wybox{\pqcd[0]{3}} & \ldots \\
\hline
& \small 0 & \small 1 & \small 2 & \small 3 & \ldots
 \end{loopsnlegs}
\end{tabular}\hspace*{-7mm}\raisebox{0.2cm}{\huge\bf +}\hspace*{-1.5mm}
\begin{tabular}{l}
{\large\bf F+1 @ LO$\times$LL} (\Hw\ Corrections)\\[2mm]
\begin{loopsnlegs}[c]{p{0.25cm}|ccccc}
 \small 2&~\wbox{\pqcd[2]{0}} & \ywbox{\pqcd[2]{1}} & \ldots & 
\\[2mm]
 \small 1&~\wbox{\pqcd[1]{0}} & \ywbox{\pqcd[1]{1}}  
   & \ywbox{\pqcd[1]{2}} & \ldots \\[2mm]
 \small 0&~\wbox{\pqcd[0]{0}} & \gwbox{\pqcd[0]{1}} 
   & \ywbox{\pqcd[0]{2}} &\ywbox{\pqcd[0]{3}} & \ldots \\
\hline
& \small 0 & \small 1 & \small 2 & \small 3 & \ldots
 \end{loopsnlegs}
\end{tabular}
\hspace*{-7mm}\raisebox{0.2cm}{\huge\bf =}\hspace*{-1.5mm}
\begin{tabular}{l}
{\large\bf F @ LO$_1\times$LL} (\Hw\ Matched)\\[2mm]
\begin{loopsnlegs}[c]{p{0.25cm}|ccccc}
 \small 2&~\ybox{\pqcd[2]{0}} & \ybox{\pqcd[2]{1}} & \ldots & 
\\[2mm]
 \small 1&~\ybox{\pqcd[1]{0}} & \ybox{\pqcd[1]{1}}  
   & \ybox{\pqcd[1]{2}} & \ldots \\[2mm]
 \small 0&~\gbox{\pqcd[0]{0}} & \gybox{\pqcd[0]{1}} 
   & \ybox{\pqcd[0]{2}} &\ybox{\pqcd[0]{3}} & \ldots \\
\hline
& \small 0 & \small 1 & \small 2 & \small 3 & \ldots
 \end{loopsnlegs}
\end{tabular}}
\caption{Illustration of the original matching scheme implemented in
  \Hw~\cite{Seymour:1994we,Seymour:1994df}, in which the dead zone of the \Hw\ shower was used as an
  effective ``matching scale'' for one emission beyond a basic hard process.
\label{fig:herwig}}
\end{center}
\end{figure}

\begin{figure}
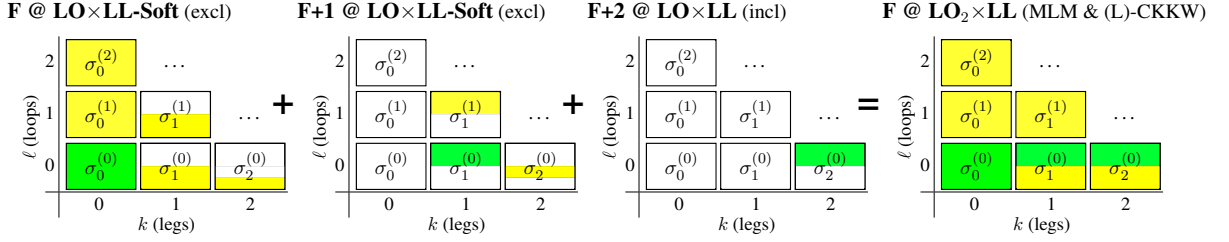

\begin{center}%
\scalebox{0.70}{\begin{tabular}{l}
{\large\bf F @ LO$\times$LL-Soft} (excl)\\[2mm]
\begin{loopsnlegs}[c]{p{0.25cm}|ccccc}
 \small 2&~\ybox{\pqcd[2]{0}} &  \ldots &
\\[2mm]
 \small 1&~\ybox{\pqcd[1]{0}} & \wybox{\pqcd[1]{1}}  
   & \ldots \\[2mm]
 \small 0&~\gbox{\pqcd[0]{0}} & \wybox{\pqcd[0]{1}} 
   & \wwybox{\pqcd[0]{2}} & \\
\hline
& \small 0 & \small 1 & \small 2 &
 \end{loopsnlegs}
\end{tabular}\hspace*{-6mm}\raisebox{0.2cm}{\huge\bf +}\hspace*{-2.5mm}
\begin{tabular}{l}
{\large\bf F+1 @ LO$\times$LL-Soft} (excl)\\[2mm]
\begin{loopsnlegs}[c]{p{0.25cm}|ccccc}
 \small 2&~\wbox{\pqcd[2]{0}} & \ldots & 
\\[2mm]
 \small 1&~\wbox{\pqcd[1]{0}} & \ywbox{\pqcd[1]{1}}  
   &  \ldots \\[2mm]
 \small 0&~\wbox{\pqcd[0]{0}} & \gwbox{\pqcd[0]{1}} 
   & \wywbox{\pqcd[0]{2}} &  \\
\hline
& \small 0 & \small 1 & \small 2 & 
 \end{loopsnlegs}
\end{tabular}\hspace*{-6mm}\raisebox{0.2cm}{\huge\bf +}\hspace*{-2.5mm}
\begin{tabular}{l}
{\large\bf F+2 @ LO$\times$LL} (incl)\\[2mm]
\begin{loopsnlegs}[c]{p{0.25cm}|ccccc}
 \small 2&~\wbox{\pqcd[2]{0}} & \ldots & 
\\[2mm]
 \small 1&~\wbox{\pqcd[1]{0}} & \wbox{\pqcd[1]{1}}  
   &\ldots \\[2mm]
 \small 0&~\wbox{\pqcd[0]{0}} & \wbox{\pqcd[0]{1}} 
   & \gwbox{\pqcd[0]{2}} &  \\
\hline
& \small 0 & \small 1 & \small 2 & 
 \end{loopsnlegs}
\end{tabular}
\hspace*{-6mm}\raisebox{0.2cm}{\huge\bf =}\hspace*{-2.5mm}
\begin{tabular}{l}
{\large\bf F @ LO$_2\times$LL} (MLM \& (L)-CKKW)\\[2mm]
\begin{loopsnlegs}[c]{p{0.25cm}|ccccc}
 \small 2&~\ybox{\pqcd[2]{0}} &  \ldots & 
\\[2mm]
 \small 1&~\ybox{\pqcd[1]{0}} & \ybox{\pqcd[1]{1}}  
   & \ldots \\[2mm]
 \small 0&~\gbox{\pqcd[0]{0}} & \gybox{\pqcd[0]{1}} 
   & \gybox{\pqcd[0]{2}} & \\
\hline
& \small 0 & \small 1 & \small 2 &
 \end{loopsnlegs}
\end{tabular}}
\caption{Illustration of slicing approaches to matching, with up
  to two additional emissions beyond the basic process. The showers off
  $F$ and $F+1$ are set to zero above a specific ``matching
  scale''. (The number of coefficients 
  shown was reduced a bit in these plots to make them fit in one row.)
\label{fig:slicing}}
\end{center}
\end{figure}

Given a parton shower and a matrix-element generator, there are
fundamentally three different ways in which we can consider matching
the two \cite{Giele:2011cb}:

{\bf 1. \sl Slicing:} The most commonly encountered matching type is
  currently based on separating (slicing)
  phase space into two regions, one of which is supposed to be
  mainly described by hard matrix elements and the other of which is
  supposed to be described by the shower. This type of 
  approach was first 
  used in \Hw~\cite{Corcella:2000bw}, to
  include matrix-element corrections for one emission beyond the 
  basic hard process \cite{Seymour:1994we,Seymour:1994df}.
  This is illustrated in \figRef{fig:herwig}.
  The method has since been generalized by several
  independent groups to include
  arbitrary numbers of additional legs
  \cite{Catani:2001cc,Lonnblad:2001iq,Mrenna:2003if,Lavesson:2005xu,Mangano:2006rw}. 
  Effectively,   
  the shower approximation is set to zero
  above some scale (either due to the presence of explicit dead zones
  in the shower, as in \Hw, or by vetoing any emissions above a certain
  \emph{matching scale}, as in the (L)-CKKW 
  \cite{Catani:2001cc,Lonnblad:2001iq,Lavesson:2005xu} 
  and MLM \cite{Mangano:2006rw,Mrenna:2003if} approaches), 
  causing the matched result to be identical
  to the matrix element (ME) in that region, modulo higher-order
  corrections. We may sketch this as 
\begin{equation}
\mbox{Matched (above matching scale)} =
\color{gray}\overbrace{\color{black}\mbox{Exact}}^{\small\mbox{ME}}~
\ {\color{black}\times} \ \overbrace{\color{black}(1 +
\mathcal{O}(\alpha_s))}^{\small\mbox{corrections}} \label{eq:scalebased1}~,
\end{equation}
where the ``shower-corrections'' include approximate Sudakov factors and $\alpha_s$
reweighting factors applied to the matrix elements in order to obtain
a smooth transition to the shower-dominated region.
Below the matching scale, the
small difference between the matrix elements and the shower
approximation can be dropped (since their leading singularities are
identical and this region by definition includes no hard jets), 
yielding the pure shower answer in that region,
\begin{eqnarray}
\mbox{Matched (below matching scale)} & = &
\color{gray}\overbrace{\color{black}\mbox{Approximate}}^{\mbox{shower}}
\ {\color{black}+} \ \overbrace{\color{black}(\mbox{Exact} -
\mbox{Approximate})}^{\mbox{correction}}\nonumber\\ 
& = & \mbox{Approximate} \ + \ \mbox{non-singular} \nonumber \\
& \to & \mbox{Approximate}~.
\end{eqnarray}
This type of strategy is illustrated in \figRef{fig:slicing}.
Since this strategy is discontinuous across phase space, a main point
here is to ensure that the behaviour across the matching scale be as
smooth as possible. CKKW showed \cite{Catani:2001cc} that it is
possible to remove any  dependence on the matching scale through  NLL
precision by careful choices of all ingredients in the matching;
technical details of the implementation 
(affecting the 
$\mathcal{O}(\alpha_s)$ terms in eq.~(\ref{eq:scalebased1}))
are important, and the dependence on the unphysical matching scale
may be larger than NLL unless the implementation
matches the theoretical algorithm
precisely~\cite{Lonnblad:2001iq,Lavesson:2005xu,Lavesson:2008ah}. 
One should also be aware that all strategies of this type are 
quite computing intensive. This is basically due to the fact that 
a separate phase-space generator is required for each of the
$n$-parton correction terms, with each such sample a priori consisting
of weighted events such that a separate unweighting step (often with
quite low efficiency) is needed before an
unweighted sample can be produced. 

\begin{figure}
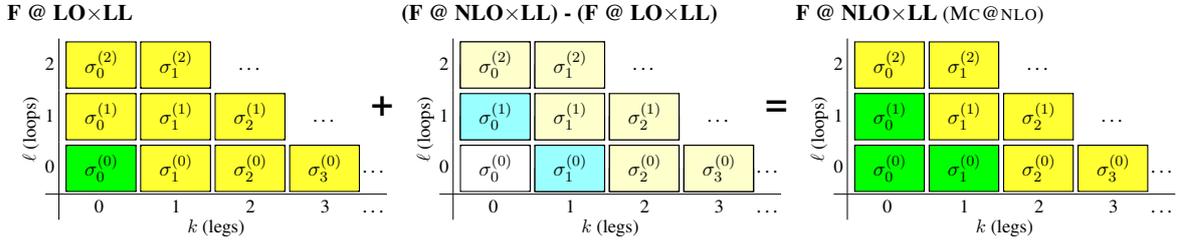

\begin{center}%
\scalebox{0.70}{\begin{tabular}{l}
{\large\bf F @ LO$\times$LL}\\[2mm]
\begin{loopsnlegs}[c]{p{0.25cm}|ccccc}
 \small 2&~\ybox{\pqcd[2]{0}} & \ybox{\pqcd[2]{1}} & \ldots &
\\[2mm]
 \small 1&~\ybox{\pqcd[1]{0}} & \ybox{\pqcd[1]{1}}  
   & \ybox{\pqcd[1]{2}} & \ldots \\[2mm]
 \small 0&~\gbox{\pqcd[0]{0}} & \ybox{\pqcd[0]{1}} 
   & \ybox{\pqcd[0]{2}} &\ybox{\pqcd[0]{3}} & \ldots \\
\hline
& \small 0 & \small 1 & \small 2 & \small 3 & \ldots
 \end{loopsnlegs}
\end{tabular}
\hspace*{-7mm}\raisebox{0.2cm}{\huge\bf +}\hspace*{-1.5mm}
\begin{tabular}{l}
{\large\bf (F @ NLO$\times$LL) - (F @ LO$\times$LL)}\\[2mm]
\begin{loopsnlegs}[c]{p{0.25cm}|ccccc}
 \small 2&~\eggbox{\pqcd[2]{0}} & \eggbox{\pqcd[2]{1}} & \ldots &
\\[2mm]
 \small 1&~\cyanbox{\pqcd[1]{0}} & \eggbox{\pqcd[1]{1}}  
   & \eggbox{\pqcd[1]{2}} & \ldots \\[2mm]
 \small 0&~\wbox{\pqcd[0]{0}} & \cyanbox{\pqcd[0]{1}} 
   & \eggbox{\pqcd[0]{2}} &\eggbox{\pqcd[0]{3}} & \ldots \\
\hline
& \small 0 & \small 1 & \small 2 & \small 3 & \ldots
 \end{loopsnlegs}
\end{tabular}
\hspace*{-7mm}\raisebox{0.2cm}{\huge\bf =}\hspace*{-1.5mm}
\begin{tabular}{l}
{\large\bf F @ NLO$\times$LL} (\Fw)\\[2mm]
\begin{loopsnlegs}[c]{p{0.25cm}|ccccc}
 \small 2&~\ybox{\pqcd[2]{0}} & \ybox{\pqcd[2]{1}} & \ldots &
\\[2mm]
 \small 1&~\gbox{\pqcd[1]{0}} & \ybox{\pqcd[1]{1}}  
   & \ybox{\pqcd[1]{2}} & \ldots \\[2mm]
 \small 0&~\gbox{\pqcd[0]{0}} & \gbox{\pqcd[0]{1}} 
   & \ybox{\pqcd[0]{2}} &\ybox{\pqcd[0]{3}} & \ldots \\
\hline
& \small 0 & \small 1 & \small 2 & \small 3 & \ldots
 \end{loopsnlegs}
\end{tabular}}
\caption{Illustration of the \Fw\ approach to matching. In the middle
  pane, cyan boxes
  denote non-singular correction terms, while the egg-coloured ones 
  denote showers off such corrections, which cannot lead to
  double-counting at the LL level.
\label{fig:fw}}
\end{center}
\end{figure}

\begin{figure}
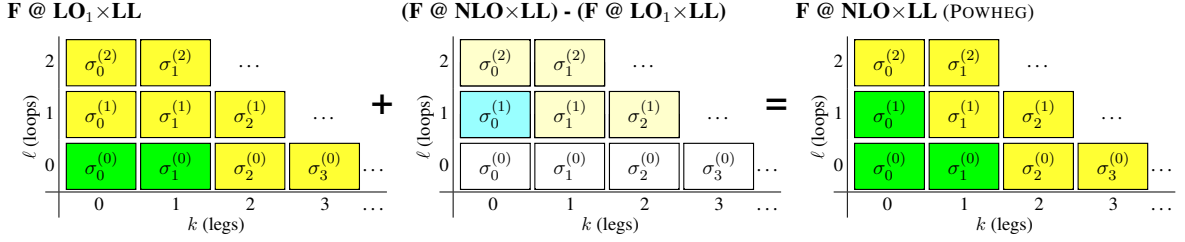

\begin{center}%
\scalebox{0.70}{\begin{tabular}{l}
{\large\bf F @ LO$_1\times$LL}\\[2mm]
\begin{loopsnlegs}[c]{p{0.25cm}|ccccc}
 \small 2&~\ybox{\pqcd[2]{0}} & \ybox{\pqcd[2]{1}} & \ldots &
\\[2mm]
 \small 1&~\ybox{\pqcd[1]{0}} & \ybox{\pqcd[1]{1}}  
   & \ybox{\pqcd[1]{2}} & \ldots \\[2mm]
 \small 0&~\gbox{\pqcd[0]{0}} & \gbox{\pqcd[0]{1}} 
   & \ybox{\pqcd[0]{2}} &\ybox{\pqcd[0]{3}} & \ldots \\
\hline
& \small 0 & \small 1 & \small 2 & \small 3 & \ldots
 \end{loopsnlegs}
\end{tabular}
\hspace*{-7mm}\raisebox{0.2cm}{\huge\bf +}\hspace*{-1.5mm}
\begin{tabular}{l}
{\large\bf (F @ NLO$\times$LL) - (F @ LO$_1\times$LL)}\\[2mm]
\begin{loopsnlegs}[c]{p{0.25cm}|ccccc}
 \small 2&~\eggbox{\pqcd[2]{0}} & \eggbox{\pqcd[2]{1}} & \ldots &
\\[2mm]
 \small 1&~\cyanbox{\pqcd[1]{0}} & \eggbox{\pqcd[1]{1}}  
   & \eggbox{\pqcd[1]{2}} & \ldots \\[2mm]
 \small 0&~\wbox{\pqcd[0]{0}} & \wbox{\pqcd[0]{1}} 
   & \wbox{\pqcd[0]{2}} &\wbox{\pqcd[0]{3}} & \ldots \\
\hline
& \small 0 & \small 1 & \small 2 & \small 3 & \ldots
 \end{loopsnlegs}
\end{tabular}
\hspace*{-7mm}\raisebox{0.2cm}{\huge\bf =}\hspace*{-1.5mm}
\begin{tabular}{l}
{\large\bf F @ NLO$\times$LL} (\Pw)\\[2mm]
\begin{loopsnlegs}[c]{p{0.25cm}|ccccc}
 \small 2&~\ybox{\pqcd[2]{0}} & \ybox{\pqcd[2]{1}} & \ldots &
\\[2mm]
 \small 1&~\gbox{\pqcd[1]{0}} & \ybox{\pqcd[1]{1}}  
   & \ybox{\pqcd[1]{2}} & \ldots \\[2mm]
 \small 0&~\gbox{\pqcd[0]{0}} & \gbox{\pqcd[0]{1}} 
   & \ybox{\pqcd[0]{2}} &\ybox{\pqcd[0]{3}} & \ldots \\
\hline
& \small 0 & \small 1 & \small 2 & \small 3 & \ldots
 \end{loopsnlegs}
\end{tabular}}
\caption{Illustration of the \Pw\ approach to matching. In the middle
  pane, cyan boxes
  denote non-singular correction terms, while the egg-coloured ones 
  denote showers off such corrections, which cannot lead to
  double-counting at the LL level.
\label{fig:powheg}}
\end{center}
\end{figure}

\begin{figure}
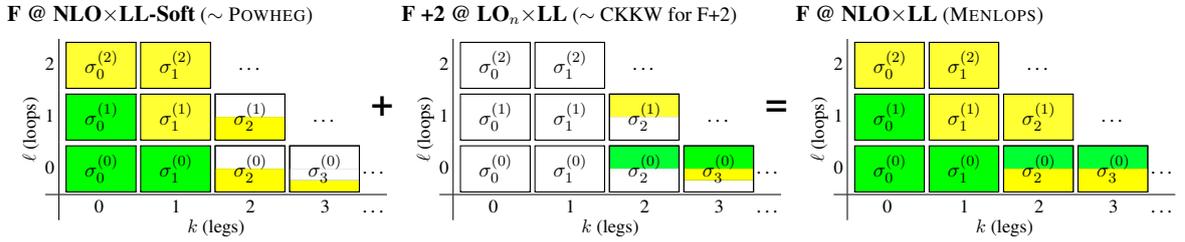

\begin{center}%
\scalebox{0.70}{\begin{tabular}{l}
{\large\bf F @ NLO$\times$LL-Soft} ($\sim$ \Pw)\\[2mm]
\begin{loopsnlegs}[c]{p{0.25cm}|ccccc}
 \small 2&~\ybox{\pqcd[2]{0}} & \ybox{\pqcd[2]{1}} & \ldots &
\\[2mm]
 \small 1&~\gbox{\pqcd[1]{0}} & \ybox{\pqcd[1]{1}}  
   & \wybox{\pqcd[1]{2}} & \ldots \\[2mm]
 \small 0&~\gbox{\pqcd[0]{0}} & \gbox{\pqcd[0]{1}} 
   & \wybox{\pqcd[0]{2}} &\wwybox{\pqcd[0]{3}} & \ldots \\
\hline
& \small 0 & \small 1 & \small 2 & \small 3 & \ldots
 \end{loopsnlegs}
\end{tabular}
\hspace*{-7mm}\raisebox{0.2cm}{\huge\bf +}\hspace*{-1.5mm}
\begin{tabular}{l}
{\large\bf F +2 @ LO$_n\times$LL} ($\sim$ CKKW for F+2)\\[2mm]
\begin{loopsnlegs}[c]{p{0.25cm}|ccccc}
 \small 2&~\wbox{\pqcd[2]{0}} & \wbox{\pqcd[2]{1}} & \ldots &
\\[2mm]
 \small 1&~\wbox{\pqcd[1]{0}} & \wbox{\pqcd[1]{1}}  
   & \ywbox{\pqcd[1]{2}} & \ldots \\[2mm]
 \small 0&~\wbox{\pqcd[0]{0}} & \wbox{\pqcd[0]{1}} 
   & \gwbox{\pqcd[0]{2}} &\gywbox{\pqcd[0]{3}} & \ldots \\
\hline
& \small 0 & \small 1 & \small 2 & \small 3 & \ldots
 \end{loopsnlegs}
\end{tabular}
\hspace*{-7mm}\raisebox{0.2cm}{\huge\bf =}\hspace*{-1.5mm}
\begin{tabular}{l}
{\large\bf F @ NLO$\times$LL} (\textsc{Menlops})\\[2mm]
\begin{loopsnlegs}[c]{p{0.25cm}|ccccc}
 \small 2&~\ybox{\pqcd[2]{0}} & \ybox{\pqcd[2]{1}} & \ldots &
\\[2mm]
 \small 1&~\gbox{\pqcd[1]{0}} & \ybox{\pqcd[1]{1}}  
   & \ybox{\pqcd[1]{2}} & \ldots \\[2mm]
 \small 0&~\gbox{\pqcd[0]{0}} & \gbox{\pqcd[0]{1}} 
   & \gybox{\pqcd[0]{2}} &\gybox{\pqcd[0]{3}} & \ldots \\
\hline
& \small 0 & \small 1 & \small 2 & \small 3 & \ldots
 \end{loopsnlegs}
\end{tabular}}
\caption{Illustration of the \textsc{Menlops} approach to
  matching. Note that each of the \Pw\ and CKKW samples are composed
  of separate sub-samples, as illustrated in \figsRef{fig:slicing} and
  \ref{fig:powheg}. 
\label{fig:menlops}}
\end{center}
\end{figure}

{\bf 2. \sl Subtraction:} 
Another way of matching two calculations is by subtracting one
  from the other and correcting by the difference, schematically
\begin{equation}
\mbox{Matched} =
\color{gray}\overbrace{\color{black}\mbox{Approximate}}^{\mbox{shower}}
\color{black} \ + \ \color{gray}\overbrace{\color{black}
(\mbox{Exact}-\mbox{Approximate})}^{\mbox{correction}}~. \label{eq:additive}
\end{equation}
This looks very much like the structure of an NLO
fixed-order calculation, in which the shower approximation plays the
role of subtraction terms, and indeed this is what is used  in
strategies like \Fw{}
\cite{Frixione:2002ik,Frixione:2003ei,Frixione:2008ym}, illustrated in
\figRef{fig:fw}.
In this type of approach, however, negative-weight events will
generally occur, for instance in  
phase-space points where the approximation is larger than the exact
answer. This motivated the development of the so-called \Pw\ approach 
\cite{Frixione:2007vw}, illustrated in \figRef{fig:powheg}, 
which is
constructed specifically to prevent negative-weight events from
occurring and simultaneously to be more independent of which
parton-shower algorithm it is used with. The advantage of these methods
is obviously that NLO corrections to the Born level can be
systematically incorporated. However, a systematic way of
extending this strategy beyond the first additional emission is not
available, save for combining them with a slicing-based strategy
for the additional legs, as in \textsc{Menlops}
\cite{Hamilton:2010wh}, illustrated in \figRef{fig:menlops}. 
These issues are, however, no more severe than
in ordinary fixed-order NLO 
approaches, and hence they are not viewed as disadvantages if the
point of reference is an NLO computation. 

{\bf 3. \sl Unitarity:} 
The oldest, and in our view most attractive, 
approach \cite{Bengtsson:1986et,Bengtsson:1986hr}
  consists of working out the 
 shower approximation to a given fixed order, and correcting
 the shower splitting functions at that order by a multiplicative
 factor given by the ratio of the matrix element
 to the shower approximation, phase-space point by phase-space
 point. We may sketch this as 
\begin{equation}
\mbox{Matched} =
\color{gray}\overbrace{\color{black}\mbox{Approximate}}^{\mbox{shower}}
\ {\color{black}\times} \
\overbrace{\color{black}\frac{\mbox{Exact}}{\mbox{Approximate}}}^{\mbox{correction}}~. \label{eq:multiplicative}
\end{equation}
When these correction factors are inserted back into the
shower evolution, they guarantee that the shower evolution off $n-1$
partons correctly reproduces the $n$-parton matrix elements, 
without the need to generate a separate $n$-parton sample. 
That is, the shower approximation is essentially used as a 
pre-weighted (stratified) all-orders phase-space generator, on which a
more exact answer can subsequently be
imprinted order by order in perturbation theory. 
In the original
approach \cite{Bengtsson:1986et,Bengtsson:1986hr}, used by \Py
\cite{Sjostrand:2006za,Sjostrand:2007gs}, this was only 
worked out for one additional emission beyond the basic hard
process. In \Pw~\cite{Frixione:2007vw}, 
it was extended to include also virtual corrections to the Born-level
matrix element. Finally, in \Vc~\cite{Giele:2011cb}, it has been 
extended to include arbitrary numbers of emissions at tree
level, though that method has so far only been applied to
final-state showers.
\begin{figure}[t]
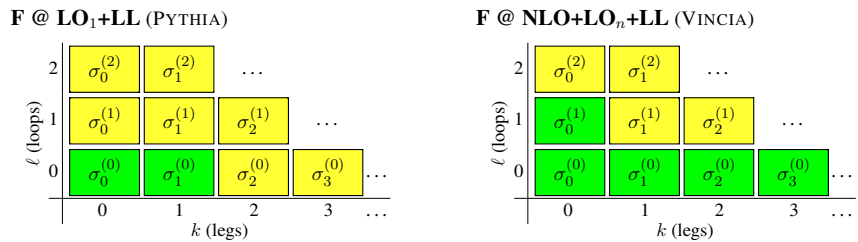

\begin{center}%
\scalebox{0.70}{
\begin{tabular}{l}
{\large\bf F @ LO$_{1}$+LL} (\Py)\\[2mm]
\begin{loopsnlegs}[c]{p{0.25cm}|ccccc}
 \small 2&~\ybox{\pqcd[2]{0}} & \ybox{\pqcd[2]{1}} & \ldots &
\\[2mm]
 \small 1&~\ybox{\pqcd[1]{0}} & \ybox{\pqcd[1]{1}}  
   & \ybox{\pqcd[1]{2}} & \ldots \\[2mm]
 \small 0&~\gbox{\pqcd[0]{0}} & \gbox{\pqcd[0]{1}} 
   & \ybox{\pqcd[0]{2}} &\ybox{\pqcd[0]{3}} & \ldots \\
\hline
& \small 0 & \small 1 & \small 2 & \small 3 & \ldots
 \end{loopsnlegs}
\end{tabular}\hspace*{1cm}
\begin{tabular}{l}
{\large\bf F @ NLO+LO$_{n}$+LL} (\Vc)\\[2mm]
\begin{loopsnlegs}[c]{p{0.25cm}|ccccc}
 \small 2&~\ybox{\pqcd[2]{0}} & \ybox{\pqcd[2]{1}} & \ldots & 
\\[2mm]
 \small 1&~\gbox{\pqcd[1]{0}} & \ybox{\pqcd[1]{1}}  
   & \ybox{\pqcd[1]{2}} & \ldots \\[2mm]
 \small 0&~\gbox{\pqcd[0]{0}} & \gbox{\pqcd[0]{1}} 
   & \gbox{\pqcd[0]{2}} &\gbox{\pqcd[0]{3}} & \ldots \\
\hline
& \small 0 & \small 1 & \small 2 & \small 3 & \ldots
 \end{loopsnlegs}
\end{tabular}}
\caption{Illustration of the two purely unitarity-based approaches to
  matching discussed in the text. Only one event sample is produced by
  each of these methods, and hence no sub-components are shown. 
\label{fig:match-unitary}}
\end{center}
\end{figure}
An illustration of the perturbative coefficients that can be 
included in each of  these approaches is illustrated in
\figRef{fig:match-unitary}, as usual with green (darker shaded) boxes
representing exact coefficients and yellow (light shaded) boxes
representing logarithmic approximations. Finally, two more
properties unique to this method deserve mention. Firstly, 
since the corrections modify the actual shower evolution kernels, the
corrections are automatically \emph{resummed} in the Sudakov
exponential, which should improve the logarithmic precision once
$k\ge2$ is included, 
and secondly, since the shower is \emph{unitary},  an initially unweighted sample
of $(n-1)$-parton configurations remains unweighted, with 
no need for a separate event-unweighting or event-rejection
step. 

\subsection{The String Model of Hadronization}

In the context of event generators,  \emph{hadronization} 
denotes the process by which a set of post-shower partons is
transformed into a set of  \emph{primary} hadrons, which may then
subsequently decay further. 
This non-perturbative transition takes place at the \emph{hadronization
scale}, which by construction 
is identical to the infrared cutoff of the parton
shower.  
In the absence of a first-principles solution to the relevant
dynamics, event generators use QCD-inspired phenomenological models to
describe this transition. 

Although non-perturbative QCD is not solved, we do have some 
knowledge of the properties that such a solution must have. 
For instance, Poincar\'e invariance, unitarity, and 
causality are all concepts that apply beyond 
perturbation theory. 
In addition, lattice QCD provides us a means
of making explicit quantitative studies in a genuinely non-perturbative
setting, albeit only of certain questions. 

An important result in ``quenched'' lattice
QCD\footnote{Quenched QCD implies no ``dynamical'' quarks, i.e., no
  $g\to q\bar{q}$ splittings  allowed.} is that the potential
of the colour dipole field between a charge and an anticharge 
appears to grow linearly with the separation of the charges, when the
separation is greater than about a
femtometer.This is known as ``linear confinement'', and it forms the starting
point for the \emph{string model of hadronization}. 

Starting from early concepts developed by Artru and Mennessier 
\cite{Artru:1974hr}, several hadronization models based on strings
were proposed in the late 1970'ies and early 80'ies. 
Of these, the most sophisticated and widely used today is
the so-called Lund model, implemented in the \Py\ code. 
We shall therefore concentrate on that particular model here, though
many of the overall concepts would be shared by any string-inspired
method. (A more extended discussion can be found in the 
very complete and pedagogical review of the Lund model by
Andersson \cite{Andersson:1998tv}.)

\begin{figure}[t]
\begin{center}
\scalebox{0.9}{
\begin{tabular}{c}
\rotatebox{90}{\small $1\,$fm}
\end{tabular}\hspace*{-7mm}
\begin{tabular}{c}
\includegraphics*[scale=1.0]{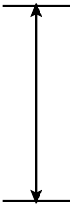}
\end{tabular}
\begin{tabular}{c}
\includegraphics*[scale=0.75]{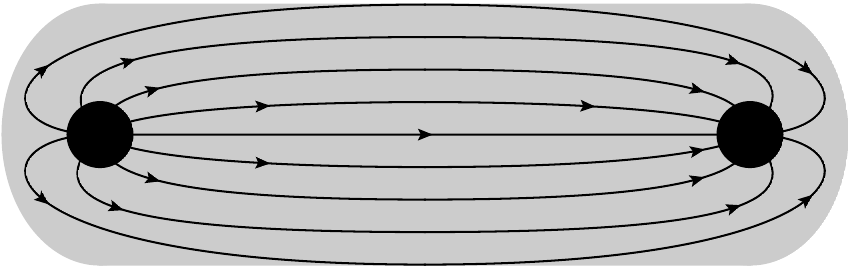}
\end{tabular}}
\caption{Illustration of the transition between a Coulomb potential
  at short distances to the string-like one of \eqRef{eq:string} at
  large $q\bar{q}$ separations.\label{fig:string}}
\end{center}
\end{figure}

Consider the production of a $q\bar{q}$ pair from vacuum, for instance
in the process $e^+e^-\to \gamma^*/Z\to q\bar{q} \to \mbox{hadrons}$. As the quarks
move apart, linear confinement implies that a potential 
\begin{equation}
V(r) = \kappa\, r \label{eq:string}
\end{equation}
is asymptotically reached for large distances, $r$. (At short
distances, there is a Coulomb term proportional to $1/r$ as well, but
this is neglected in the Lund model.) This potential describes a 
string with tension (energy per unit length) $\kappa$.  
The physical picture is that of a colour flux tube being 
stretched between the $q$ and the $\bar{q}$, \figRef{fig:string}.
From hadron mass spectroscopy the string tension $\kappa$, 
is known to be
\begin{equation} 
\kappa~\sim~1\,\mbox{GeV/fm}~\sim~0.2\,\mbox{GeV}^2.
\end{equation}
A straightforward Lorentz-invariant description of this
object is provided by the massless relativistic string in 1+1
dimensions, with no transverse degrees of freedom. 
The mathematical, one-dimensional string can be thought of as
parameterizing the position of the axis of 
a cylindrically symmetric flux tube. (Note that the expression
``massless'' is somewhat of a misnomer, since $\kappa$ effectively
corresponds to a ``mass density'' along the string.)

\begin{figure}[t]
\begin{center}
\scalebox{0.9}{
\begin{tabular}{ccc}
\begin{tabular}{c}
\includegraphics*[scale=0.75]{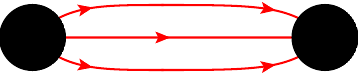} \ \
\includegraphics*[scale=0.75]{strBrk3.pdf}\\[1mm]
\includegraphics*[scale=0.75]{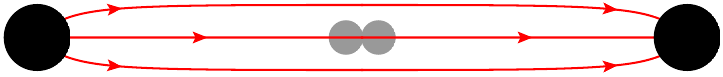}\\[1mm]
\includegraphics*[scale=0.75]{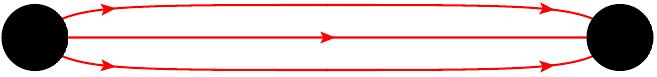}
\end{tabular}\hspace*{1.2cm}
& 
\begin{tabular}{c}
\rotatebox{90}{\small time}
\end{tabular}\hspace*{-2.2cm}
\begin{tabular}{r}
\includegraphics*[scale=1.4]{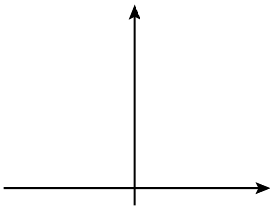}\\[-3mm]
{\small space}
\end{tabular}\hspace*{-0.5cm}&
\raisebox{0.9cm}{\begin{tabular}{c}\includegraphics*[scale=0.6]{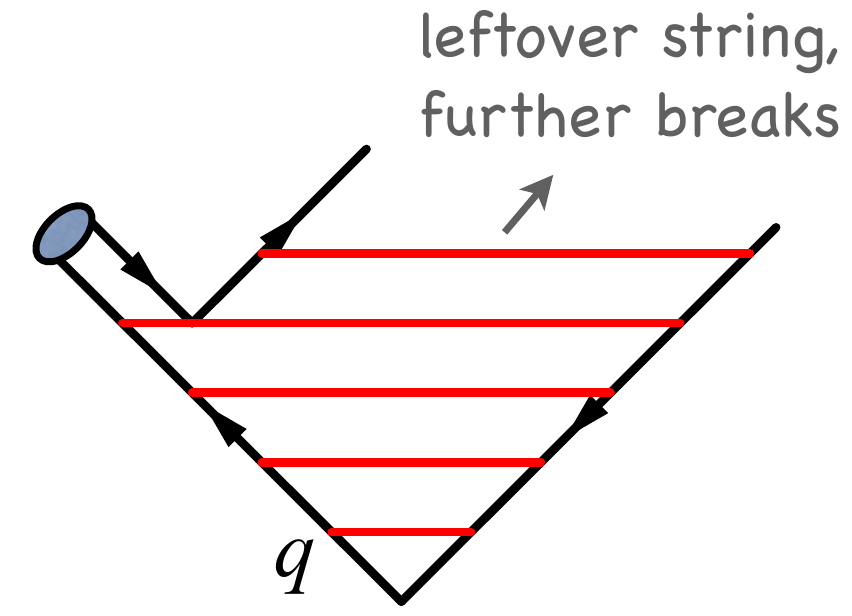}
\end{tabular}}\\
\hspace*{-1cm} a) & & b)
\end{tabular}
}
\caption{{\sl a)} Illustration of string breaking by quark pair creation in the
  string field. {\sl b)} Illustration of the algorithmic choice to
  process the fragmentation from the outside-in, splitting off a
  single on-shell hadron in each step.\label{fig:stringbreak}}
\end{center}
\end{figure}

As the $q$ and $\bar{q}$ move apart, their kinetic energy is gradually
converted to potential energy, stored in the growing string spanned
between them. In the ``quenched'' approximation, in which $g\to
q\bar{q}$ splittings are not allowed, this process would continue
until the endpoint quarks have lost \emph{all} their
momentum, at which point they would reverse direction and be
accelerated by the now shrinking string. In the real world,
 quark-antiquark fluctuations inside the string field
can make the transition to become real particles by absorbing 
energy from the string, thereby screening the original endpoint 
charges from each other and breaking the string into two separate
colour-singlet pieces, $(q\bar{q}) \to
(q\bar{q}')+(q'\bar{q})$, illustrated in \figRef{fig:stringbreak}\!a.
This process then continues until only ordinary hadrons remain. (We
will give more details on the individual string breaks below.)
More complicated multi-parton topologies including gluons are treated by 
representing gluons as transverse ``kinks''. Thus soft gluons
effectively ``build up'' a transverse structure in the originally
one-dimensional object, with infinitely soft ones absorbed into the
string without leading to modifications. For strings with
finite-energy kinks, the space-time evolution is
then slightly more involved \cite{Andersson:1998tv}, and modifications
to the fragmentation 
model to handle stepping across gluon corners have to be included,
but the main point is that there are no separate free parameters for
gluon jets. Differences with respect to 
quark  fragmentation arise simply because quarks are only
connected to a single string piece, while gluons have one on either
side, increasing the energy loss per unit (invariant) time 
from a gluon to the string by a
factor of 2 relative to quarks, which can be compared to the ratio of
colour Casimirs $C_A/C_F = 2.25$. 

Since the string breaks are causally disconnected
(as can easily be realized from space-time diagrams 
\cite{Andersson:1998tv}), they do not have to be considered in any
specific time-ordered sequence. In the Lund model, the string breaks are
instead generated starting 
with the leading hadrons, containing the endpoint quarks, and iterating inwards
towards the centre of the string, alternating randomly between
fragmentation off the left- and right-hand sides, respectively,
\figRef{fig:stringbreak}\!b.  
This has the advantage that a single on-shell
hadron can be split off in each step, making it straightforward
to ensure that only states consistent with  the known spectrum of hadron
resonances are produced, as will be discussed below.

The details of the individual string breaks 
are not known from first principles. The
Lund model invokes the idea of quantum mechanical tunneling, which
leads to a Gaussian suppression of the energies and masses imparted to
the produced quarks, 
\begin{equation}
\mathrm{Prob}(m_q^2,\pt[q]^2)~\propto~\exp\left(\frac{-\pi
  m_q^2}{\kappa}\right)  \exp\left(\frac{-\pi
  \pt[q]^2}{\kappa}\right)~, \label{eq:tunnelling}
\end{equation}
where $m_q$ is the mass of the produced quark and $\pt$
is the transverse momentum imparted to it by the
breakup process (the antiquark obviously has the same mass and opposite
\pt). 

Due to the factorization of the \pt and $m$ dependence
implied by \eqRef{eq:tunnelling}, the \pt spectrum of produced
quarks in this model is independent of the quark flavour, with a
universal average value of 
\begin{equation}
\left<\pt[q]^2\right> = \sigma^2 = \kappa/\pi \sim
(250\,\mbox{MeV})^2~. 
\end{equation}
Bear in mind that ``transverse'' is here defined
with respect to the string axis.  Thus, the
\pt in a frame where the string is moving is modified by a Lorentz
boost factor. Also bear in mind that $\sigma^2$ is here 
a purely non-perturbative parameter. In a Monte Carlo model
with a fixed shower cutoff, the effective amount of ``non-perturbative'' \pt 
may be larger than this,  due to effects of additional unresolved
soft-gluon radiation below the shower cutoff scale. 
In principle, the magnitude of this additional component 
should scale with the cutoff, but in practice it is up to
the user to enforce this by retuning the effective $\sigma$ parameter 
when changing the hadronization scale. Since hadrons
receive $\pt$ contributions from two breakups, one on either side, 
their average transverse momentum squared will be twice as large,
\begin{equation}
\left<\pt[h]^2\right> = 2\sigma^2~. 
\end{equation}

The mass suppression implied by \eqRef{eq:tunnelling} is less 
straightforward to interpret. Since quark masses
are notoriously difficult to define for light quarks, the value of the
strangeness suppression must effectively be extracted from experimental
measurements, e.g., of the $K/\pi$ ratio, with a
resulting suppression of roughly $s/u \sim s/d \sim$ 0.2 -- 0.3.
Inserting even comparatively low values for the charm quark mass in
\eqRef{eq:tunnelling}, however, one obtains a relative suppression of
charm of the order of $10^{-11}$. Heavy quarks can therefore safely be
considered to be produced only in the perturbative stages 
and not by the soft fragmentation.

Baryon production can be incorporated in the same basic picture
\cite{Andersson:1981ce},  
by  allowing  string breaks to occur also by the production of
pairs of so-called \emph{diquarks}, 
loosely bound states of two quarks in an overall $\bar{3}$
representation (e.g., red + blue = antigreen). Again, the relative
rate of diquark-to-quark production is not known a priori and must be
extracted from experimental measurements, e.g., of the $p/\pi$ ratio. 
More advanced
scenarios for baryon production have also been proposed, in particular
the so-called popcorn model \cite{Andersson:1984af,Eden:1996xi}, which is normally used in addition
to the diquark picture and then acts to decrease the correlations
among neighbouring baryon-antibaryon pairs by allowing mesons to be formed
inbetween them. Within the \Py\ framework, 
a fragmentation model including explicit \emph{string
  junctions}  \cite{Sjostrand:2002ip} has so far only been applied to
baryon-number-violating new-physics processes  and to the 
description of beam remnants (and then acts to increase baryon
stopping  \cite{Sjostrand:2004pf}). 

This brings us to the next step of the algorithm, 
assignment of the produced quarks within hadron
multiplets. 
The fragmenting quark (antiquark) may combine with the antiquark
(quark) from a newly created breakup to produce either a vector or
a pseudoscalar meson, or, if diquarks are involved, either a spin-1/2
or spin-3/2 baryon. Unfortunately, the string model is  entirely
unpredictive in this respect, and this is therefore the sector that
contains the largest amount of free parameters. 
 From spin counting alone, one would expect the ratio $V/S$ of
vectors to pseudoscalars to be 3, but in practice this is only
approximately true for $B^*/B$. For lighter flavours, the difference
in phase space caused by the $V$--$S$ mass splittings 
implies a suppression of vector production. Thus, for $D^*/D$, the effective 
ratio is already reduced to about $\sim$ 1.0~--~2.0, while for $K^*/K$
and $\rho/\pi$,  extracted values range from 0.3~--~1.0. Recall, as 
always, that these are production ratios of \emph{primary hadrons},
hence feed-down complicates the extraction of these parameters from
experimental data, in  
particular for the lighter hadron species.  
The production of higher meson resonances is assumed to be low in a string
framework\footnote{The four $L = 1$ multiplets are implemented in \Py, 
but are disabled by default, largely because several states are poorly
known and thus may result in a worse overall description when
included.}. For diquarks, separate parameters control the relative
rates of spin-1 diquarks vs.\ spin-0 ones and, likewise, have to
extracted from data, with resulting values of order $(qq)_1/(qq)_0
\sim$ 0.075~--~0.15. 

With $\pt^2$ and $m^2$ now fixed, the final step is to select the
fraction, $z$, of the fragmenting endpoint quark's longitudinal
momentum that is carried by the created hadron. In this respect, the
string picture is substantially more predictive than for the flavour
selection. Firstly, the requirement  that the
fragmentation be independent of the sequence in which breakups are
considered (causality) imposes a ``left-right symmetry'' on the
possible form of the fragmentation function, $f(z)$, 
with the solution 
\begin{equation}
f(z) \propto \frac{1}{z} (1-z)^a \exp\left(-\frac{b\,(m_h^2 +
  \pt[h]^2)}{z}\right)~, \label{eq:lund-symm}
\end{equation}
which is known as the Lund symmetric fragmentation function
(normalized to unit integral). 
As a by-product, the probability distribution in invariant time $\tau$
of $q'\bar{q}′$ breakup vertices, or equivalently $\Gamma = (\kappa
\tau)^2$, is also obtained, with $\mathrm{d}P/\mathrm{d}\Gamma \propto
\Gamma^a \exp(-b\Gamma)$ implying an area law for the colour flux, and
the average breakup time lying along a 
hyperbola of constant invariant time $\tau_0 \sim
10^{-23}\mathrm{s}$ \cite{Andersson:1998tv}. 
The $a$ and $b$ parameters are the only free
parameters of the fragmentation function, though $a$ may in principle
be flavour-dependent. Note that the explicit mass dependence in $f(z)$ 
implies a harder fragmentation function for heavier hadrons (in the
rest frame of the string). 

\begin{figure}[t]
\begin{center}
\includegraphics*[scale=0.65]{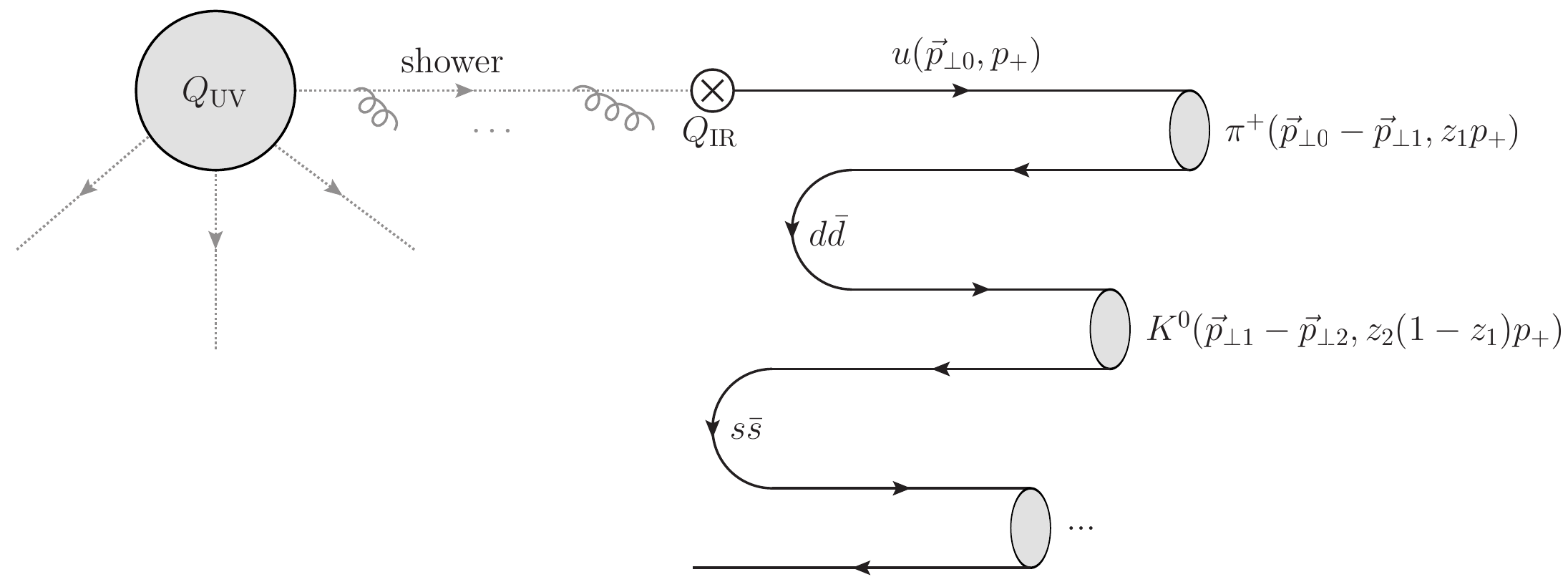}
\caption{Illustration of the iterative selection of flavours and
  momenta in the Lund string fragmentation model. \label{fig:iterative}}
\end{center}
\end{figure}

The iterative selection of flavours, \pt,
and $z$ values is illustrated in \figRef{fig:iterative}.
A parton produced in a hard process at some high scale $Q_{\mathrm{UV}}$ emerges from
  the parton shower, at the hadronization scale $Q_{\mathrm{IR}}$, 
  with 3-momentum $\vec{p}=(\vec{p}_{\perp 0},p_+)$, where the ``+'' on the third
  component denotes ``light-cone'' momentum, $p_\pm = E\pm p_z$. 
  Next, an adjacent $d\bar{d}$ pair from the
  vacuum is created, with relative transverse momenta $\pm\pt[1]$. 
  The fragmenting quark combines with the $\bar{d}$ from the breakup
  to form a $\pi^+$, which carries off a fraction $z_1$ of the total
  lightcone momentum $p_+$. The next hadron carries off a fraction
  $z_2$ of the remaining momentum, etc. 

For massive endpoints (e.g., $c$ and $b$ quarks, or hypothetical hadronizing
new-physics particles, generally called $R$-hadrons), which do not
move along straight lightcone sections, the exponential suppression
with string area leads to modifications of the form
\cite{Bowler:1981sb}, $f(z) \to f(z)/z^{b\,m_Q^2}$, with $m_Q$ 
the mass of the heavy quark. Strictly speaking, this is 
the only fragmentation function that is consistent with causality in
the string model, though a few alternative forms are typically
provided as well. 

Note, however, that the term \emph{fragmentation
  function} in the context of non-perturbative hadronization models is
used to denote only the corrections originating from scales below the
infrared cutoff scale of the parton shower. 
That is, the fragmentation functions introduced here 
 are defined at an intrinsically low scale of order
$Q\sim 1\,$GeV. It would therefore be highly inconsistent and
misleading to compare them directly to those that are 
used in fixed-order and/or analytical-resummation contexts, which are
typically defined at a factorization scale of order the scale of the
hard process.

\subsection{Multiple Parton Interactions \label{sec:mc-ue}}

In Monte Carlo modeling contexts, \emph{multiple parton interactions} (MPI)
denote the possibility of having 
multiple partonic $2\to 2$ interactions occurring within a single
hadron-hadron collision. The most striking and easily
identifiable consequence of MPI
is thus arguably the possibility of observing several distinct (i.e.,
hard) parton-parton interactions in some fraction of hadron-hadron events. 
Additional jet pairs produced in this way are sometimes
referred to as ``minijets'', but in the interest
of maintaining a compact terminology, we shall here just call them MPI
jets. The main distinguishing feature of such jets is that they tend to form 
back-to-back pairs, with little total \pt. For comparison, jets from 
bremsstrahlung tend to be aligned with the direction of their
``parent'' partons. The fraction of multiple interactions that give
rise to additional reconstructible jets is, however, quite small (how
small depends on the exact jet definition used). 
Additional soft interactions, below the
jet cutoff, are much more plentiful, and can give significant
corrections to the colour flow and total scattered energy 
of the event. 
This affects the final-state activity in a more global way, increasing
multiplicity and summed $E_T$ distributions, and contributing to the
break-up of the beam remnant in the forward direction. 

The first detailed Monte Carlo model for 
perturbative MPI was proposed by Sj\"ostrand and van Zijl in 
\cite{Sjostrand:1987su}, and with some variation 
this still forms the basis for most modern
implementations. Here, we therefore focus on that model and on its more
recent ``interleaved'' version \cite{Sjostrand:2004ef}. 
Some discussion of alternative models as well as additional 
references to the history and development of the subject of  multiple
interactions can be found in \cite{Buckley:2011ms}.

An intuitive way of arriving at the idea of multiple interactions
is to view hadrons simply as `bunches' of incoming
partons. No physical law then  prevents several distinct pairs of partons
from undergoing scattering processes within one and the same hadron-hadron
collision.  The other key idea to bear in mind is that the exchanged QCD 
particles are coloured, and hence such multiple interactions --- even when soft --- 
can cause non-trivial changes to the
colour topology of the colliding system as a whole, with potentially
major consequences for the particle multiplicity in the final state. 

To begin to construct a model for this, we first observe that, at low \pt{}, 
$t$-channel propagators almost go on shell (reminiscent of the case of
bremsstrahlung, described in detail in 
 \secRef{sec:parton-showers}), which causes the differential QCD
parton-parton scattering cross sections (such as the Rutherford one
illustrated in \secRef{sec:pQCD}) 
to become very large, behaving roughly as:  
\begin{equation}
\mrm{d}\hat{\sigma}_{2\to 2} \propto \ 
\frac{\mrm{d}\hat{t} }{\hat{t}^2} \ \sim \ 
 \frac{\drm{{\pthat}^2}}{{\pthat}^4}~, \label{eq:dpt4}
\end{equation}

\begin{figure}
\begin{center}
\vspace*{-30mm}\includegraphics*[scale=0.35]{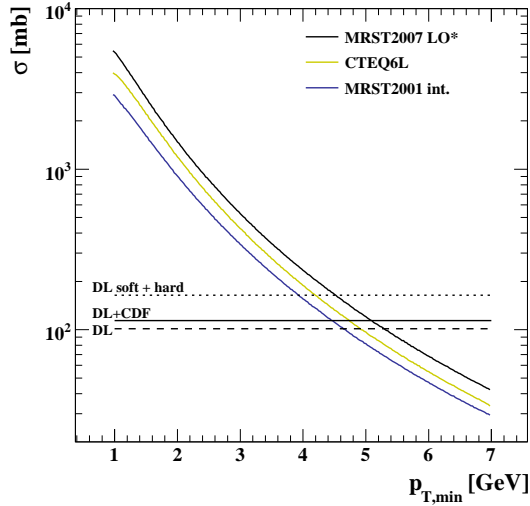}
\caption{The inclusive jet cross section calculated at LO for three
different proton PDFs, compared to various extrapolations
of the non-perturbative fits to the total pp
cross section at 14 TeV centre-of-mass
energy. From \cite{Bahr:2008wk}. \label{fig:sigma2to2}}
\end{center}
\end{figure}
An integration of this cross section from a lower cutoff
$\ptmin$ to $\sqrt{s}$, \
using the full (leading-order) QCD $2\to 2$ matrix elements 
folded with some recent parton-density sets, is
shown in \figRef{fig:sigma2to2}, for $pp$ collisions at 14 TeV
\cite{Bahr:2008wk}.  
The solid curves, representing the calculated 
cross sections as functions of $\ptmin$, are 
compared to a few different predictions for $\sigma_{\mrm{tot}}$ 
(the total $pp$ cross section \cite{Donnachie:1992ny}), 
shown as horizontal lines with  different dashing styles on the same
plot. Physically, the jet cross section can of course not 
exceed the total $pp$ one, yet this is what appears to be happening 
at scales of order 4--5 GeV in \figRef{fig:sigma2to2}. 
How to interpret this behaviour?

Recall that the interaction cross section is an inclusive
number. Thus, if a single hadron-hadron event contains two
parton-parton interactions, it will count
twice in $\sigma_{2\to2}$ but only once in $\sigma_{\mrm{tot}}$,
and so on for higher parton-parton interaction numbers.
 In the limit that all the
individual parton-parton interactions are independent and equivalent
(to be improved on below), we have 
\begin{equation}
\sigma_{2\to2}(\ptmin) = \langle n \rangle(\ptmin)
\ \sigma_{\mrm{tot}} ~,
\end{equation}
with $ \langle n \rangle(\ptmin)$ giving the 
average of a Poisson distribution in the number of parton-parton interactions
above $\ptmin$ per hadron-hadron collision,
\begin{equation}
\mathcal{P}_n(\ptmin) = \left[\langle n
  \rangle(\ptmin)\right]^n \frac{\exp\left[ - \langle n
    \rangle(\ptmin)\right]}{n!}~,
\label{eq:Poisson}
\end{equation}
and that number may well be above unity. 
This simple argument in fact expresses unitarity;
instead of the total interaction cross
section diverging as $\ptmin\to 0$ (which would violate
unitarity), we have restated the problem so
that it is now the \emph{number of interactions per collision} that
diverges.

Two important ingredients remain to be introduced in order to fully
regulate the remaining divergence. Firstly, the interactions cannot use up more
momentum than is available in the parent hadron. This 
will suppress the large-$n$ tail of the na\"ive estimate
above. Obviously, exact momentum conservation is included in all Monte
Carlo models currently on the market, although the details vary
somewhat from model to model. 
In the \Py-based models
\cite{Sjostrand:1987su,Sjostrand:2004ef,Corke:2009tk}, 
the multiple interactions are ordered in $\pt{}$, and the parton
distributions for each successive 
interaction are explicitly constructed so
that the sum of $x$ fractions can never be greater than unity. In
the \Hw models \cite{Butterworth:1996zw,Bahr:2009ek}, instead the
uncorrelated estimate of  
$\langle n \rangle$ above is used directly as an initial guess, but the actual
generation of interactions stop once the energy-momentum conservation
limit is exceeded (with the last ``offending'' interaction also
removed from consideration).  

The second ingredient suppressing the number of interactions,
at low \pt{} and $x$, is colour screening; if 
the wavelength $\sim$ $1/\pt{}$ of an exchanged coloured parton 
becomes larger than a typical colour-anticolour separation distance, 
it will only see an \emph{average} colour charge that vanishes in the limit
$p_{\perp} \to 0$, hence leading to suppressed interactions. 
This screening effectively provides an infrared cutoff for
MPI similar to that provided by the hadronization
scale for parton showers. 
A first estimate of an effective lower cutoff due to colour screening would be
the proton size 
\begin{equation}
\ptmin \simeq \frac{\hbar}{r_{p}} \approx
\frac{0.2~\mrm{GeV}\cdot\mrm{fm}}{0.7~\mrm{fm}} \approx
0.3~\mrm{GeV} \simeq \Lambda_{\mrm{QCD}} ~, \label{eq:cutoff}
\end{equation}
but empirically this appears to be too low. In current models, one
replaces the proton radius $r_{p}$ in the above formula by a ``typical
colour screening distance'' $d$, i.e.\ an average size of a region within
which the net compensation of a given colour  
charge occurs. This number is not known from first principles, so
effectively this is simply a cutoff parameter, which can then just as
well be put in transverse momentum space. 
The simplest choice is to introduce a step function
$\Theta(\pt{} - \ptmin)$, such that the perturbative cross section
completely vanishes below the $\ptmin$ scale.
Alternatively, one may note that the jet cross section is divergent
like $\alpha_s^2(\pt^2)/\pt^4$, cf.~eq.~(\ref{eq:dpt4}), 
and that therefore a factor
\begin{equation}
\frac{\alpha_s^2(\pt[0]^2 + \pt^2)}{\alpha_s^2(\pt^2)} \,
\frac{\pt^4}{(\pt[0]^2 + \pt^2)^2}
\label{eq:ptzerodampen}
\end{equation}
would smoothly regularize the divergences, now with $\pt[0]$ as the
free parameter to be tuned to data. Regardless of whether it is 
imposed as a smooth (\Py\ and \Sh) or steep (\Hw++) function, this is
one of the main ``tuning'' parameters in such models. Note also that 
this parameter does not have to be energy-independent. 
Higher energies imply that parton densities can be probed at smaller
$x$ values, where the number of partons rapidly increases. Partons
then become closer packed and the colour screening distance $d$
decreases. The uncertainty on the
energy and/or $x$ scaling of the cutoff is a major concern when
extrapolating between different collider energies.

We now turn to the origin of the so-called ``pedestal
effect'', the observational fact that hard jets appear to sit on top
of a higher ``pedestal'' of underlying activity than events with no
hard jets. This is interpreted as a consequence of
impact-parameter-dependence, as follows. In peripheral
collisions, only a small fraction of events contain any
high-\pt\ activity, whereas central collisions are more
likely to contain at least one hard scattering; a high-\pt\ triggered
sample will therefore be biased towards small impact parameters. 
The ability of a model to describe the shape of the pedestal (e.g., to
describe both ``minimum-bias'' data and underlying-event distributions
simultaneously) therefore depends upon its modeling of the
impact-parameter dependence, and correspondingly the impact-parameter
shape constitutes another main tuning parameter for models that
include this dependence.  

For each impact parameter, $b$, the number of interactions $\tilde{n}$
can then still be assumed to be distributed according to a Poissonian,
eq.~(\ref{eq:Poisson}), again modulo momentum conservation, but now
with the mean value of the Poisson distribution depending on impact
parameter, $\langle \tilde{n}(b)\rangle$. If the matter
distribution has a tail to infinity (as, e.g., Gaussians do),
one may nominally obtain events with arbitrarily large $b$ values. 
In order to obtain finite total cross sections, it is therefore 
necessary to give a separate interpretation to the ``zero bin'' of the
Poisson distribution, which corresponds to no-interaction events. In
models that attempt to describe the entire inelastic non-diffractive
cross section, this bin is simply ignored, since the events in it can
only represent elastic or diffractive scatterings, which are modeled
separately. Alternatively, in  models that pertain only to 
\emph{hard} inelastic events, it can be reinterpreted as containing
that fraction of the total inelastic cross section which do not
contain any hard interactions. 
  
Finally, we should mention that there are 
two perturbative modeling aspects which go
beyond the introduction of MPI themselves. In particular, this concerns 
\begin{enumerate}
\item Parton showers off the MPI. 
\item Perturbative parton-rescattering effects.
\end{enumerate}

Without showers, MPI models would generate very sharp peaks for back-to-back
MPI jets, caused by unshowered partons passed directly to the hadronization
model. However, with the exception of the oldest \Py 6 model
\cite{Sjostrand:1987su}, all of the general-purpose event-generator
models do include such showers, and hence should exhibit more
realistic (i.e., broader and 
more decorrelated) MPI jets. 
On the initial state side of the MPI shower issue, the main questions
are whether and how correlated multi-parton densities
are taken into account, and, as discussed
previously, how the showers are regulated at low \pt{} and/or low $x$. 
Although none of the MC models currently 
impose a rigorous correlated multi-parton evolution, all of them
include some elementary aspects. The most significant for
parton-level results is arguably momentum conservation, which is
enforced explicitly in all the models. The so-called ``interleaved'' models
\cite{Sjostrand:2004pf,Sjostrand:2004ef} attempt to go a step
further, generating an explicitly correlated multi-parton evolution
in which flavour sum rules can be imposed to conserve, e.g.,
the total numbers of valence and sea quarks across interaction chains.  

\emph{Perturbative rescattering} in the final state occurs if 
partons are allowed to undergo several distinct interactions, with
showering activity possibly taking place inbetween. This has so far
not been studied extensively, but a first 
fairly complete model and exploratory study has been presented in the
context of \Py 8 \cite{Corke:2009tk}. 
In the initial state, parton rescattering effects have so far not 
been included in any of the general-purpose Monte Carlo models.
 
\subsection{Colour (Re)-Connections and Beam Remnants\label{sec:mbmpi-np}}

Consider now a hadron-hadron collision, i.e., including MPI, 
at the parton level, equivalent to a resolution scale of about one
GeV. The system of coloured partons emerging 
from the short-distance phase (primary parton-parton interaction plus 
parton-level underlying event plus beam-remnant partons)
must now undergo the transition to colourless hadrons. 
Infrared sensitive observables, such as individual
hadron multiplicities and spectra are crucially dependent on the
parton-parton correlations in colour space, and on the properties and
parameters of the hadronization model used.  
Here, we concentrate on the specific issues connected with the
structure of the event in colour space.  

Keeping the short-distance parts unchanged, the colour structure
\emph{inside} each of the MPI systems is normally still described using 
just the ordinary leading-colour matrix-element and parton-shower
machinery described in  \secsRef{sec:fixed-order} and
\ref{sec:parton-showers}. The crucial question, in the context of MPI, 
is then how colour is neutralized \emph{between} different MPI
systems, including also the remnants. Since these systems can lie at
very different rapidities (the extreme case being the two opposite
beam remnants), the strings spanned between them can have
very large invariant masses (though normally low \pt), 
and give rise to large amounts of (soft)
particle production. Indeed, in the context of soft-inclusive physics,
it is precisely these ``inter-system'' strings which furnish the
dominant particle production mechanism, and hence their modeling 
is an essential part of the infrared physics description.

As discussed more fully in \cite{Buckley:2011ms}, there is a large
amount of ambiguity concerning how to address this, and a substantial
amount of variation between current models. 
Experimental investigations of colour reconnections at LEP 
 \cite{Abbiendi:1998jb,Abbiendi:2005es,Schael:2006ns,
Abdallah:2006ve} were only able to exclude some fairly extreme models,
 with comparatively moderate ones still allowed. 
Furthermore, in hadron collisions the initial state contains soft
colour fields with wavelengths of 
order the confinement scale. The presence of such fields,
unconstrained by LEP measurements, could impact in a
non-trivial way the process of colour  neutralization 
\cite{Buchmuller:1995qa,Edin:1995gi}. And finally, 
the MPI produce an additional amount of displaced colour charges,
translating to a larger density of hadronizing systems. It is not
known to what extent the collective hadronization of such a system
differs from a simple sum of independent systems. 
  
A new generation of colour-reconnection toy models have therefore been
developed specifically with soft-inclusive and underlying-event
physics in mind \cite{Sandhoff:2005jh,Skands:2007zg,Skands:2010ak},
and also the cluster-based \cite{Webber:1997iw} and Generalized-Area-Law
\cite{Rathsman:1998tp} models have been revisited in that context. 
Although still quite crude, these models do appear to be able to
describe significant features of the Tevatron and LHC data, such as
the $\langle \pt \rangle (N_{\mrm{ch}})$ distribution in minimum-bias
data, which appears to be quite sensitive to this effect. 
It is nonetheless clear 
that the details of the full fragmentation process in hadron-hadron
collisions are still far from completely understood.

\subsection{Tuning}

The main virtue of general-purpose Monte Carlo event generators
is their ability to
provide a complete and fully differential picture of collider final
states, down to the level of individual particles. This  
allows them to be used as detailed --- albeit approximate --- 
theoretical references for measurements performed at accelerators like the LHC, 
against which models of both known and `new' physics can be tested. 
As has been emphasized in these lectures, 
the achievable accuracy depends both on the
inclusiveness of the chosen observable and on the  
sophistication of the simulation itself. An important driver for the
latter is obviously the development of improved theoretical models,
e.g., by including matching to higher-order matrix elements, more
accurate resummations, or better non-perturbative models,  as
discussed in the previous sections; but it  
also depends crucially on the available constraints on the remaining
free parameters of the model. Using existing data to constrain these is
referred to as generator tuning.  

Although Monte Carlo models may appear to have a
bewildering array of independently adjustable parameters, it is worth
keeping at the front of one's mind that most  of these parameters only control
relatively small (exclusive) details of the event generation. The majority of the
(inclusive) physics is determined by only a few, very important ones, 
such as, e.g., the value of the strong coupling, in the perturbative
domain, and the form of the fragmentation function for massless
partons, in the non-perturbative one.

Armed with a good understanding of the underlying model, an expert would
therefore normally take a highly factorized approach to constraining
the parameters, 
first constraining the perturbative ones and thereafter the
non-perturbative ones, each ordered in a measure of their relative
significance to the overall modeling. This factorization, and carefully chosen
experimental distributions corresponding to each step,  allows one 
to concentrate on just a few parameters and distributions 
at a time, reducing the full parameter space to manageable-sized
chunks. Still, each step will often involve more than one single
parameter, and non-factorizable 
corrections still imply that changes made in 
subsequent steps can change the agreement obtained in previous ones 
by a non-negligible amount, requiring additional iterations from
the beginning to properly tune the entire generator framework. 

Recent years have seen the emergence
of automated tools that attempt to reduce the amount of both computer
and manpower required for this task, for instance 
by making full generator runs only for a
limited set of parameter points, and then interpolating between
these  to obtain approximations to what the true generator result
would have been for any intermediate parameter point, as, e.g., in the
Professor tool \cite{Buckley:2009vk,Buckley:2009bj}. 
Automating the human expert input is of course more difficult. 
In the tools currently on the market,
this question is addressed by a combination of input solicited from
the generator authors (e.g., which parameters and ranges to consider,
which observables constitute a complete set, etc)
and the elaborate construction of non-trivial weighting
functions that determine how much weight is assigned to each 
individual bin and to each distribution. The field is still
burgeoning, however, and future sophistications are to be
expected. Nevertheless, at this point the overall quality of the tunes
obtained with automated methods appear to at least be competitive with
the manual ones. 

A sketch of a reasonably complete tuning procedure, without going into
details about the parameters that control each of these sectors in
individual Monte Carlo models, would be the following:

{\bf 1) Keep in mind} that inabilities of models to
 describe data is a vital part of the feedback cycle between
 theory and experiment. Also keep in mind that
 perturbation theory at LO$\times$LL is doing \emph{very well} if it gets
 within 10\% of a given IR safe measurement. An agreement of 5\% should be
 considered the absolute sanity limit, beyond which it does not make
 any sense whatsoever to tune further. The advent of NLO Monte
 Carlos may reduce these numbers slightly, but only for quantities for which
 one expects NLO precision to hold. However, the sanity limit should be taken
 to be at least twice as large for quantities governed by
 non-perturbative physics. For some quantities, e.g., ones for which
 the underlying modeling is \emph{known} to be poor, an order-of-magnitude
  agreement or worse may have to be accepted. Attempting to force 
 Monte Carlo models to describe data far outside their domains of
 validity must be expected to produce similar side effects as
 attempting to turn a Fiat into a Ferrari merely by cranking up the
 engine revolutions.
 
\textbf{2) Final-State Radiation and Hadronization:} 
 mainly using LEP and other $e^+e^-$ collider data. On the IR safe
 side, there are event shapes and jet observables, the latter
 including rates, resolutions, masses, shapes, and jet-jet
 correlations. On the IR sensitive side, special attention
 should be paid to the high-$z$ tail of the fragmentation spectra,
 where a single hadron carries a large fraction of an entire jet's
 momentum, since this is the tail that is most likely to give ``fake
 jets''. Depending on the focus of the tuning, attention should also
 be paid to identified-particle rates and ratios, and to fragmentation
 in events containing heavy quarks and/or gluon jets. 
 Usually, more weight is given to
 those particles that are most copiously produced, though 
 this again depends on the focus. Finally, particle-particle
 correlations and baryon production are typically some of the least
 well constrained components of the overall modeling. The scaling
 properties of IR safe vs.\ IR sensitive contributions can be
 tested by comparing data at several different $e^+e^-$ collider
 energies.  
 
\textbf{3) Initial-State Radiation, and so-called
  ``Primordial\footnote{Primordial $k_T$: an
  additional soft \pt\ component that is injected on top of the
  \pt\ generated by the initial-state shower itself, see
  \cite[Section 7.1]{Buckley:2011ms}.} $k_T$'':} here, one would in principle like to use
  data from DIS reactions, which are less complicated to interpret
  than full hadron-hadron collisions. However, due to difficulties in
  translating between the $ep$ and $pp$ environments, this is normally
  \emph{not} what is done in practice. Instead, the main constraining
  distribution is the dilepton \pt distribution in Drell-Yan events in
  hadron-hadron collisions. For any observables containing explicit
  jets, be aware that the underlying event can produce small 
  horizontal shifts in jet \pt\ distributions, which may in turn 
  result in seemingly larger-than-expected 
  vertical changes if the distributions are falling sharply. Also note
  that the ISR evolution is sensitive to the choice of PDFs, with
  caveats as discussed in \secRef{sec:pdfs}.

\textbf{4) Initial-Final Connections:} e.g., radiation from
  colour lines connected to the initial state and 
  jet broadening in hadron collider environments. This is one of the
  most poorly controlled parts of most MC models. Keep in mind that it
  is \emph{not} directly constrained by pure final-state observables,
  such as LEP fragmentation, nor by pure initial-state observables,
  such as the Drell-Yan \pt\ spectrum, which is why we list it as a
  separate item here. In principle, DIS would again be a prime
  territory for placing constraints on this aspect at least for quark
  jets, but in practice
  more often inclusive-jet and other multi-jet processes (such as
  $W/Z+\,$jets) in hadron colliders are used.

\textbf{5) Underlying Event:} Good constraints on the overall level of the
  underlying event can be obtained by counting the summed transverse
  energy (more IR safe) and/or particle multiplicities and average
  transverse momenta (more IR sensitive) in regions \emph{transverse}
  to a hard trigger jet (more IR safe) or particle (more IR
  sensitive). Constraints on the \emph{fluctuations} of the underlying
  event are also important, and can be obtained, e.g., by comparing to
  measurements of the RMS of such distributions. Again, note
  that the UE is sensitive to the choice of PDFs.

\textbf{6) Colour (Re-)Connections and other Final-State Interactions:}  
 By Final-State Interactions,
  we intend a broad spectrum of possible collective effects that may
  be included to a greater or lesser extent in various models. These
  effects include: Bose-Einstein correlations, colour reconnections,
  hydrodynamics, string interactions, Cronin effect, etc. 
  As a rule, these effects
  are non-perturbative and hence should not modify IR safe observables
  appreciably. They can, 
  however, have \emph{drastic} effects on IR sensitive ones, such as
  particle multiplicities, and particle momentum distributions,
  wherefore useful constraints are typically furnished by 
  particle-particle correlations, by measurements of 
  particle momentum spectra as functions of quantities believed to
  serve as indicators of the strength of these phenomena 
  (such as event multiplicity), and/or by collective-flow-type
  measurements.  
  Finally, if the model includes a universal description of
  underlying event and soft-inclusive QCD, as many MPI-based models do, then
  minimum-bias data can also be used as a control sample, though one must
  then be careful either to address diffractive contributions properly
  or to include only data samples that minimize their impact. 

\textbf{7) Beam Remnants:} Constraints on beam
  remnant fragmentation are most easily obtained in the forward
  region, but, e.g., the amount of baryon transport from the remnant
  to a given rapidity region can also be used to probe how much the colour
  structure of the remnant was effectively disturbed, with more baryon
  transport indicating a larger amount of ``beam baryon blowup''.

We round off by emphasizing that comparisons of specific
models and tunes to data can be useful both as immediate 
tests of commonly used models, and to illustrate the current amount of
theoretical uncertainty surrounding a particular
distribution. Independently of how well the models fit the data, such
comparisons also provide a set of well-defined theoretical reference 
curves that serve as useful guidelines for future
studies. However, the conclusions that can be drawn 
from comparisons of individual tunes of specific models on single
distributions are necessarily limited. In order to obtain more general
conclusions, a strategy for a more coherent and over-arching look at both the data
and the models was recently proposed in \cite{Schulz:2011qy}. 
Specifically, rather than performing one global tune to all the
data, as is usually done,  a more systematic check on the validity of
the underlying physics model can be obtained by instead performing
several independent 
optimizations of the model parameters for a range of different 
phase space windows and/or collider environments.
In regions in which consistent parameter sets are obtained, with 
predictions that are acceptably close to the data,
the underlying 
model can then be considered as interpolating well, i.e., it is universal. 
If not, a breakdown in the ability of the model ability to span different
physical regimes has been identified, and can be addressed, with the
nature of the deviations giving clues as to the nature of the breakdown. 
With the advent of automated tools making it easier to run several
optimizations without much additional computing overhead, 
such systematic studies are now becoming feasible, with a first
example given in \cite{Schulz:2011qy}.

\section*{Acknowledgments}
Thanks to L.~Hartgring, J.~Lopez-Villarejo, P.~Nason, G.~Salam, and T.~Sj\"ostrand
whose valuable comments and sharing of insight contributed to these lectures. 
In addition, material from the ESHEP lectures by Mangano \cite{Ellis:2009zzb}, by
Salam \cite{Salam:2010zt,Grojean:2010ab}, by Sj\"ostrand \cite{Sjostrand:2006su}, 
and by Stirling \cite{Ellis:2008zzf}, as well as the recent review on
Monte Carlo event generators by the MCnet collaboration \cite{Buckley:2011ms} 
has been used in compiling these lectures. 
This work was supported in part by the Marie Curie research training
network ``MCnet'' (contract number MRTN-CT-2006-035606). 

\bibliography{eshep-skands}

\end{document}